\documentclass{aa}
\usepackage[T1]{fontenc}
\usepackage{amsmath}
\usepackage{amssymb}
\usepackage{txfonts}
\usepackage{multirow}
\usepackage{booktabs}
\usepackage{threeparttable}
\usepackage{tabularx}
\usepackage{graphicx}
\usepackage{caption}
\usepackage{subcaption}
\usepackage{float}
\usepackage{xcolor}
\usepackage{natbib,twoopt}
\bibpunct{(}{)}{;}{a}{}{,}
\usepackage[breaklinks=true]{hyperref}
\usepackage{xspace}
\usepackage{appendix}
\usepackage{comment}
\usepackage{listings}

\begin{document}

   \title{Time domain astrophysics with transient sources}

   \subtitle{Delay estimate via Cross Correlation Function techniques}

     \author{W. Leone \inst{1,3,4}
          \and
          L. Burderi\inst{2}
          \and
          T. di Salvo\inst{3}
           \and
           A. Anitra\inst{2,3}
           \and
          A. Sanna\inst{2}
          \and
          A. Riggio\inst{2}
          \and
          R. Iaria\inst{3}
          \and
          F. Fiore\inst{4}
          \and
           F. Longo\inst{5}
           \and
           M. {\v D}ur{\' i}{\v s}kov{\' a}  \inst{6}
           \and
           A. Tsvetkova \inst{2,7,8}
           \and
           C. Maraventano \inst{3}
           \and
           C. Miceli \inst{3,9,10}
          }

   \institute{Department of Physics, University of Trento, Via Sommarive 14, 38122 Povo(TN), Italy 
              \email{wladimiro.leone@unitn.it}
              \and
             Dipartimento di Fisica, Universit\'a degli Studi di Cagliari, SP Monserrato-Sestu, km 0.7, I-09042 Monserrato, Italy 
             \and
             Università degli Studi di Palermo, Dipartimento di Fisica e Chimica, via Archirafi 36 - 90123 Palermo, Italy 
             \and
             INAF—Osservatorio Astronomico di Trieste, Via G.B. Tiepolo 11, I–34143 Trieste, Italy 
             \and
             Istituto Nazionale di Fisica Nucleare (INFN), sezione di Trieste, Trieste, Italy
             \and
             Department of Theoretical Physics and Astrophysics, Faculty of Science, Masaryk University, Brno, Czech Republic
             \and
             Ioffe Institute, Politekhnicheskaya 26, 194021 St. Petersburg, Russia
             \and
             INAF -- Osservatorio di Astrofisica e Scienza dello Spazio di Bologna, Via Piero Gobetti 93/3, 40129 Bologna, Italy
             \and
             INAF/IASF Palermo, via Ugo La Malfa 153, I-90146 Palermo, Italy
             \and
             IRAP, Universitè de Toulouse, CNRS, UPS, CNES, 9, avenue du Colonel Roche BP 44346 F-31028 Toulouse, Cedex 4, France
                }

   \date{Accepted, 25 June 2025}

   \abstract{
    
The timing analysis of transient events allows for investigating numerous still open areas of modern astrophysics. The article explores all the mathematical and physical tools required to estimate delays and associated errors between two Times of Arrival (ToA) lists, by exploiting Cross Correlation Function (CCF) techniques.

The CCF permits the establishment of the delay between two observed signals and is defined on two continuous functions. A detector does not directly measure the intensity of the electromagnetic signal (interacting with its material) but rather detects each photon ToA through a probabilistic process. Since the CCF is defined on continuous functions, the crucial step is to obtain a continuous rate curve from a list of ToA. This step is treated in the article and the constructed rate functions are light curves that are continuous functions. This allows, in principle, the estimation of delays with any desired resolution.

Due to the statistical nature of the measurement process, two independent detections of the same signal yield different photon times. Consequently, light curves derived from these lists differ due to Poisson fluctuations, leading the CCF between them to fluctuate around the true theoretical delay. This article describes a Monte Carlo technique that enables reliable delay estimation by providing a robust measure of the uncertainties induced by Poissonian fluctuations. GRB data are considered as they offer optimal test cases for the proposed techniques.

The developed techniques provide a significant computational advantage and are useful analysis of data  characterized by low-count statistics (i.e., low photon count rates in c/s), as they allow overcoming the limitations associated with traditional fixed bin-size methods.
}

   \maketitle

\section{Introduction}

The delay estimate plays a pivotal role in several fields of modern astrophysics. We roughly characterize two types of delays: "spectral" lags and "temporal" delays.
 
Spectral lags might be present when observing a source in different energy bands. Several factors can lead to the formation of delays between light curves obtained by two detectors in such conditions. In the Gamma Ray Burst (GRB) case, emission mechanisms can drive such effect, spanning a range from tenths of seconds to even tens of seconds \citep{Giuliani_2008,Frontera_2000,Tsvetkova_2017}. Some quantum gravity theories predict that spectral lags depend on a dispersion law for light in vacuo \citep{Camelia98,Piran}. Delays can also be estimated between the continuum and ionized line-emission (e.g., Mg II line) light curves of bright sources such as AGNs. That allows for probing the AGN geometry and the spatial extent of the accretion disk via reverberation mapping techniques as in \cite{2019Zajacek....340..577Z}. The topic of spectral lags is thoroughly discussed in the following paper Leone W. et al., in preparation.
 
GRBs are short, intense, and unrepeatable flashes of radiation, with a spectral energy distribution peaking in the gamma-ray band \citep{2015JHEAp...7...73D}. The theoretical isotropic energy released can reach up to $10^{55}$ erg \citep{Wu_2012,Dado_2022}, over a period ranging from fractions of a second up to several thousand seconds \citep{von_Kienlin_2020}. However, jet-like emission in GRBs can reduce the required energy budget to produce the observed brightness by at least a factor of 100, as the energy is directed narrowly rather than spread isotopically \citep{1999ApJ...519L..17S}. 
The GRB generation is associated with the gravitational collapse of a massive star \citep{PIRAN1999575,CampanaS2008} or the coalescence of two neutron stars in an extremely close binary system because of the emission of gravitational waves. Indeed the GRB 170817A event \citep{Savchenko_2017,Goldstein_2017} is believed to be an example of the latter option. 
Another hypothesis proposes that the merger of a black hole and a neutron star could serve as a GRB progenitor. However, no direct observational evidence currently supports this possibility \citep{Mochkovitch1993,Gompertz_2020}.

The simultaneous detection of a GRB and an emission of gravitational waves from these events marked the beginning of multi-messenger astrophysics   \citep{M_sz_ros_2019}. However, this remains the first and only GRB whose detection is associated with a Gravitational Wave (GW) counterpart \citep{Abbott_2017} up to now.
 
An all-sky X-ray monitor working in tandem with a sensitive GW detector is crucial for maximizing the probability of observing multi-messenger events like GRB 170817A \citep{Ghirlanda_2024}. Precise localization of the X-ray source is essential to associate the electromagnetic event with its gravitational wave counterpart accurately. Additionally, an accurate position for such events (in coincidence with gravitational wave detection) enables targeted searches for counterparts in lower energy bands, such as optical or IR, where a large number of observable objects and an eventual low spatial resolution make precise localization even more critical.
 
HERMES-TP/SP (High Energy Rapid Modular Ensemble of Satellites Technologic and Scientific Pathfinder) is a 3U nano-satellites project based on the distributed architecture concept mission \citep{Fiore_2020}. The six-unit formation is designed for monitoring and localizing high energy transient events via triangulation method \citep{Hurley_2013,Sanna_2020,burderi2021grailquesthermeshunting}.
 
Temporal delays are crucial for triangulating the position of transient events, which is the purpose of the HERMES-SP mission. Such delays arise between detectors located at different positions in space while observing the same event.

The accurate and rapid localization of the events is key to a rapid and effective follow-up of the source by another in-orbit or ground-based instrument along several energy bands.

This article covers the physical and mathematical tools that enable the estimation of this type of delay between two Time of Arrival (ToA) lists.

\section{The Cross Correlation method}
Electromagnetic waves transport (at the speed of light) an amount of energy per square centimeter per second (flux $\phi$) along the propagation direction. Plane waves in vacuo are related to an energy flow whose intensity is equal to the modulus of the pointing vector \Vec{s}: 

\begin{equation}
    {|\Vec{\rm{S}}| = \rm{\phi} =  \frac{\rm{E}^2}{4\pi} \rm{c}} .
    \label{pointing_vec}
\end{equation}

It is quite evident that a transitory phenomenon is characterized by a variable flux during the occurrence of a transient source. 

Let's consider a series of theoretically identical detectors that are positioned on the wavefront of \autoref{pointing_vec}. Each detector measures the same intensity at the same time.

On the other hand, if the detectors are displaced in space on arbitrary positions each detector measures the same intensity at a delayed time $\tau$ which is equal to the scalar product of the line of propagation and the vector connecting the positions of the detectors, divided by the speed of light.

By measuring $\tau$ we can deduce the projected distances along the line of propagation and therefore, determine the direction of the wave. At least three detectors are required to determine the direction of propagation, from geometrical considerations. This can be intuitively understood by considering that three points are sufficient to define a plane in space and, consequently, its perpendicular direction \citep{Sanna_2020}. This method is well known as the temporal triangulation technique (see, i.e., \cite{2013ApJS..207...39H})  and is so based on the experimental determination of time delays between signals observed by different detectors.

Delays can be obtained by cross-correlating two light curves (the product of $\phi$ and the instrument's effective area projected along the line of sight) obtained from detectors' photons lists. To perform the Cross Correlation Function (CCF), a continuous function $f(t)$ must be derived from each ToA list.

Once two continuous f(t) are obtained for a couple of detectors (1 and 2) the delay can be computed by the CCF:

\begin{equation}
\label{CCF_int}
\rm{CCF}_{1,2}(\rm{\rm{\tau}})= \int_{-\infty}^{+\infty} \rm{f}_1(t) \rm{f}_2(t+\rm{\tau}) \mathrm{d} \rm{t}.
\end{equation}

The value of $\tau$  where $CCF_{1,2} (\rm{\tau})$ reaches its maximum, is the expected delay between the two light curves \citep{MIT_CCF}.
 
It is important to note that the detector does not directly measure the intensity of the observed signal, making the derivation of the light curve from a ToA list a non-trivial task.

\subsection{Statistic of Times of Arrival}

When using a counting device (detector) the energy is recorded discretely, as a list of ToA of photons (quanta of energy). If the wave is monochromatic (single-frequency $\nu$) each energy grain transports the same amount of energy $\rm E=\rm h\nu$. In the case of multi-frequency electromagnetic spectra, the same argument can be applied to the "average quanta":
\begin{equation}
    <\rm h\nu>=\frac{\int_{\rm \nu_{min}}^{\nu_{max}} h \nu\, f(\nu) \,d\nu}{\int_{\rm \nu_{min}}^{\rm \nu_{max}} f(\nu) d\nu}.
\end{equation}
 
Since we are detecting photons, we do not directly measure the variation of $\phi$ over time. Instead, we measure the ToA of photons associated with a given rate $\rm{r}(t)$, where $\rm{r}(t)$ represents the continuous rate at which photons are detected by the detector. The clear relation between $\phi$ and $r$ is:

\begin{equation}\label{flux}
    \rm{r}(t) = \begin{cases}
    \frac{\rm\phi(t)}{\rm h\nu} , & \text{multi-frequency spectra} \\
    \frac{\rm \phi(t)}{\rm \langle h\nu \rangle} , & \text{mono-frequency spectra}
    \end{cases}
\end{equation}

Following the theorem 5.2 in \cite{Park2018} and \autoref{Norm_Poiss_function} we derive the Normalized Poisson probability function associated to the detection of $N$ photons within a time interval $\rm{\Delta t}$, given a specific photon arrival rate $\rm{r}(t)$:

\begin{equation}
    \rm {Q_{N,\rm{\Delta t}}(r)= \rm{\Delta t} \frac{(\rm{r}\,\rm{\Delta t})^N \rm{e}^{-r \rm{\Delta t}}}{N!}}.
    \label{er:norm_poisson prob}
\end{equation}

\begin{figure}[h!]
    \centering
    \includegraphics[scale=0.33]{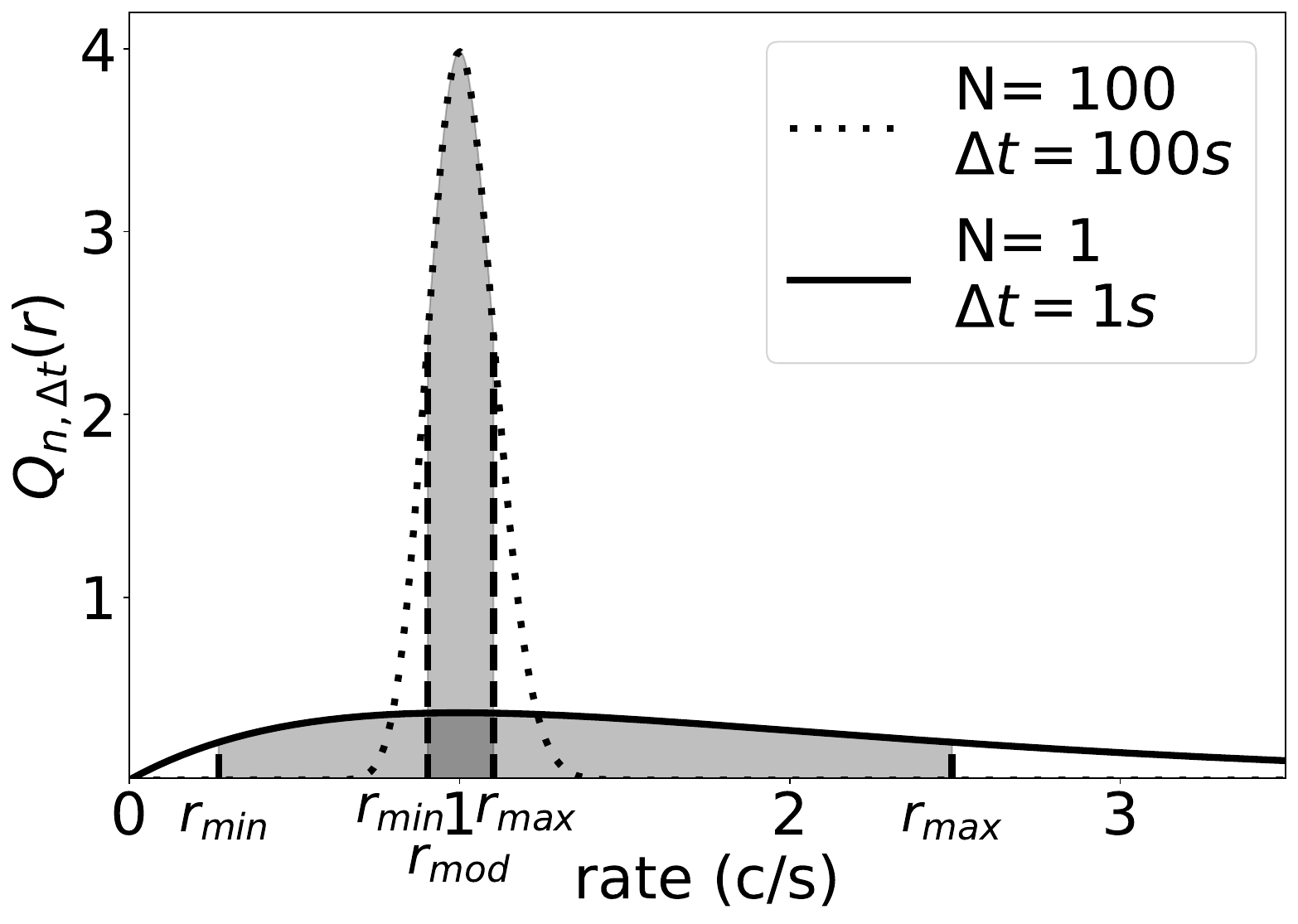}
    \caption{$Q_{N,\rm{\Delta t}}$ as a function of the rate r for N=1 (solid line) and n=100 (dotted line). The gray areas are obtained by fixing a CL=0.68 corresponding to 1 $\sigma$ CL of a Gaussian distribution.}
    \label{fig:Poisson_figure}
\end{figure}

Since the detection of N photons depends on a specific rate chosen among all possible rates, we determine the corresponding confidence interval for the rate at a given confidence level (CL), in accordance with the condition described in \autoref{Statistical Confidence Level}. As illustrated in \autoref{fig:Poisson_figure}, the same average rate (1 c/s) corresponds to a broader or narrower confidence interval depending on the number of observed counts. These two cases highlight how statistical regimes, defined by low or high count, influence the accuracy of an otherwise identical rate measurement.

As usual, we can define the mode (best value), average, and median of the distribution. The mode is given by$\frac{\rm{n}}{\rm{\Delta t}}$ (see \autoref{mean,mod,med}) and differs from the median and the mean (defined in \autoref{mean,mod,med}). We note that, for the case $n\rightarrow +\infty$ the mode, mean, and median converge to the same value.
 
In general $r_{min}$ and $r_{max}$ depend on the chosen CL and can be numerically evaluated using \autoref{condition 1} and \autoref{condition 2}, as in \autoref{app:conditions}.
\begin{figure}[h!]
    \centering
    \includegraphics[scale=0.29]{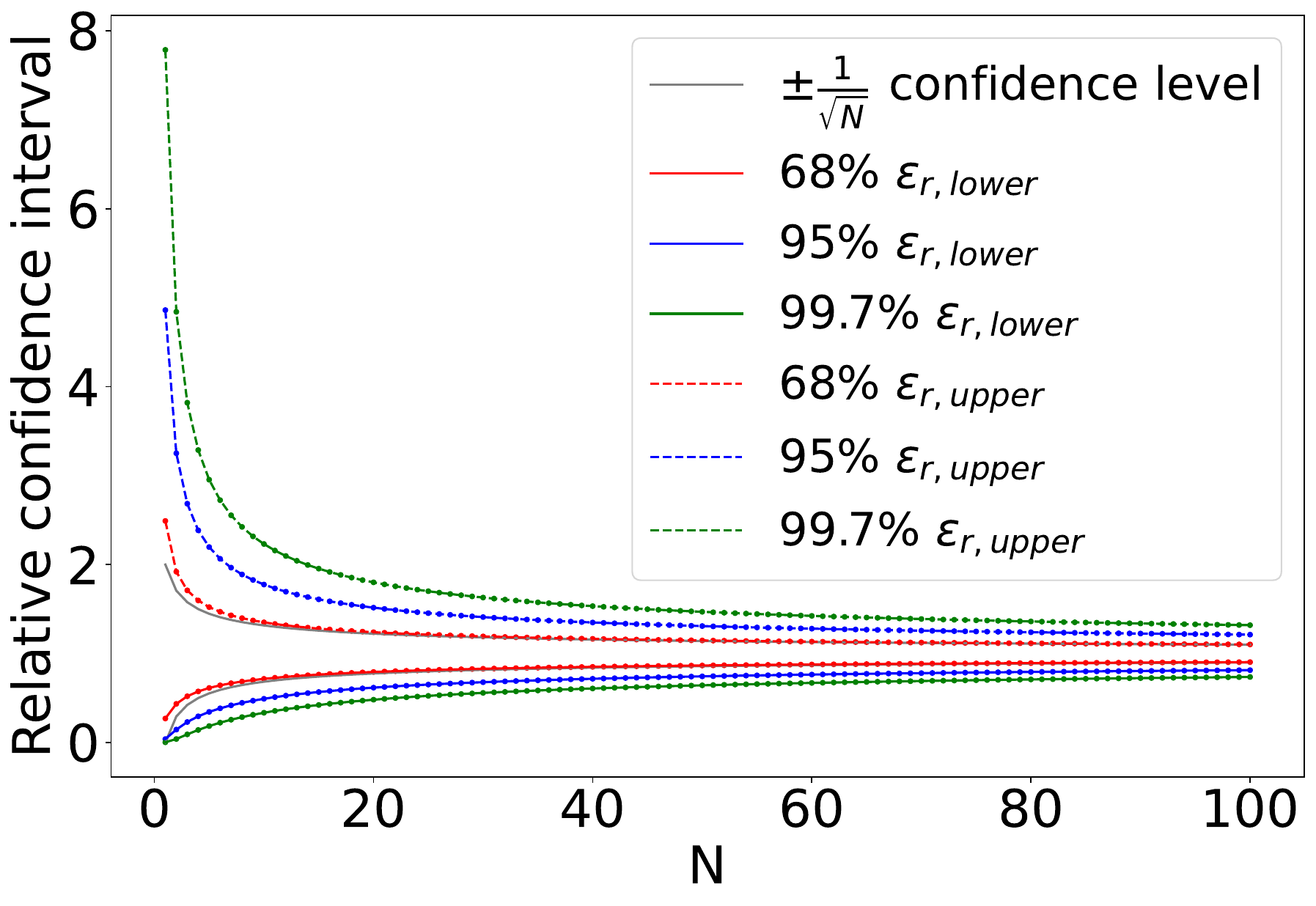}
    \caption{Relative errors (for different CLs) as a function of the observed number of events. The upper and lower limits are expressed in units of the mode $\rm{r_{mod}}={\rm{N}}/{\rm{\Delta t}}$ (see \autoref{mean,mod,med} for a detailed computation). The 1,2 and 3 $\sigma$ confidence levels are respectively associated to $68\%$, $95\%$ and $99.7\%$. The gray lines represent the $\pm{1}/{\sqrt{\rm{N}}}$ confidence level as a function of the observed number of events.}
    \label{fig:rel_rate_error}
\end{figure}

\autoref{fig:rel_rate_error} shows the relative upper and lower errors associated with several CLs. The gray lines in \autoref{fig:rel_rate_error} are associated with $1 \pm \frac{1}{\sqrt{\rm{N}}}$ symmetric Gaussian relative error, for a 1 sigma confidence level. It is clear then, for $N\gtrsim10$ the lower relative limit $\epsilon^- (N,\;1\sigma \;CL)$ and upper relative limit $\epsilon^+ (N,\;1\sigma \;CL)$ are approximately the same values of the Gaussian relative symmetric error (see, i.e., \autoref{tab:relative_err_comparison}), confirming what it has been discussed so far.

\subsection{How to built a light curve }\label{rate_build}

If $N$ photons are detected in a certain $\rm{\Delta t}$, the mode of the Poisson distribution, $\rm{r}= {\rm{N}}/{\rm{\Delta t}}$, is the most probable rate value in such interval. This value is an average rate associated with a confidence interval which depends on the chosen  CL and the number of photons $N$, as illustrated in \autoref{fig:rel_rate_error}. 
 
We can achieve a statistically uniform representation of a light curve if each rate point is derived by fixing the number of photons N. Examining a ToA list, N photons are measured during a $\rm{\Delta t}_i$:
\begin{equation}
    \rm{\Delta t}_i = \rm{t}_{i\cdot N-1} - \rm{t}_{(i-1)\cdot N},
\end{equation}
and the corresponding rate is:
\begin{equation}
    r_i=\frac{\rm{N}}{\rm{\Delta t}_i}.
\end{equation}

The time $\rm{t}_i$ associated with each rate point $r_i$ is the "barycenter" of the ToA in the time interval [$\rm{t}_{i\cdot N-1}$,$ \rm{t}_{(i-1)\cdot N}$] i.e.:
\begin{equation}
    \rm{t}_i=\frac{1}{\rm{N}}\sum_{k=(i-1)\cdot N}^{i\cdot N-1} \rm{t}_k.
\end{equation}

The relative confidence interval is [$\epsilon^-(N,CL)$ , $\epsilon^+ (N,CL)$] as shown in \autoref{fig:rel_rate_error} and in \autoref{app:conditions} and the absolute errors confidence interval is:

\begin{equation}
\begin{split}
    \rm r^-_{i} (\rm N,\rm{\Delta t})&=\rm \epsilon^- (N,CL) \cdot r_i,
 \\
    \rm r^+_{i} (\rm N,\rm{\Delta t})&= \rm \epsilon^+ \rm (N,CL) \cdot r_i.
\end{split}
\end{equation}
Once N is fixed this method guarantees the same relative accuracy for each estimated rate point.
 
We note that the rate $r_i$ depends on the $\rm{\Delta t}_i$ required to collect $N$ photons. By increasing the number of photons N we get smaller confidence intervals (see \autoref{fig:rel_rate_error}). For $\rm N>>1$ $\epsilon^-\approx\epsilon^+\approx\frac{1}{\sqrt{\rm{N}}}$. We note that increasing the accuracy of the rate measurement requires increasing $N$, which consequently reduces temporal resolution.
 
On the other hand, the number of detected photons increases with the detector's effective area ($A_{eff}$). Therefore, keeping a fixed $N$, an increase of the $A_{eff}$ allows us to explore smaller time scales with the same accuracy.
 
The method described above allows to obtain a continuous function by linearly connecting the rate points $\rm r_i(\rm{t}_i)$. The light curve obtained in this way is a continuous function of the generic variable $t$.

\begin{table}[h!]
\centering
\caption{Poisson vs.\ Gaussian $1\sigma$ relative confidence intervals}
\label{tab:relative_err_comparison}
\begin{tabular}{|c|c|c|c|c|}
\hline
$\rm N$ &
$\epsilon^-(N,\;1\sigma\;{\rm CL})$ &
$\epsilon^+(N,\;1\sigma\;{\rm CL})$ &
$1-\dfrac{1}{\sqrt{N}}$ &
$1+\dfrac{1}{\sqrt{N}}$ \\
\hline
 1   & 0.29 & 2.49 & 0     & 2    \\
 5   & 0.61 & 1.52 & 0.55  & 1.45 \\
 10  & 0.72 & 1.35 & 0.68  & 1.32 \\
 50  & 0.86 & 1.15 & 0.86  & 1.14 \\
 100 & 0.90 & 1.10 & 0.90  & 1.10 \\
\hline
\end{tabular}
\tablefoot{Comparison between the Poisson confidence interval
[$\epsilon^-(N,1\sigma),\,\epsilon^+(N,1\sigma)$] and the Gaussian
approximation $1\pm 1/\sqrt{N}$ as a function of the number of counts
$N$.}
\end{table}

\subsection{Light curve variability} \label{section:Light curve variability}

The variability of the light curve obtained by the method above is the result of three different phenomena: 
\begin{enumerate}
    \item The intrinsic variability of the {\it unknown} light curve that represents the genuine variability of the source.
    \item The variability induced by the detection Poissonian process. As shown in \autoref{rate_build} and in \autoref{tab:relative_err_comparison}, for $\rm N>>1$, the relative weight of this variability scales as the inverse of the square root of the total number N of photons adopted to build the light curve. In particular, in \autoref{tab:relative_err_comparison} is shown how
    for $\rm N=1$ the Gaussian and the [$\rm \epsilon^-(N,CL)$ , $\rm \epsilon^+ (N,CL)$] relative confidence intervals differs. Note that, for N=1, the asymptotic formula $\rm \epsilon^\pm_a(N)= 1 \pm \frac{1}{\rm \sqrt{\rm{N}}}$ overestimates the upper limit of the CL and underestimates the lower limit of the CL because of the intrinsic skewness of the Poisson distribution.
    \item The spurious variability introduced by the linear interpolation between rate points to generate a continuous light curve from a ToA list (see \autoref{rate_build}). This variability is independent of the chosen $N$. In any case, linear interpolation between rate points introduces the minimum possible spurious variability, as it minimizes the ‘necessary’ distance to connect each rate point to the next. 
    
\end{enumerate}

Therefore, we face a dilemma. On the one hand, we want to keep N (the number of photons used to build up each point of the light curve) as small as possible to exploit our detector's minimal temporal resolution for observing the shortest intrinsic variability of the light curve. On the other hand, we need a larger $N$ to minimize the variability induced by Poisson fluctuations.
For instance, N=10 represents a good compromise between achieving an approximately symmetric confidence interval, with reasonably small relative errors (around $30\%$), and exploring finer temporal resolutions.

\subsection{Cross Correlation Function}
\label{subsection:CCF}
Once the rate $\rm{r}(t)$ is obtained from a ToA list, we can perform CCF between two rates, $\rm r_1(t)$ and $\rm r_2(t)$ defined for the same time interval $\rm{t}_1<t<\rm{t}_2$:
\begin{equation}
\rm CCF_{1,2}(\rm \theta)= \int_{\rm{t}_1}^{\rm{t}_2} \rm r_1(t) r_2(\rm t+\rm \theta) \mathrm{d} \rm t
\end{equation}
The best fit maximum of $\rm CCF_{1,2}(\theta)$ corresponds to the best estimate of the delay $\rm{\tau}$ between the two light curves:
\begin{equation}
\rm{max}\{CCF_{1,2}(\theta)\}_{\rm{for}\;\rm{t}_1<\theta<\rm{t}_2} = \rm CCF_{1,2} (\rm{\tau})
\end{equation}
Since the CCFs are nearly symmetric functions (as in \autoref{fig:CCF_grid}), we adopt a Gaussian profile to estimate this parameter.

\begin{figure*}
    \centering
    \includegraphics[scale=0.12]{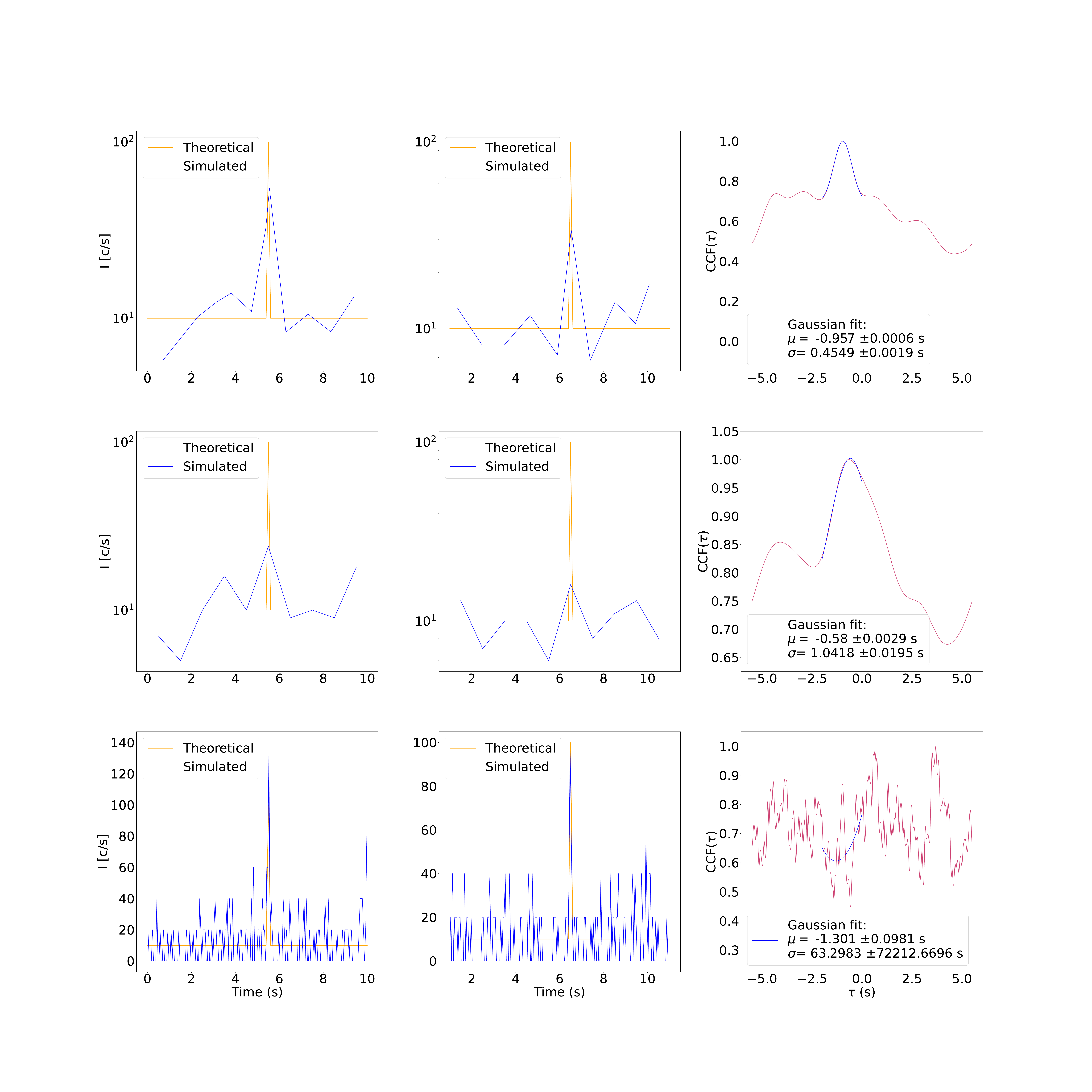}
    \caption{Top row: Light curves (left and central panels) and corresponding CCF (right panel) obtained using the adaptive rebinning method (considering 10 photons per bin). Middle and bottom rows: Same configuration, but using fixed bin sizes of 1 s and 0.05 s, respectively. In each row, the left panel shows in blue the simulated signal obtained by rebinning the ToA list generated by simulating the theoretical profile shown in orange. The central panel displays the same signal, delayed by 1 s before rebinning. The right panel presents the CCF between the two simulated light curves shown in the left and central panels of the corresponding row.}
    \label{fig:fixed_vs_variable_comparison}
\end{figure*}

\section{Comparing fixed and adaptive binning for CCF}

Several studies employ fixed bin-size light curves to estimate time lags using the CCF \citep{Sanna_2020} or the discrete correlation function \citep{Castignani_2014}. However, adaptive binning becomes particularly advantageous in low-count regimes. While the techniques presented here yield results practically indistinguishable from fixed bin-size methods when applied to high-count-rate signals (e.g., >10$^3$ c/s), their benefits become evident when dealing with sparser data.

\autoref{fig:fixed_vs_variable_comparison} illustrates how fixed bin-size techniques exhibit clear limitations in low-count scenarios, as commonly encountered in high-energy astrophysics.

Suppose we observe a signal with an average count rate of 10 cts/s, featuring a sharp spike lasting 0.1 s with a peak of 100 cts/s . As shown in Figure 4, fixed bin-size light curves may inadequately represent the source's temporal variability. Using this theoretical signal, we simulate ToA lists following the procedure detailed in the next section. The resulting ToA lists are simulated under the same conditions and resemble realistic observations from an X-ray detector.
If a fixed bin size of 1 s is chosen (much larger than the spike duration) the rebinning process effectively smooths out the intrinsic variability, rendering the spike invisible. Conversely, selecting a bin size of 0.05 s, comparable to the spike duration, allows the spike to be captured. However, this comes at the cost of introducing significant statistical noise: many bins contain only 0 or 1 photon, leading to substantial spurious variability unrelated to the original signal. In such cases, detecting a single photon corresponds to a 100\% uncertainty in the measured rate. These issues directly impact the reliability of CCF-based delay measurements.

To further demonstrate this, we inject a 1-second delay into the theoretical signal and simulate the corresponding ToA lists. In order to estimate the expected delay, a Gaussian fit is applied to each CCF in \autoref{fig:fixed_vs_variable_comparison}, centered at -1 s and spanning a width of 1 s. As clearly shown, only the adaptively binned light curves recover the expected delay. 

\section{Errors treatment}
\label{section:Errors treatment}
Due to the probabilistic nature of the process, when two identical detectors observe the same GRB, the obtained rates, $r_1$ and $r_2$, differ due to Poisson fluctuations. Consequently, the delay estimate $\tau$ generally differs from the expected value $\rm{\tau} = 0 \text{s}$ when cross-correlating light curves of the same event observed by two detectors positioned side by side. 

\begin{figure}[h!]
    \centering
    \includegraphics[scale=0.18]{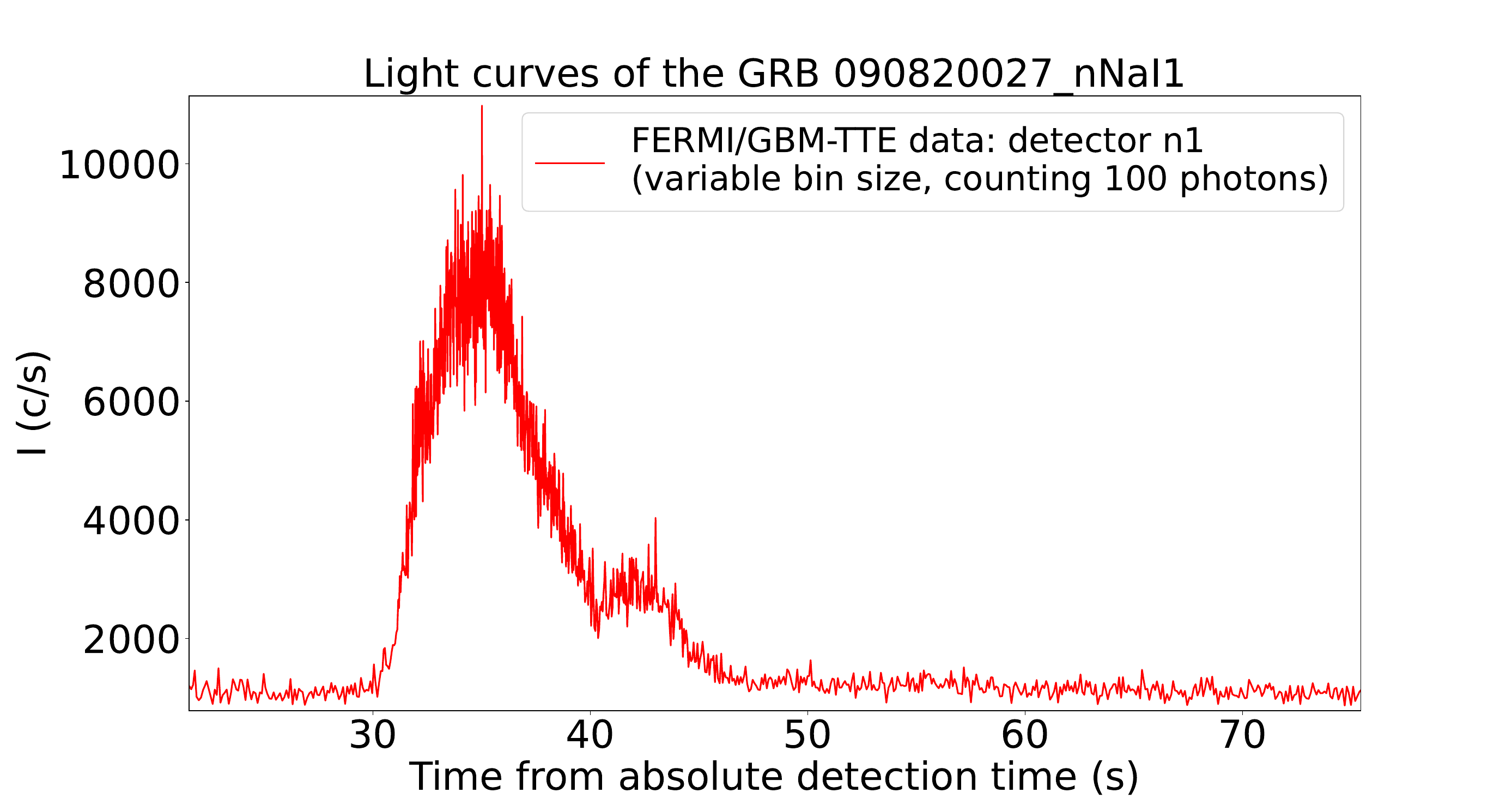}
    \includegraphics[scale=0.18]{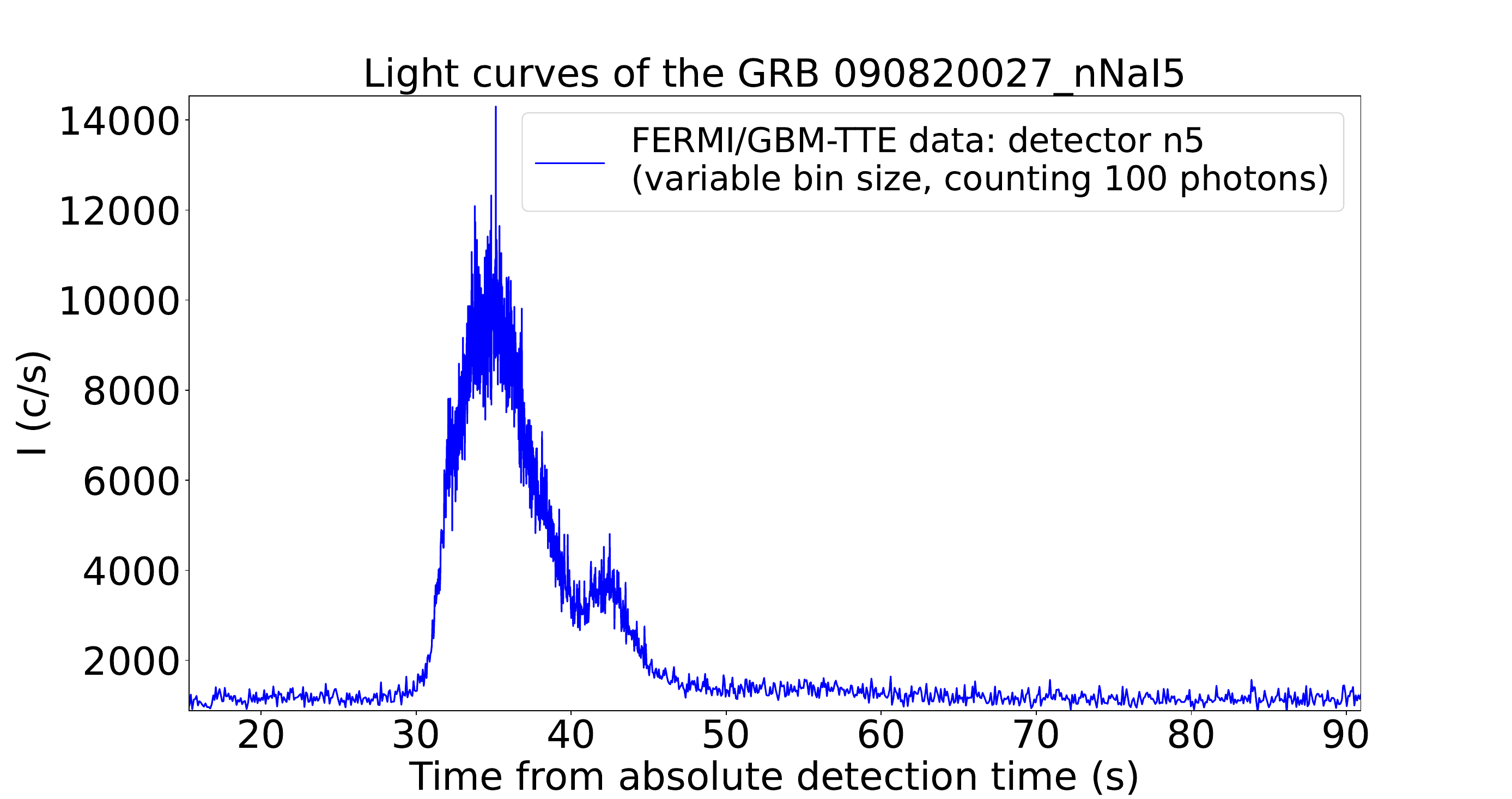}
    \caption{GRB 090820 light curves obtained by counting N=100 photons per bin. The upper plot is the n1 detector light curve and on the lower panel the n5 detector light curve.}
    \label{fig:GRB090820027_lc}
\end{figure}

For instance, we consider the ToA lists associated with the GRB 0908207 (see \autoref{fig:GRB090820027_lc}) as observed by two Fermi/GBM detectors (NaI detector 1 and NaI detector 5, 10 keV - 900 keV energy band). These are physically separated by a maximum distance of 5 m, corresponding to the diagonal of the almost cubic shape of the Fermi satellite, which implies a maximum theoretical delay of $\rm{\tau}_{th} = 15 \rm{ns}$ \citep{Bissaldi_2009,Meegan_2009}. We must also consider that the two detectors had different pointing directions at the time of the burst. Consequently, the observed photon count and respective rate vary depending on the off-axis angle relative to the source.
 
\begin{figure}[h!]
    \centering
    \includegraphics[scale=0.2]{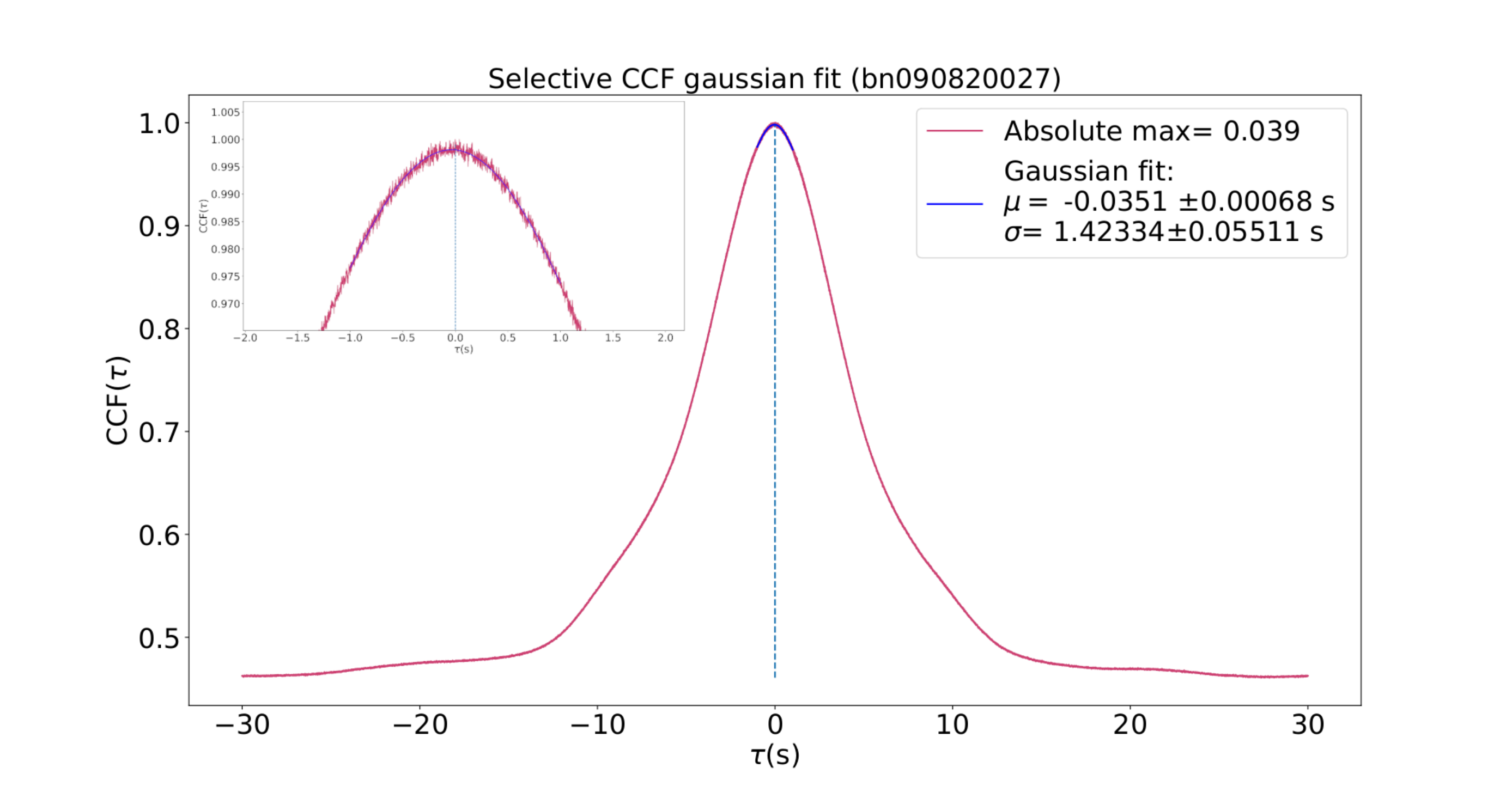}
    \caption{Upper plot shows the CCF between light curves from n1 and n5 detectors ToA lists,  with a variable bin size of 10 photons per bin and a $1\rm{\mu}\rm{s}$ resolution. The zoom in the lower panel shows the Gaussian fit centroid fluctuation concerning the vertical green dashed line that indicates the null theoretical delays.}
    \label{fig:CCF_n1_n5}
\end{figure}

On the other hand, we estimate the delay between the two detector rate curves, obtained with $\rm N =10$ and a sampling resolution of $1 \rm\mu s$. The CCF in \autoref{fig:CCF_n1_n5} is computed by using the procedure above, and the CCF upper region is fitted with a Gaussian profile over a 1-second baseline. The lag estimate $\rm{\tau}_{exp}= (-3.5 \pm 0.068) \times 10^{-2} \rm{s}$ corresponds to the Gaussian best-fit parameter $\mu$ and its associated error, as shown in \autoref{fig:CCF_n1_n5}.
It is important to note that the delay estimation result, based on the procedures discussed in the next section, is independent of the Gaussian profile's width.
 
Taken at face value, this result would imply that some unknown systematic effect has biased the measurement. The single CCF formally yields a significant lag with a minimal uncertainty, as it inherently captures the particular statistical fluctuation in the pair of detector measurements. This small uncertainty, however, is purely mathematical and pertains only to the statistical variation specific to that individual realization. Repeating the measurement under identical conditions would result in a different lag estimate due to random 
fluctuations.

Therefore, a Monte Carlo (MC) simulation approach is necessary to accurately estimate the overall uncertainty, incorporating the full range of possible statistical fluctuations \citep{Zhang_1999}.
The MC distribution of delays is centered around the best experimental estimate $\rm{\mu}$ and the associated error is the standard deviation of the distribution $\sigma$.

\subsection{The methods}

Standard MC methods are based on simulating light curves by the 'flux-randomization' process, as stressed by \cite{1998PASP..110..660P}. We explore two alternative methods for delay estimation: the Double Pool (DP) method revisits the concept of flux randomization, while the Modified Double Pool (MDP) method departs from this approach entirely, providing an experimental delay estimate without relying on simulations.

We note that we intentionally retain the background, particularly when it is comparable to the GRB signal, since background fluctuations significantly impact the observed variability. Background subtraction would artificially enhance statistical fluctuations, potentially causing random variations to be mistaken for genuine source variability. Thus, preserving the background allows us to evaluate variability under realistic observational conditions, avoiding the attribution of false significance to statistically insignificant features. The cross-correlation functions (CCFs) are computed over the T$_{90}$ and the background data intervals of 1.5 $\cdot$ T$_{90}$ before and after the T$_{90}$ interval. These intervals ensure that the resulting CCFs exhibit the characteristic "wings," thus enabling a correct interpretation of any potential physical delay.

\subsection{"Double Pool" Method - Light curves simulation}

We propose an alternative method that is conceptually consistent with the real detection process of a detector. This is based on the generalization of the inversion method in \cite{Klein1984ATP} for variable light curves. Instead of using flux randomization, the proposed method allows the generation of a simulated ToA list from a given rate curve \rm{r}(t) defined over an interval $\rm{t}_1<t<\rm{t}_2$:

\begin{equation}
    \int_{\rm{T\_SIM[N-1]}}^{\rm{T\_SIM[N]}} \rm{r (t') dt'} = -\rm{ln}\{ 1 - \rm{RND}(0,1) \},
    \label{generilzed eq TSIM}
\end{equation}

where RND(0,1) denotes a value drawn from a uniform distribution in the interval between 0 and 1. ${\rm {T\_SIM[N]}}$ is the ToA to be recursively simulated, starting with a previous time $\rm T\_SIM[N-1]$. In the first step, $\rm T\_SIM[N-1]$ is $\rm{t}_1$.
This approach emulates the detector measurement process, using a Poisson arrival process applied to the rate of the observed signal. 
 
\begin{figure}[h!]
    \centering
    \includegraphics[scale=0.2]{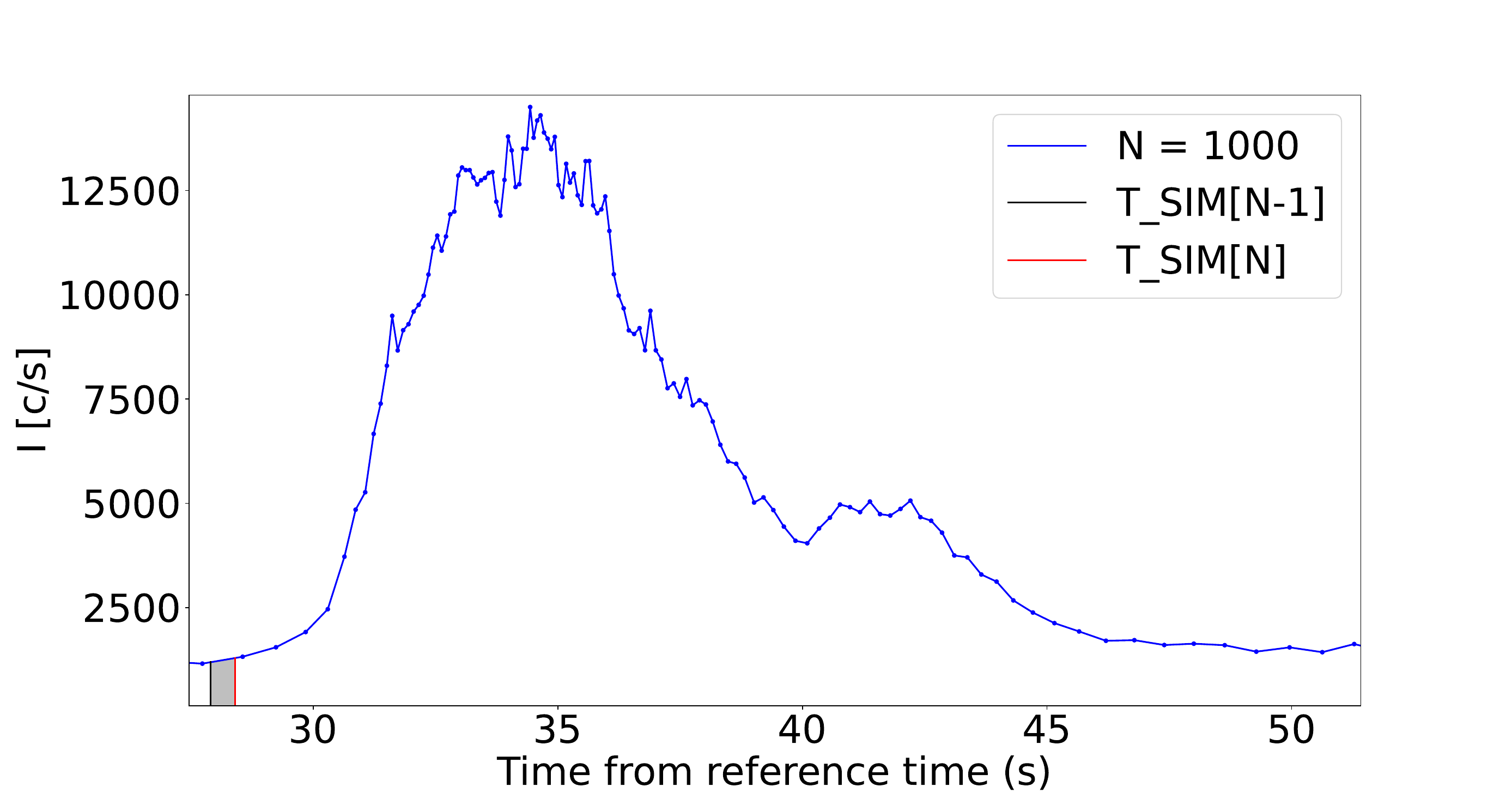}
    \caption{The luminosity curve displays a graphical example of the trapezoidal integral of \autoref{generilzed eq TSIM}. The black point represents a generic TOA that was just simulated in the previous step, and the red point represents the TOA to be simulated in the current step.}
    \label{lc_1000_integral}
\end{figure}

In the case of a constant rate $\rm{r}(t)=\rm \lambda$ the integral in \autoref{generilzed eq TSIM} is:
\begin{equation}\lambda \cdot(\rm T\_SIM[N]-\rm T\_SIM[N-1]) = -\rm ln\{ 1 - \rm RND(0,1) \},
\end{equation}
and each simulated time $T\_SIM[N]$ is:
\begin{equation} \rm T\_SIM[N] = \rm T\_SIM[N-1] - \frac{\rm ln\{ 1 - \rm RND(0,1) \}}{\rm \lambda}.
\end{equation}
As shown in the sketch of \autoref{lc_1000_integral}, the integral in \autoref{generilzed eq TSIM} is the area of the trapezoid between $\rm T\_SIM[N-1]$ and $\rm T\_SIM[N]$, under the given rate function. According to these considerations, the Poisson arrival process $\rm T\_SIM[N]$ can be analytically solved as in \autoref{GIM_solution}.

The net result is the generation of a ToA list, produced by a counter subject to Poissonian (quantum) fluctuations that observe the rate $\rm{r}(t)$.

\subsubsection{"Double Pool" - The method}

The DP method exploits this simulation technique to perform the required MC analysis between two detectors ToA lists.
In principle, one can use a light curve derived from a ToA list, as a "theoretical" template to generate a large number of simulated light curves. However, as discussed in \autoref{section:Light curve variability} our "theoretical" template is affected by Poisson fluctuations that depend on the chosen N. This implies that, when using as a template the light curve derived by a particular ToA list, Poisson fluctuations can not be distinguished from the genuine variability of the source.

This issue can be mitigated in several ways. When estimating the CCF between two detectors, an effective approach is to use the two ToA lists to build two different templates. Each of these templates is plagued with the Poisson fluctuations discussed above. However, since these fluctuations are randomly distributed around the true rate value, any correlation between the Poisson-induced variability between the two curves is washed out. For a fixed reasonable value of N (e.g., N=10) we use each of these two templates to generate several simulated ToA lists (Pool of ToA lists) by using the Inversion Method described above. 
We therefore perform CCF with the method described in \autoref{subsection:CCF} between a pair of ToA lists, each extracted only once from each of the two pools. In this way, Poisson fluctuations variability, imprinted on each template, does not correlate between the light curves and, therefore, will not bias the CCF result.
 
By generating a large number of couples of ToA, each belonging to one Pool, and cross-correlating the light curves obtained from them, we obtain a large number of delays with an approximately Gaussian distribution because of the central limit theorem. The mean of this distribution represents the expected delay, while the sigma indicates the uncertainty.

\subsection{Modified Double Pool method}

The MDP method allows for obtaining the required delay distribution, granting an exceptional computational time gain. No simulations are indeed required to obtain such distribution.
\label{section:MDP_method}
Let's consider a list of ToA obtained by a detector. This list can be splitted into two independent lists of ToA by calling an RND(0,1) for each ToA. The ToA belongs to one of the two lists depending on the exit of the RND(0,1). In particular for RND(0,1)<0.5 the ToA belongs to the first list, otherwise to the second.
Since the spatial position of the photon on the detector area is randomly distributed with a flat distribution over the entire detection area, this splitting procedure will furnish two ToA lists as obtained by two identical detectors observing the same GRB in the same spatial position, each with an effective area that is half the original one.
This means that cross-correlating the two light curves derived from these ToA lists will yield a temporal delay that fluctuates around the expected null value. These fluctuations are purely of statistical origin.
 
By repeating the splitting procedure with different random realizations two new ToA lists are obtained. It should be noted that this second couple is not fully statistically independent of the first one, as the original ToA list remains the same. However, this statistical dependence is weak as each point of the light curve is built using a large number of photons ($N\sim10$) and it does not significantly affect the computation of the sigma of the distribution, as demonstrated by numerical computation (see \autoref{HXMT_demonstration}).
 
Indeed, by averaging each rate point over $N$ photons, each resulting light curve within the same pool represents a distinct Poissonian realization, with each rate value and associated time being approximately statistically independent of any other realization.
 
By recursively applying this method we obtain a Pool of almost statistically independent ToA lists from the detector. We explicitly note that the splitting is necessary only to obtain statistically independent ToA lists and the fact that this method produces a couple on each step is irrelevant to the statistical independence of the final sample of ToA lists in the Pool.

Now consider a second detector observing the source. The procedure described above can be applied to obtain, also in this case, a Pool of almost statistically independent ToA lists.
 
\begin{figure}[h!]
    \centering
    \includegraphics[scale=0.35]{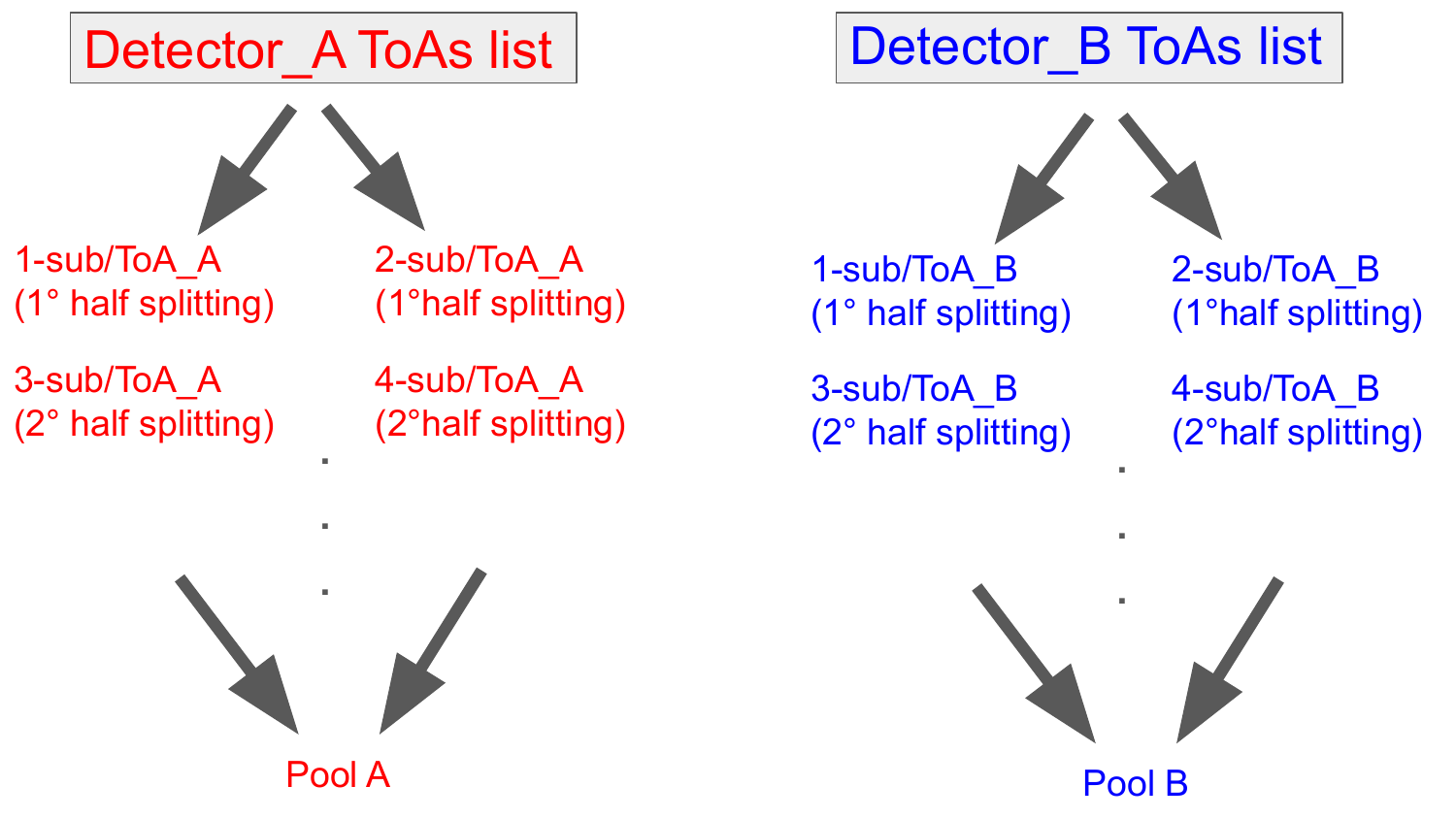}
    \caption{Scheme of the MDP splitting procedure.}
    \label{fig:MDP_schematic}
\end{figure}

We can now cross-correlate light curves each one extracted from each of the two pools as depicted in \autoref{fig:MDP_schematic}. The average value of this distribution is the expected delay and the sigma is the associated uncertainty.

\section{DP and MDP testing}
\label{section:testing}

GRBs are optimal candidates for testing the capability of the developed tools. We aim to obtain experimental delays that are compatible with the expected true delays. The GRB sample is randomly selected from bursts observed by Fermi/GBM, considering a broad range of fluence values. As shown in \autoref{fig:CCF_grid}, the discussed CCF techniques ensure a Gaussian profile under all flux and background conditions.
 
To this end, the test allows us to discriminate between the proposed procedures, defining the most effective one.
During the test, Gaussian fit guess parameters are fixed for both methods. The number of photons per bin, used to construct the light curves, is set to N=10.

\subsection{MDP and DP methods comparison}

To demonstrate the effectiveness of the two methods, we consider a representative sample of 20 GRBs observed by one Fermi/GBM detector. For each GRB, two independent ToA lists are generated by randomly splitting the considered Fermi/GBM detector data. This approximately corresponds to having at our disposal two ToA lists that have almost half the GBM detector effective area (e.g., as observed by two HERMES detectors). We apply MDP and DP methods to each pair of ToA lists. The expected theoretical temporal delay between the two ToA lists must be null (see the discussion in \autoref{section:Errors treatment}).

\autoref{fig:MDP_DP_comparison} compares the results of the two MC methods applied to the considered sample.  Both methods accurately estimate delays statistically compatible with the true null delay, considering the standard deviation of each distribution. The residual histograms are indeed compatible with zero.

As intuitively evident from \autoref{fig:MDP_DP_comparison} and also from \autoref{fig:CCF_grid}, the precision of the estimated delays decreases as the number of photons associated with the source diminishes relative to the background.
 
The Gaussian fit of the DP residual distribution shows that the DP method is probably less accurate in ensuring compatibility with the true delay. This discrepancy may stem from the simulation process used in the MC procedure. Specifically, the ToA lists from the two starting detectors are employed to define the initial templates for the DP method, which are then fixed during the MC simulations. Since templates remain fixed throughout the MC simulations, any injected Poisson variability may propagate through all the simulated light curves. 
This may result in an MC distribution influenced by the Poissonian variability of the initially generated templates. On the other hand, in the MDP method, the reshuffling of the ToA guarantees that no privileged light curves are considered.

We note that the MDP method effectively mitigates intrinsic Poisson fluctuations in the input templates used by standard flux randomization methods or the DP method. These fluctuations would otherwise propagate and amplify across all Monte Carlo realizations, with a stronger effect as the chosen number of photons per bin is lower. While the method requires halving the available data at each step, resulting in an average loss of precision of approximately $\sqrt{2}$, it prevents the artificial amplification of Poisson noise. Consequently, it removes eventual bias in delay estimation introduced by Poissonian fluctuations in the original templates.

\subsubsection{Proving that MDP method is fully independent}
\label{HXMT_demonstration}
We emphasize that each MDP step is statistically independent, even though the split ToA lists are always derived from the same set of events. Due to the random nature of the halving procedure, each generated light curve represents a specific Poisson realization of the true signal light curve. As a result, each delay constitutes an independent estimate, forming a delay distribution with the correct associated error, as shown in \autoref{fig:MDP_DP_comparison}.
 
\begin{figure}[h!]
    \centering
    \includegraphics[scale=0.25]{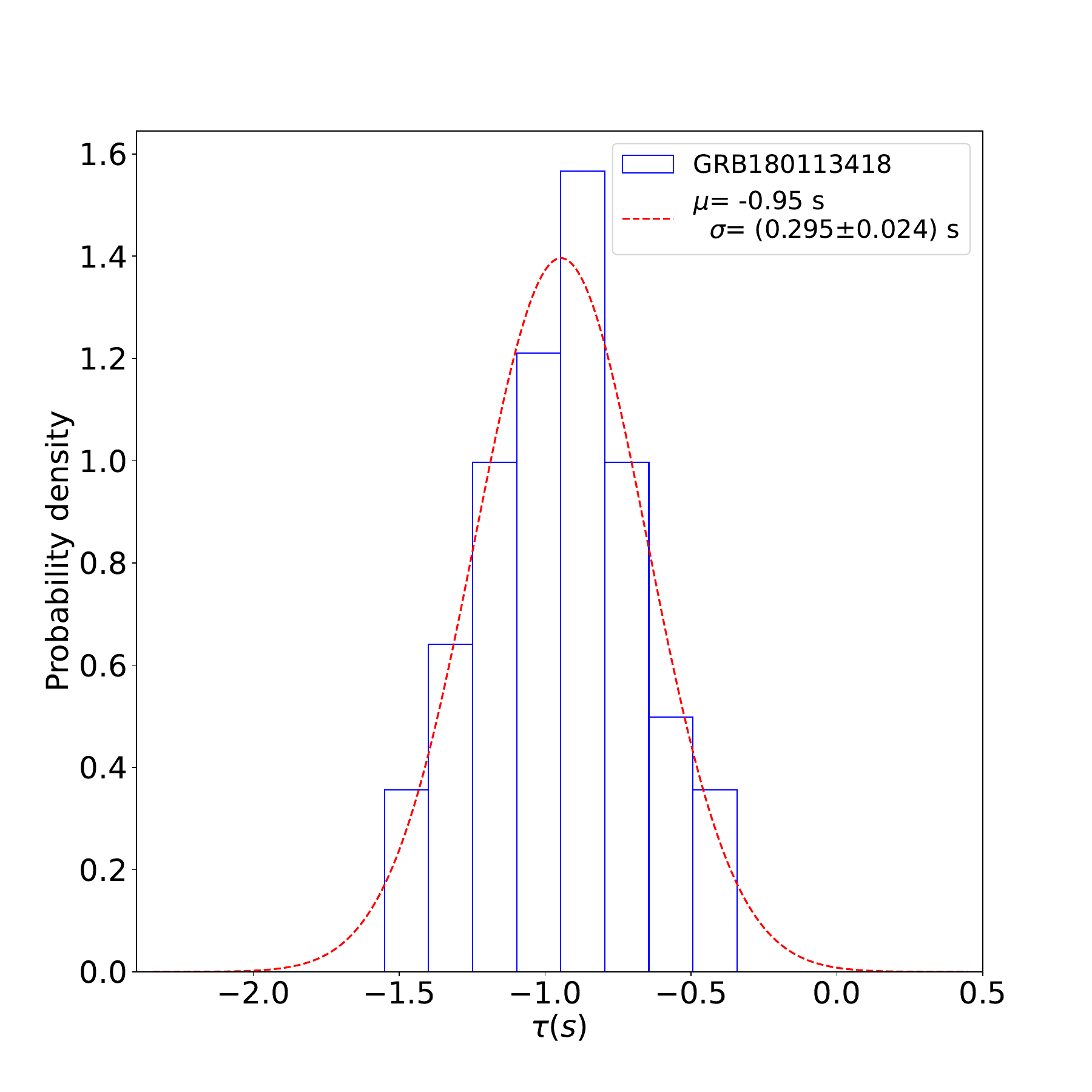}
    \caption{The delay distribution is obtained by cross-correlating 100 pairs of ToA lists, derived from the random division of the GRB 180113418 event file. A 1-second delay is injected into one of the ToA lists in each pair.}
    \label{fig:200_split}
\end{figure}

To demonstrate this, we use data from the Insight/HXMT instrument \citep{Insight, Insight1}, specifically focusing on GRB 180113C in the 1 keV - 600 keV energy band. With the instrument's effective area of approximately 2000 ${\rm cm^{2}}$, we can randomly split the initial ToA list into 200 independent sub-lists. These truly correspond to the lists obtained by 200 detectors that observe the same GRB under the same conditions  (effective area, detector response, attitude, off-axis angle) and spatial location. We inject a 1 s delay in 100 ToA and apply CCF techniques to estimate delays between the delayed and not-delayed groups. The resulting distribution of 100 values is shown in \autoref{fig:200_split}, with an average of $\rm{\mu}=-0.95s$ with an associated error of $\sigma=0.295s$.

This experiment remains conceptual, as 200 identical detectors observing the same source are not available. Typically, we want to estimate the delay between two instruments, so for this analysis, we randomly select two lists from the sample of 200. We again injected a 1 s delay in one of the two lists and the MDP method in \autoref{fig:MDP_schematic} is applied to estimate such delay.

\begin{figure}[h!]
    \centering
    \includegraphics[scale=0.25]{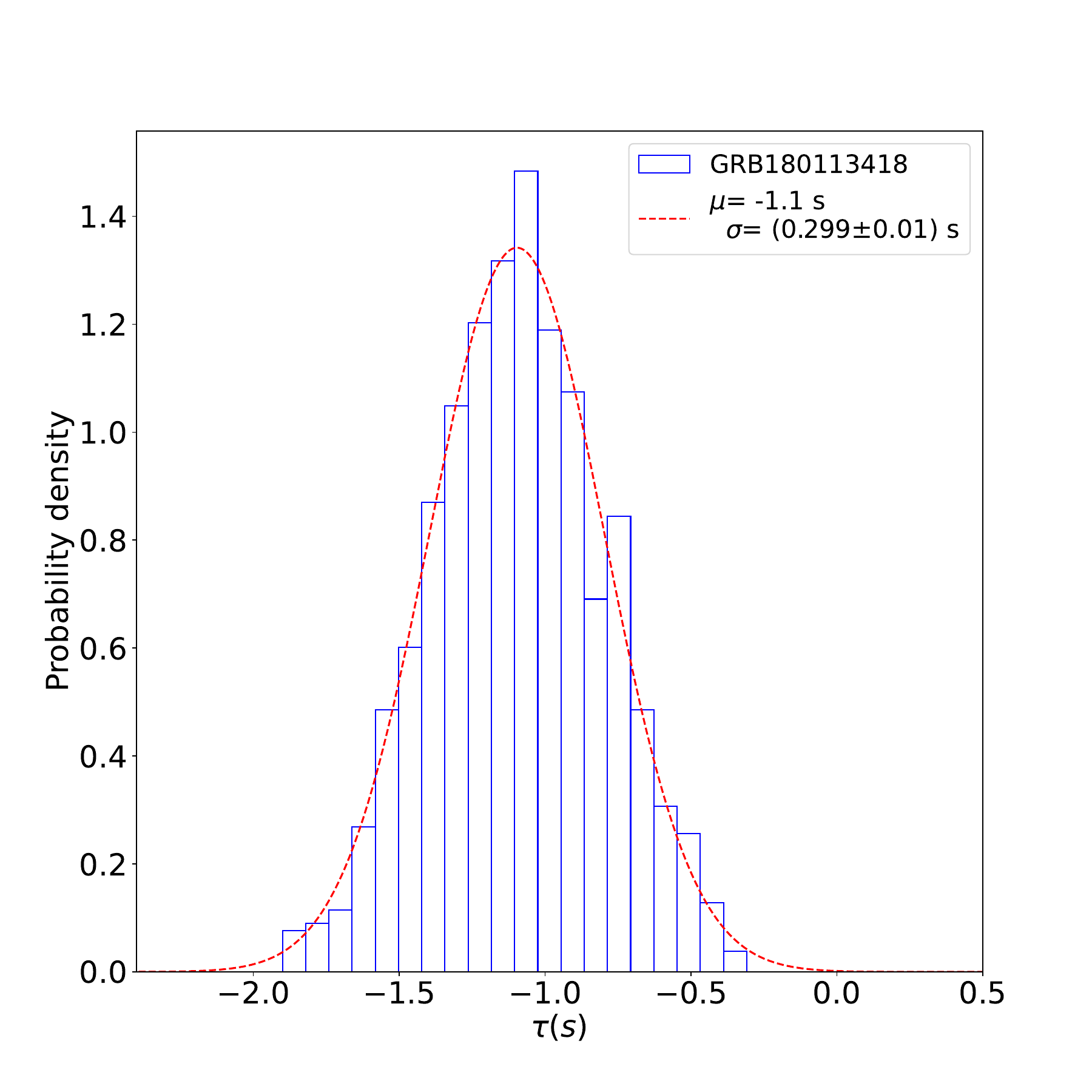}
    \caption{The delay distribution is obtained by applying the MDP method to two of the 200 ToA lists derived from the random division of the GRB 180113418 event file. A 1-second delay is introduced into one ToA list. The MDP procedure is carried out by randomly splitting the initial ToA lists 500 times, resulting in two Pools of 1000 light curves each.}
    \label{fig:MDP200_split}
\end{figure}

\begin{figure*}[t]
    \centering
    \includegraphics[scale=0.18]{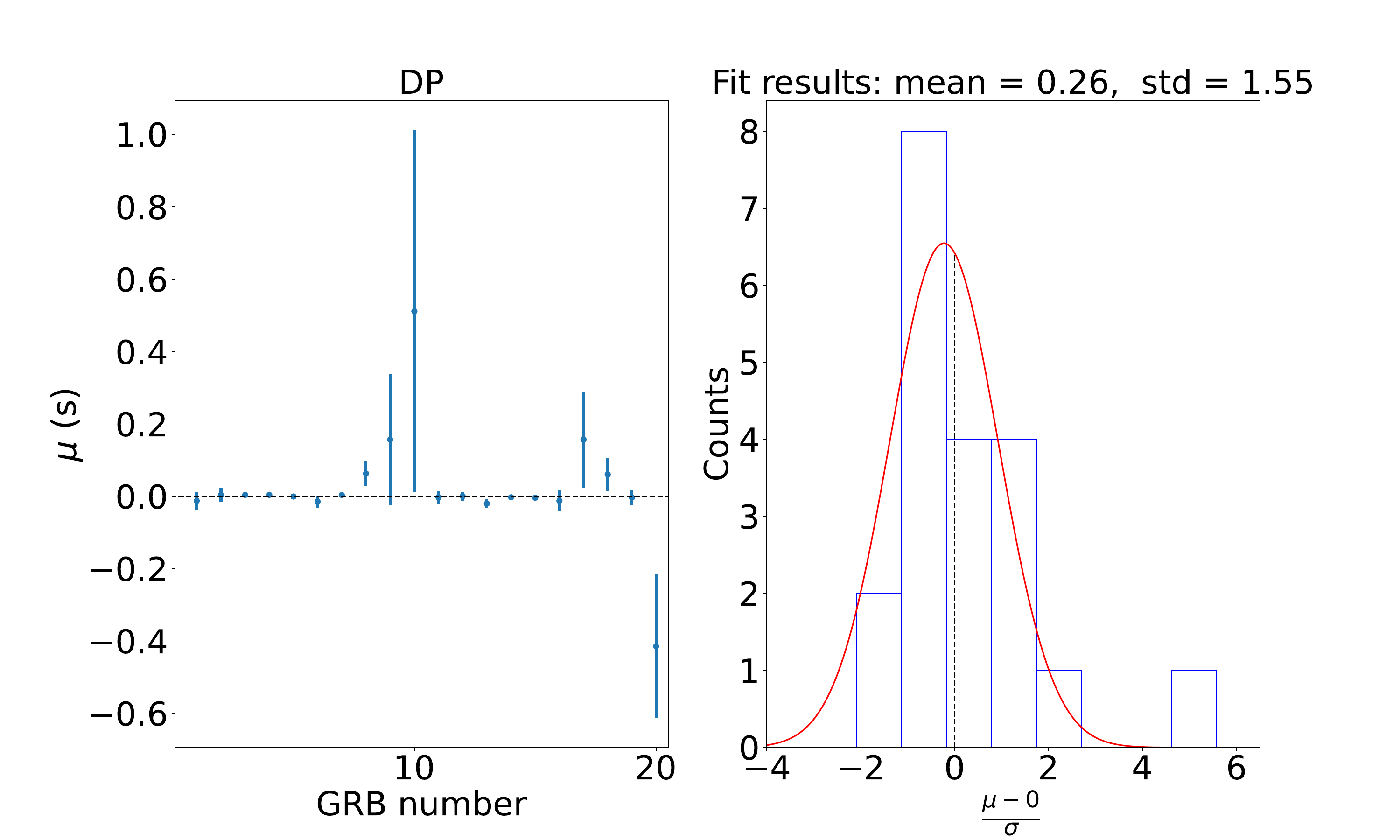}
    \includegraphics[scale=0.18]{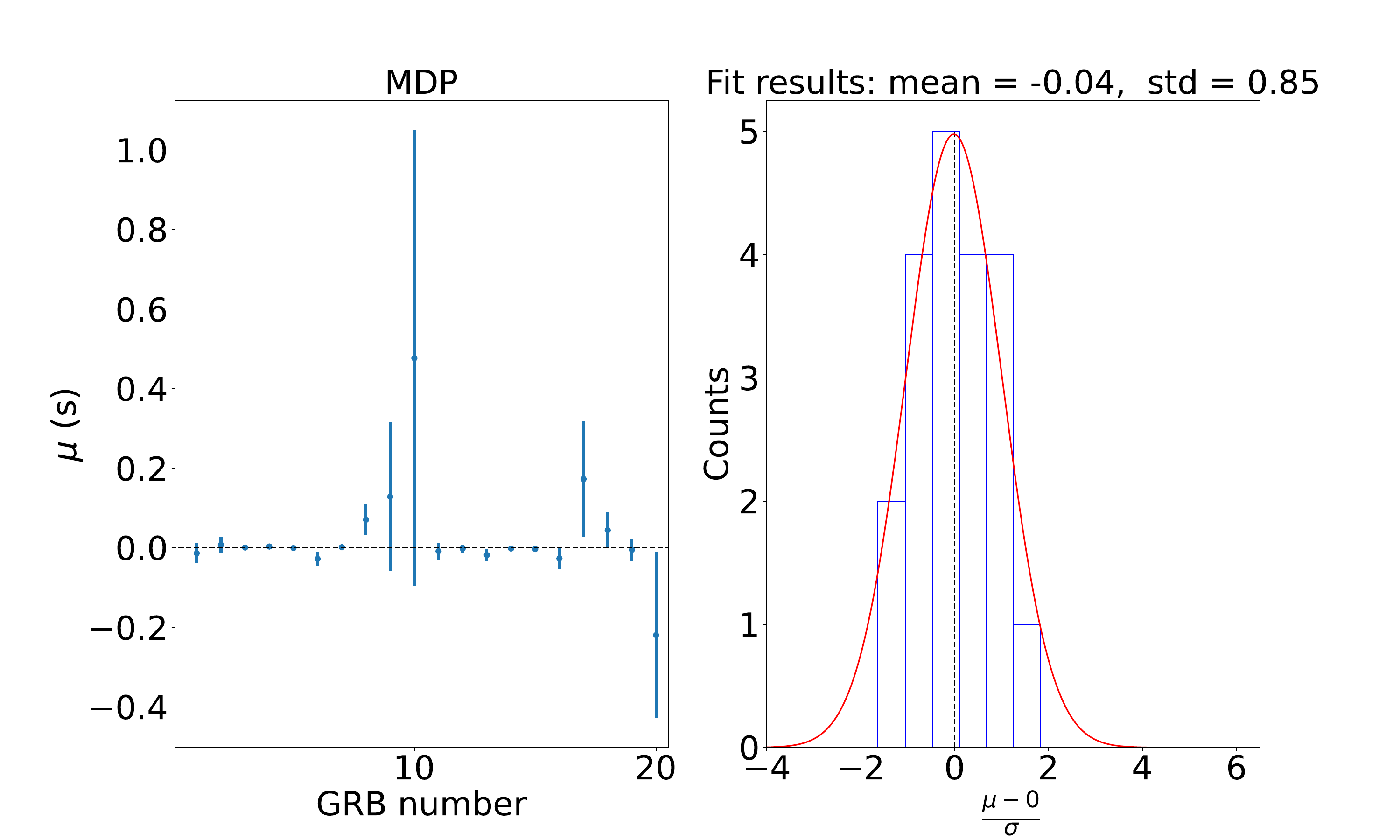}
    \caption{DP and MDP methods comparison. The plots on the left (DP and MDP) show the experimental delays estimated via MC procedures and the associated error as the vertical error bar. The plots on the right show the residual distribution for each method in units of sigma.}
    \label{fig:MDP_DP_comparison}
\end{figure*}

The distribution in \autoref{fig:MDP200_split} is centered in $\rm{\mu}=-1.1s$ with an associated error of $\sigma= 0.299s$. This result demonstrates the efficiency of the MDP method in estimating the true existing delays and the associated error. On the other hand, the standard deviation is approximately the same as in the previous case. That indicates that the MDP estimates are accurate even though the generated lists
are not statistically truly independent because of the reshuffle of the same ToA.

\section{Conclusion}

Deriving the light curve associated with the observation of a cosmic source is not a trivial task and requires careful handling of the data obtained from the detector. This step is crucial in timing astronomy when estimating delays between the ToA lists from different detectors via cross-correlation since this tool is defined on continuous functions.
The proposed variable bin size method facilitates the construction of "averaged" light curves from a list of ToA, sampling the observed electromagnetic signal with uniform statistical accuracy. This approach allows sampling the light curve at a finer temporal resolution when the intensity is higher. By linearly interpolating between the rate points of the sampled curves, it is possible to obtain a rate function, enabling the estimation of delays via cross-correlation on continuous functions. It is important to note that linear interpolation introduces minimal variability between consecutive rate points.
 
The MDP Monte Carlo procedure enables the generation of a distribution of delays, where the mean is the experimental value of the delay and the sigma is the experimental associated error.

We can therefore confirm the capability of the MDP to deliver reliable scientific results, providing a significant increase in both accuracy and computational efficiency. 
We conclude that the MDP method reduces the impact of intrinsic Poisson fluctuations in input templates, avoiding their amplification in Monte Carlo simulations. Despite a $\sqrt{2}$ loss in precision due to halved statistics, it ensures unbiased delay estimates.

Furthermore, the developed techniques demonstrate crucial effectiveness in low-statistics regimes, where traditional methods may struggle to yield consistent results. These techniques remain effective regardless of the transient signal luminosity, although the precision of the estimated lag improves with the increasing number of source-associated photons.
 
The entire package, written in Python, is publicly available on the GitHub platform.

\begin{acknowledgements}
      This article was produced while attending the PhD program in  PhD fin Space Science and Technology at the University of Trento, Cycle XXXVIII.
      A.T. acknowledges financial support from “ASI-INAF Accordo Attuativo HERMES Pathinder operazioni n. 2022-25-HH.0” and the basic funding program of the Ioffe Institute FFUG-2024-0002. We sincerely thanks Cristiano Guidorzi for his valuable comments and discussions.
      
\end{acknowledgements}
\bibliography{biblio}{}
\bibliographystyle{aa}

\begin{appendices}

\section{Normalized Poisson Probability Function}
\label{Norm_Poiss_function}
The detection process is a probabilistic process in which the infinitesimal probability of detecting a photon within an infinitesimal time interval $dt$ is:

\begin{equation}
    {\rm  {dP(t)} = \rm{r}(t) dt.}
\end{equation}

The probability to detect $N$ photons in a time $\rm{\Delta t}$ for a given rate $\rm{r}(t)$ is \citep{Kingman_Taylor_1966,pishro2014introduction,Dekking2005}:
\begin{equation}\label{er:poisson prob}
    {\rm {P(\rm{\mu},N)}= \frac{\rm{\mu}^N \rm{e}^{-\rm{\mu}}}{N!} = P_{N,\rm{\Delta t}}(r),}
\end{equation}

where $\rm{\mu}(t)=\rm{r}(t) \rm{\Delta t}$ and we assumed that $\rm{\Delta t}$ is small with respect to the time scale on which $\rm{\mu}(t)$ varies. Therefore:
\begin{equation}\label{er:poisson prob}
    {\rm {P_{N,\rm{\Delta t}}(r)}= \rm {\frac{(\rm{r}\,\rm{\Delta t})^N \rm{e}^{-r \rm{\Delta t}}}{N!}}.}
\end{equation}

\autoref{er:poisson prob} represents the probability of detecting $N$ photons in a time interval $\rm{\Delta t}$ given a rate $\rm{r}(t)$. Apart from the overall normalization factor, \autoref{er:poisson prob} can be interpreted as the probability that $N$ photons, detected in a time interval $\rm{\Delta t}$, derive from a particular rate $\rm{r}(t)$. Since the detection of $N$ photons must depend on a rate, among all the possible rates, the normalization factor A is obtained by integrating \autoref{er:poisson prob} in rate between 0 and $\infty$:
\begin{equation}
    {\rm {A \int_{0}^{+\infty} P_{N,\rm{\Delta t}}(r) dr} = 1,}
    \label{norm_fact}
\end{equation}

which gives $A=\frac{1}{\rm{\Delta t}}$.

Therefore, given that $N$ photons are detected within a time interval $\rm{\Delta t}$, the probability distribution of the rate $r$ is:
\begin{equation}
    {\rm {Q_{N,\rm{\Delta t}}(r)= \rm{\Delta t} \frac{(\rm{r}\,\rm{\Delta t})^N \rm{e}^{-r \rm{\Delta t}}}{N!}}.}
\end{equation}

\subsection{Statistical Confidence Level}
\label{Statistical Confidence Level}
To evaluate the confidence level (CL) for the rate (as in any statistical treatment), we must integrate \autoref{er:norm_poisson prob} between two points of equal probability, one below and one above the (unique) maximum of the distribution. Notably, in the case of $N=0$, the function simplifies to $\rm{e}^{-\rm{r}\rm{\Delta t}}$, which is monotonically decreasing, allowing us to determine an upper limit.
 
By this definition, the CL corresponds to the area under the normalized probability distribution $Q_{N,\rm{\Delta t}}(r)$, enclosed between the upper and lower bounds of the rate confidence interval:
\begin{equation} \label{condition 1}
    {\rm {CL = \rm {\int_{r_{min}}^{r_{max}} Q_{N,\rm{\Delta t}}(r) dr}},}
\end{equation}

with the constraint:

\begin{equation}\label{condition 2}
    {\rm {Q_{N,\rm{\Delta t}}(r_{min}) = Q_{N,\rm{\Delta t}}(r_{max})}.}
\end{equation}

\section{Poisson characteristic values} \label{mean,mod,med}
\subsection{Mode}
The Poisson mode is the rate value where the $\rm{Q}(\rm{r} \,\rm{\Delta t}; \rm N)$ is maximum, $\frac{\partial Q_{\rm N;\rm{\Delta t}}(\rm r)}{\partial \rm r} |_{\rm r=r_{mode}} = 0$ :

\begin{equation}
    \frac{\rm{\Delta t}^2}{\rm N!}\,\rm{e}^{-\rm{r_{mod}}\,\rm{\Delta t}}\Bigg[N(\rm{r_{mod}}\,\rm{\Delta t})^{N-1}-(\rm{r_{mod}}\,\rm{\Delta t})^N\bigg] = 0
\end{equation}

that leads to $\rm N(\rm{r_{mod}}\,\rm{\Delta t})^{N-1}=(\rm{r_{mod}}\,\rm{\Delta t})^N$, so the mode value is 
$$\rm{r_{mod}}(N,\rm{\Delta t})=\frac{N}{\rm{\Delta t}}$$

\subsection{Median}

The Poisson median divides the area under the Poisson distribution into two identical parts ($\rm x=r \Delta t$):

\begin{equation}
    \int_{0}^{\rm{r_{med}}} \rm{Q}(\rm{r} \,\rm{\Delta t}; N) dr = \frac{\rm{\Delta t}}{N!} \int_{0}^{\rm{r_{med}}} x^N \,\rm{e}^{-x} dx = \frac{1}{2}
\end{equation}
The integral $\int (\rm xk)^N \, \rm e^{-xk} \, \mathrm{d}k = - \frac{\rm e^{-xk}}{k} \sum_{l=0}^{N} \frac{N!}{l!}(xk)^l
$

So the median value can be numerically solved by imposing:

\begin{equation}
    \bigg[- \rm{e}^{-r \,\rm{\Delta t}} \sum_{l=0}^{N} \frac{(\rm{r}\, \rm{\Delta t})^l}{\rm{l!}}\bigg]_{0}^{\rm{r_{med}}} = \frac{1}{2}
\end{equation}

\subsection{Mean}

The Poisson mean can be evaluated by considering the expectation value formula:

\begin{equation}
    \rm{r_{mean}} = \frac{\int_{0}^{\infty}\rm{r}\, Q_{N;\rm{\Delta t}} (r) dr}{\int_{0}^{\infty} Q_{N;\rm{\Delta t}} (r) dr} = \frac{\rm{\Delta t}^{N+1}}{N!} \int_{0}^{\infty}r^{N+1} \rm{e}^{-r \,\rm{\Delta t}} dr
\end{equation}

but $\int_{0}^{\infty}x^{\rm N+1} \rm{e}^{-\rm k \,\rm x} \rm dx = \frac{(\rm n+1)!}{\rm k^{\rm N+2}}$, therefore:

\begin{equation}
    \rm{r_{mean}}(\rm N,\rm{\Delta t}) = \frac{\rm{\Delta t}^{N+1}}{\rm N!} \frac{(\rm N+1)!}{\rm{\Delta t}^{N+2}} = \frac{N+1}{\rm{\Delta t}}
\end{equation}

\section{CL analytical solution} \label{app:conditions}

By substituting the expression of $\rm Q_{\rm N,\,\rm{\Delta t}}\, (r)$ in \autoref{er:norm_poisson prob}, the the confidence level condition in \autoref{condition 1} is:
\begin{equation}
    \frac{\rm{\Delta t}^{N+1}}{\rm N!} \int_{\rm r_{min}}^{\rm r_{max}} \,\rm  (\rm r \Delta t)^n \, \rm{\rm e}^{-\rm r \,\rm{\rm \Delta t}} dr = CL
\end{equation}

but $\int (\rm  xk)^N \,\rm{e}^{-xk} dk = 
- \frac{\rm{e}^{-\rm  xk}}{k} \sum_{l=0}^{N} \rm  \frac{N!}{l!}(xk)^l$, so we can write the previous equation as:

\begin{equation}
    \bigg[ \rm{e}^{-r \,\rm{\Delta t}} \sum_{l=0}^{N} \frac{(\rm{r}\, \rm{\Delta t})^l}{\rm{l!}}\bigg]_{r_{max}}^{r_{min}} = CL
\end{equation}

By considering the probability condition in \autoref{condition 2} we obtain:

\begin{equation}
    \frac{\rm{\Delta t}}{\rm  N!} (\rm  r_{min}\,\rm{\Delta t})^N\,\rm{e}^{-\rm  r_{min}\,\rm{\Delta t}} = \frac{\rm{\Delta t}}{N!} (\rm  r_{max}\,\rm{\Delta t})^N\,\rm{e}^{-\rm  r_{max}\,\rm{\Delta t}}
\end{equation}

therefore:

\begin{equation}
    \bigg(\frac{\rm  r_{max}}{\rm 
 r_{min}}\bigg)^N=\rm{e}^{(\rm  r_{max}-r_{min})\rm{\Delta t}}
\end{equation}

\subsection{Numerical solution}

This 2 equation and 2 variables system can be numerically solved. By defining $x=(r_{max}-r_{min}) \cdot \rm{\Delta t}$ and considering the condition above.

\begin{equation}
    \rm  F_2(x,y(x,n),z(x,n),n) = \exp(-\rm 
 y)\bigg[\sum_{l=0}^{N} \frac{y^l}{\rm{l!}} \bigg] - \exp(\rm 
 yz) \bigg[\sum_{l=0}^{N} \frac{(yz)^l}{\rm{l!}} \bigg] - CL
\end{equation}
where $\rm  y(x,n)$ is:
\begin{equation}
    \rm  y(x,n) = \frac{x}{\exp(\frac{x}{n})-1}
\end{equation}
and $\rm  z(x,n)$ is:
\begin{equation}
    \rm  z(x,n)=\exp\bigg(\frac{x}{n}\bigg)
\end{equation}

By imposing $\rm  F_2(x,y(x,n),z(x,n),n)=0$, we can find a numerical solution by solving this equation in the variable x. This yields as a result a x that depends on the number of events considered. The $x_n$ corresponds to the $x_{N,CL}$ that satisfies the confidence level conditions for a certain number of counts in a given $\rm{\Delta t}$. At this point, we can express the relative confidence interval [$\rm \epsilon^-(N,CL)$ , $\rm \epsilon^+ (N,CL)$]  as

\begin{equation}
    \rm \epsilon^-(N,CL) = \rm \epsilon^- (x_{N,cl})= \frac{\frac{x_{N,CL}}{N}}{\exp{\frac{x_{N,cl}}{N}}-1}
\end{equation}
\begin{equation}
    \epsilon^+(\rm N,CL) = \epsilon^+ (x_{\rm N,cl})= \frac{\frac{x_{N,CL}}{N}\cdot\exp{\frac{x_{N,CL}}{n}}}{\exp{\frac{x_{N,CL}}{n}}-1}
\end{equation}

Therefore the absolute confidence interval [$\rm r^- (N,CL,\rm{\Delta t})$ , $\rm r^+ (N,CL,\rm{\Delta t})$]

\begin{equation}\label{r_min}
    \rm r^- (N,CL,\rm{\Delta t})= \rm \epsilon^- (N,CL) \cdot \rm{r_{mod}} \rm (N,\rm{\Delta t})
\end{equation}
\begin{equation}\label{r_max}
    \rm r^+ (N,CL,\rm{\Delta t})= \rm \epsilon^+ (N,CL) \cdot \rm{r_{mod}} (N,\rm{\Delta t})
\end{equation}

\section{Generalized inversion method analytical solution}

\label{GIM_solution}
The inversion method integral in \autoref{generilzed eq TSIM} can be considered as a trapezoidal integral when the rate curve \rm{r}(t) is a continuous piecewise linear function. By looking at \autoref{lc_1000_integral} the integral can be rewritten as:
\begin{equation}
\begin{split}
   &\frac{\rm{r}(T\_SIM[N]) + \rm{r}(T\_SIM[N-1])}{2}\cdot  \\
   &\cdot (\rm T\_SIM[N] - \rm T\_SIM[N-1]) =-ln\{ 1 - \rm RND(0,1) \}
\end{split}
\end{equation}

In the most general case, $T\_SIM[N-1]$ is between two rate points of \rm{r}(t) as in \autoref{lc_1000_integral}. The rate $\rm{r}( T\_SIM[N-1])$ can be linearly extrapolated by considering the intensities of the two rate points $r_1$ and $r_2$ that are before and after $T\rm \_SIM[N-1])$, as well as their respective associated times $\rm{t}_1$ and $\rm{t}_2$:

\begin{equation}
\begin{split}
    \rm m &=\frac{\rm{r}_2 - r_1}{\rm{t}_2 - \rm{t}_1}
     \\
    \rm{r}(\rm T\_SIM[N-1]) & = \rm r_1 + \rm m\cdot(\rm \rm T\_SIM[N-1] - \rm{t}_1) 
\end{split}
\end{equation}

The same procedure can be performed for the unknown $\rm T\_SIM[N]$ from the $\rm T\_SIM[N-1]$ where $\rm{r}(\rm T\_SIM[N-1])$ is known. Let us define $\rm T\_SIM[N-1]$ as as $\Bar{t}$ and $\rm T\_SIM[N]$ as x: 

\begin{equation}
\begin{split}
    \rm{r}(x)  = & \rm r_1 + \rm m\cdot(\rm x - \rm{t}_1)
     \\
    \frac{\rm{r}(\Bar{t}) + \rm{r}(\rm x)}{2} \cdot (\rm x - \Bar{t})  = & -\rm ln\{ 1 - \rm RND(0,1) \}
     \\
    \frac{\rm{r}(\Bar{t}) + \rm r_1 + \rm mx - \rm m\rm{t}_1}{2} \cdot (\rm x - \Bar{t})  = & -\rm ln\{ 1 - \rm RND(0,1) \}
\end{split}
\end{equation}

The equation can then be rewritten by rearranging the random terms with $\rm ZETA  \equiv -ln\{ 1 - \rm RND(0,1) \}$:

\begin{equation}
    \rm m x^2 + x (\rm{r}(\Bar{t}) + r_1 - m\rm{t}_1 - m\Bar{t}) + m\rm{t}_1\Bar{t} -\Bar{t}\rm{r}(\Bar{t}) -\Bar{t}r_1  -2 ZETA = 0 
\end{equation}

The only possible solution is therefore when $\rm x>\Bar{t}$:

\begin{equation}
\begin{split}
        B &= \rm{r}(\Bar{t}) + r_1 - m\rm{t}_1 - m\Bar{t}
         \\
        C &= \rm m\rm{t}_1\Bar{t} -\Bar{t}\rm{r}(\Bar{t}) -\Bar{t}r_1  -2 ZETA
         \\
        \rm x &= \frac{-\rm B + \sqrt{\rm B^2 - 4 m C}}{2\rm m}
\end{split}
\end{equation}

\captionsetup{labelformat=empty}

\section{CCF examples}
\autoref{fig:CCF_grid} from page 12 to 16 shows the GRB considered in the comparison in \autoref{fig:MDP_DP_comparison} between DP and MDP method.
\begin{figure*}[t]
    \centering
    \includegraphics[width=0.49\textwidth]{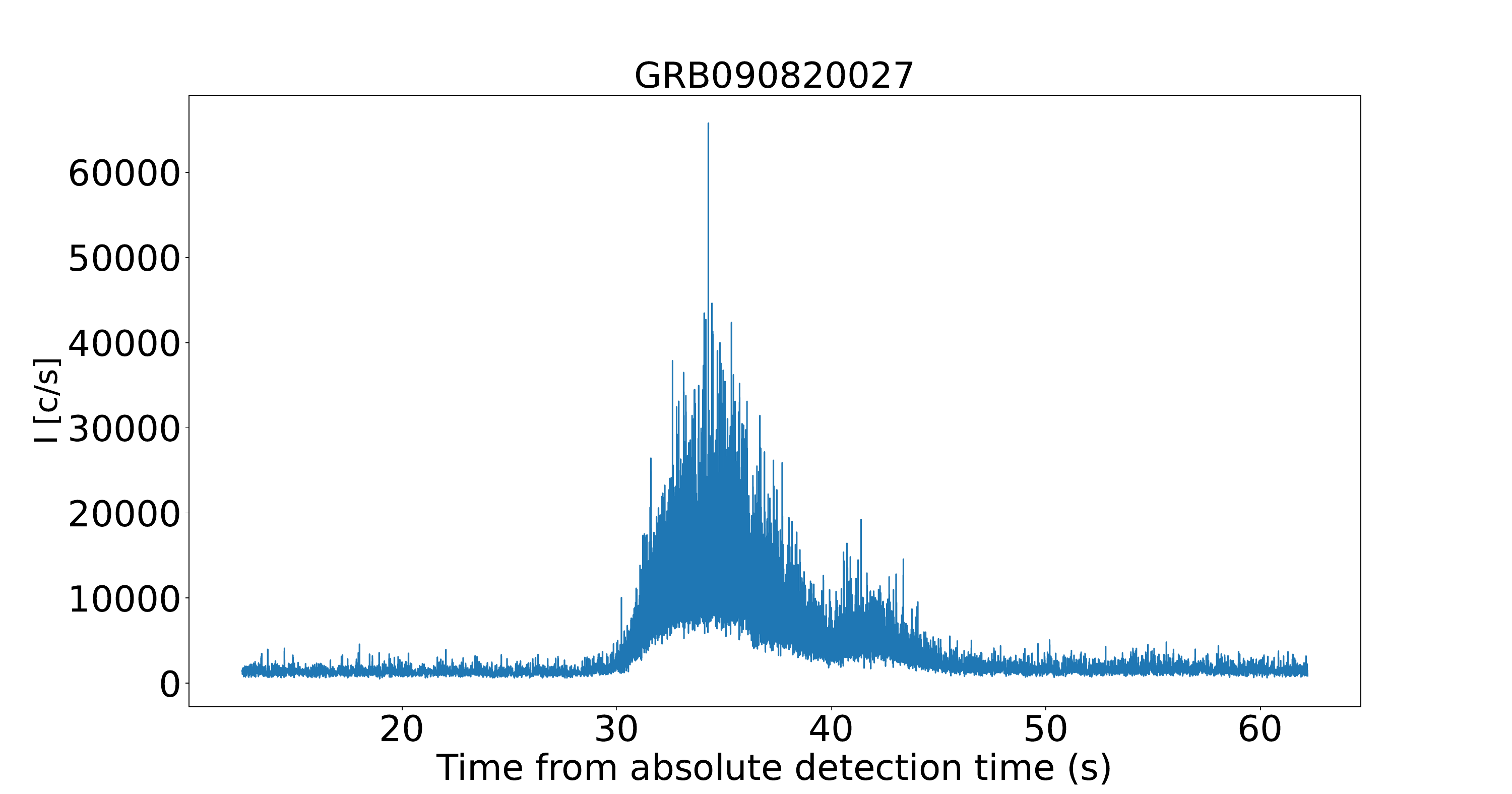}
    \includegraphics[width=0.49\textwidth]{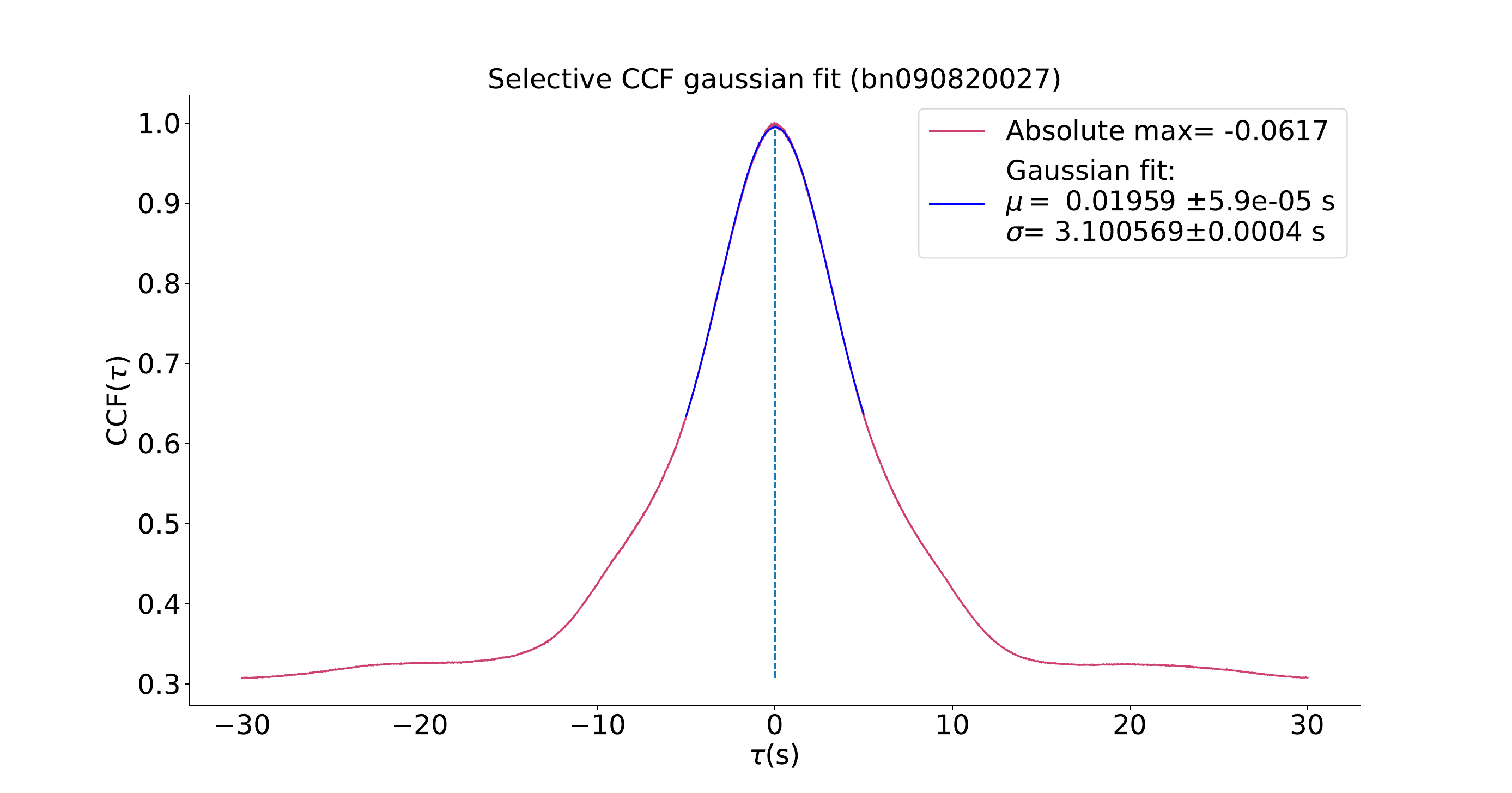}
    \includegraphics[width=0.49\textwidth]{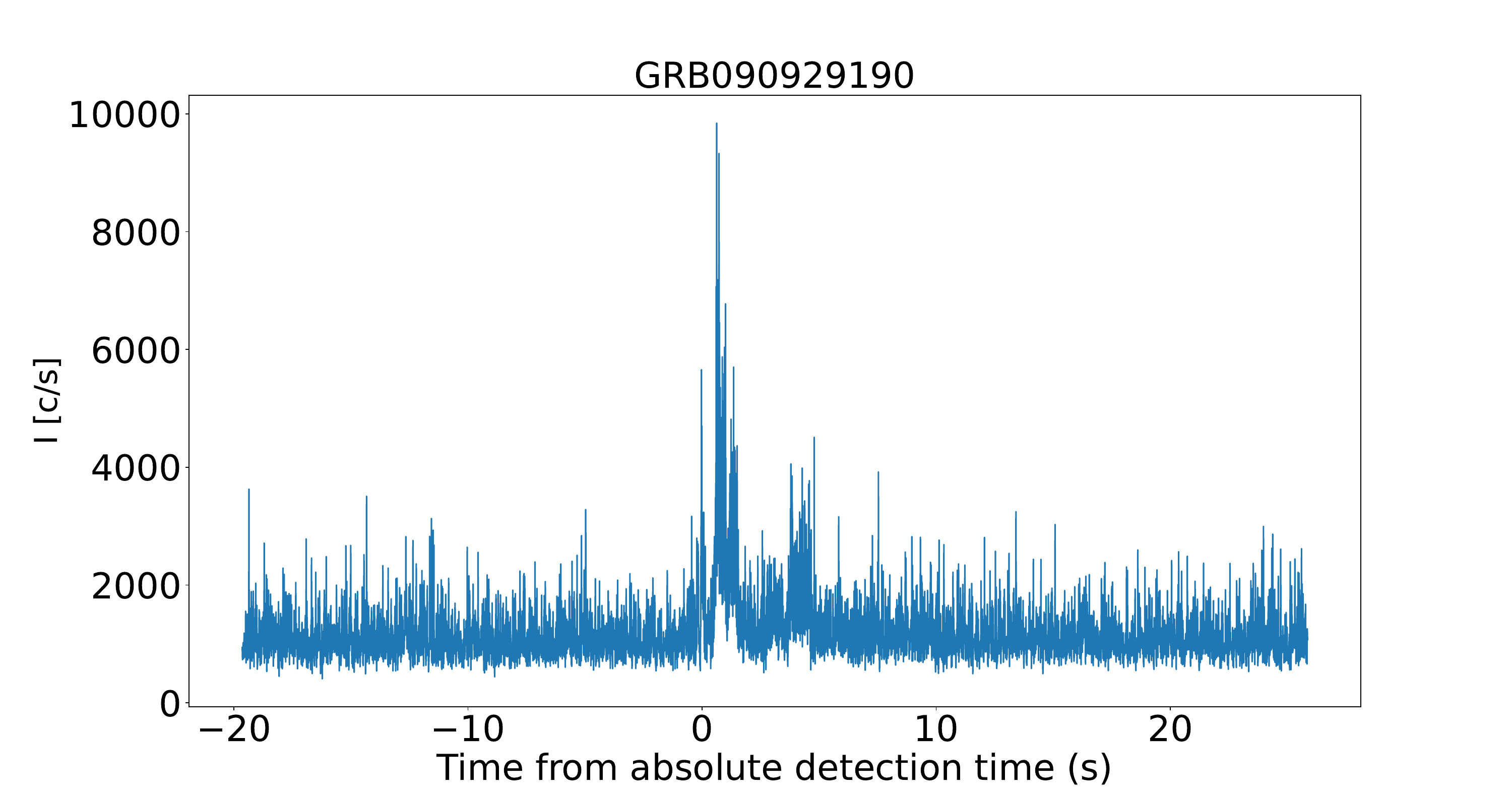}
    \includegraphics[width=0.49\textwidth]{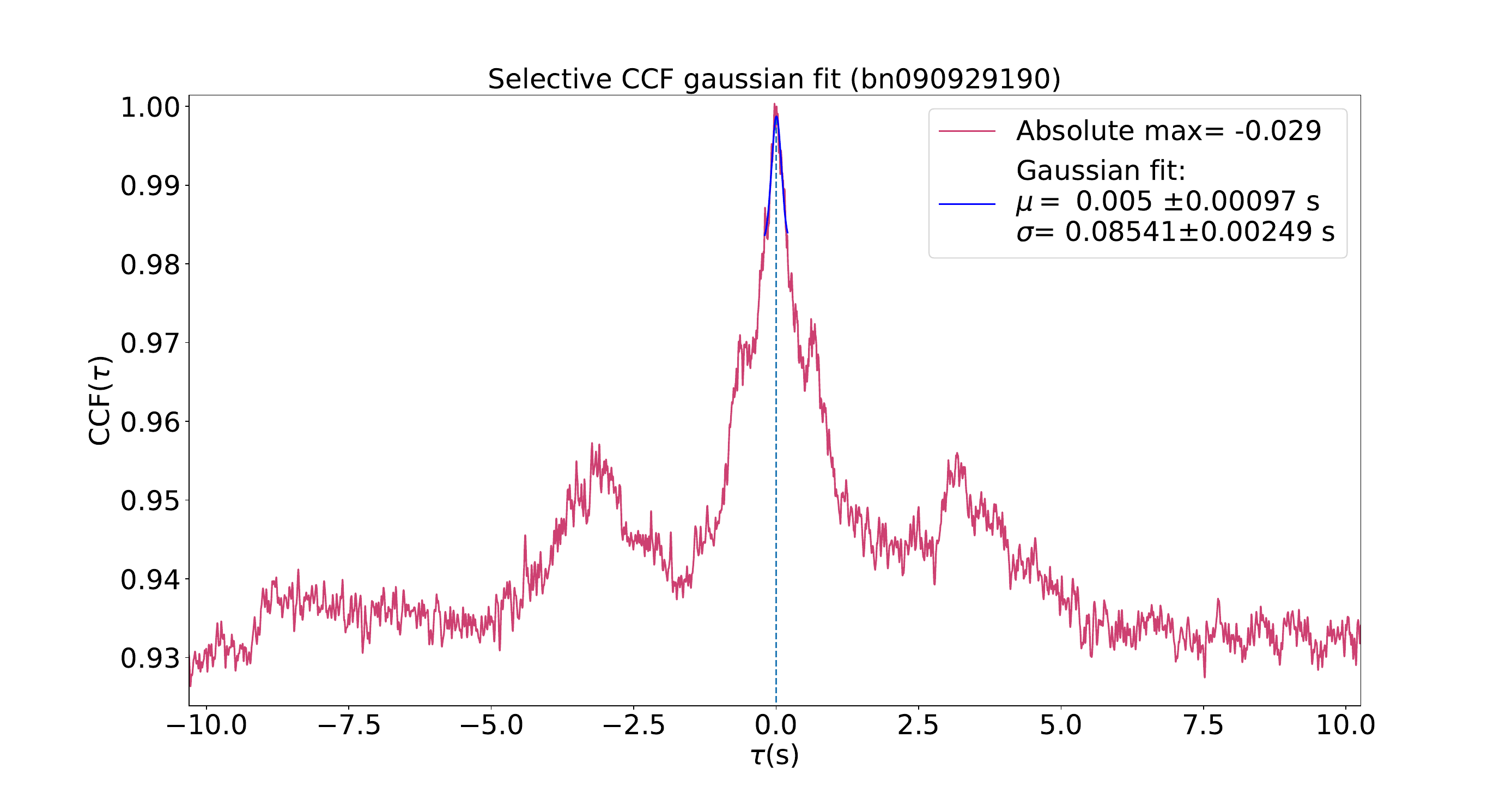}
\end{figure*}

\begin{figure*}[t]
    \centering
    \includegraphics[width=0.49\textwidth]{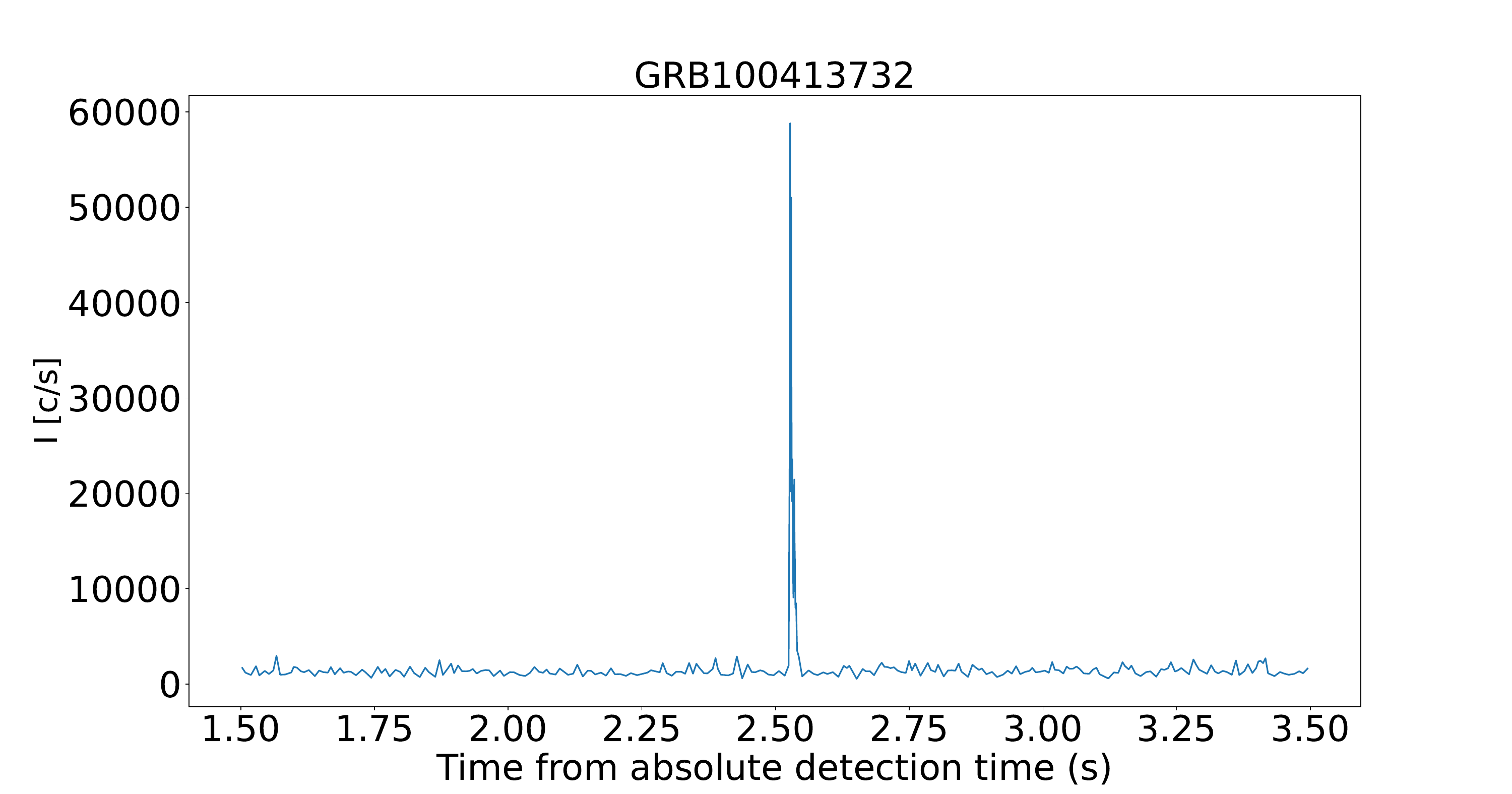}
    \includegraphics[width=0.49\textwidth]{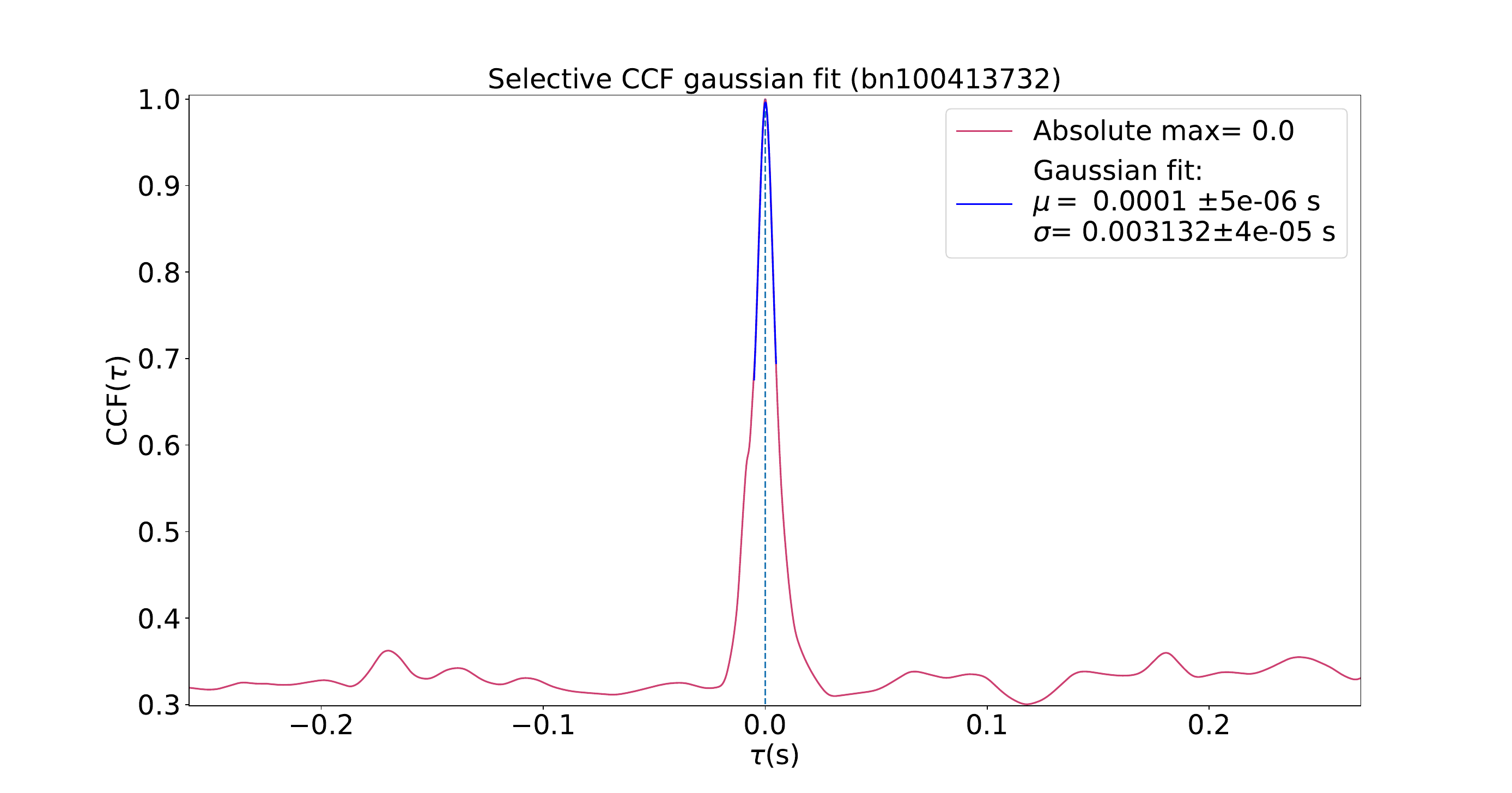}
    \includegraphics[width=0.49\textwidth]{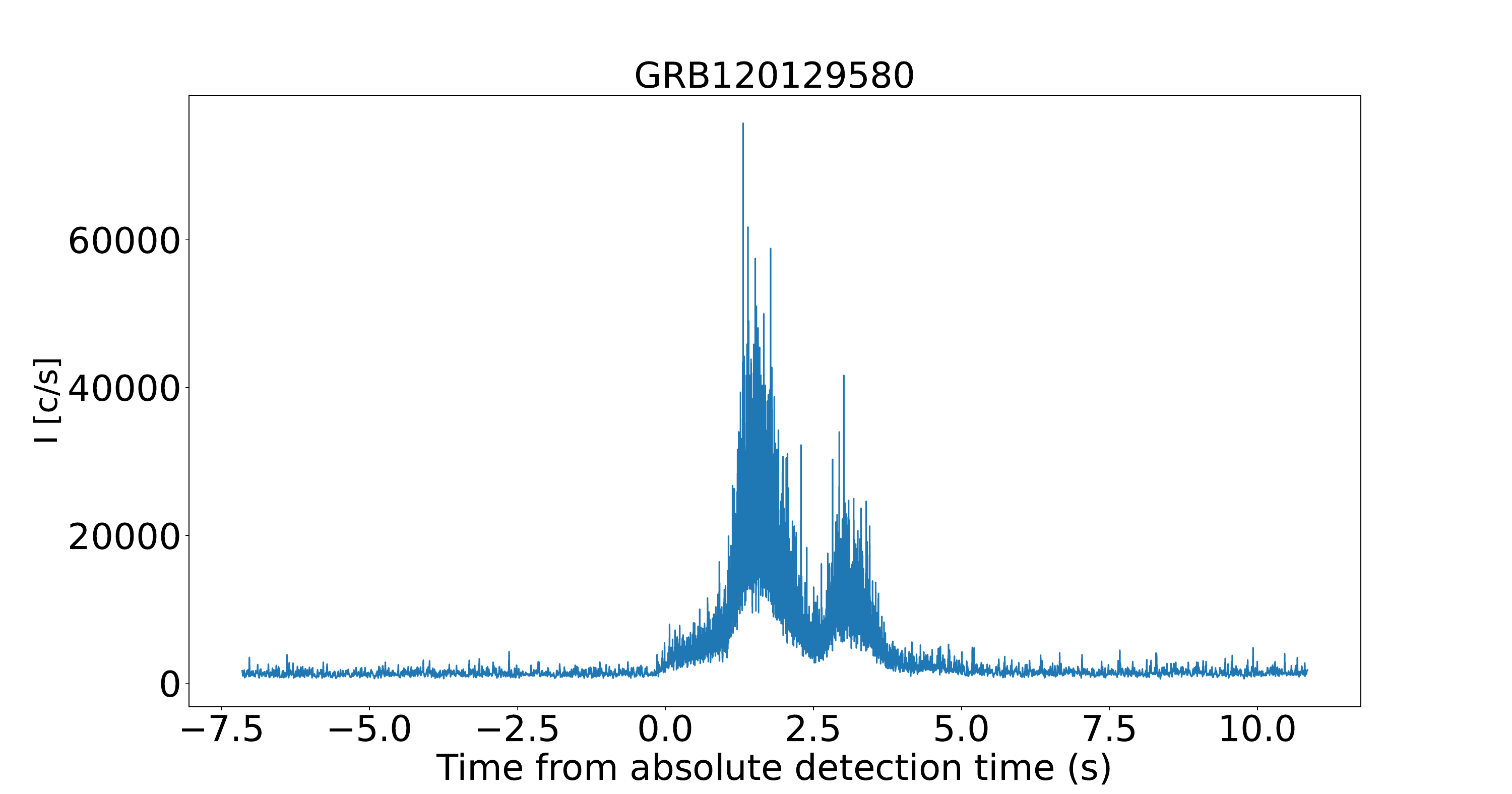}
    \includegraphics[width=0.49\textwidth]{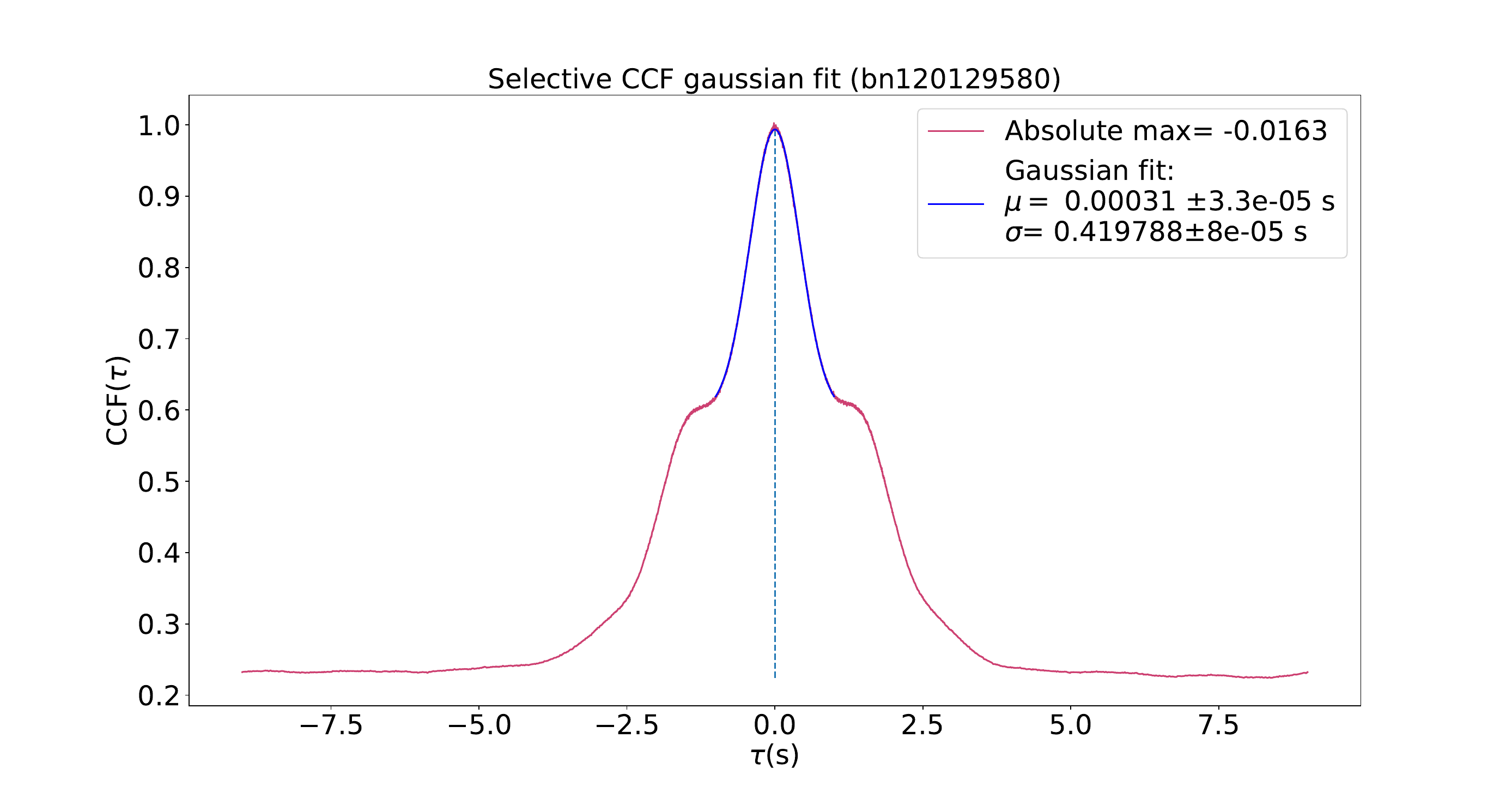}
\end{figure*}

\begin{figure*}[t]
    \centering
    \includegraphics[width=0.49\textwidth]{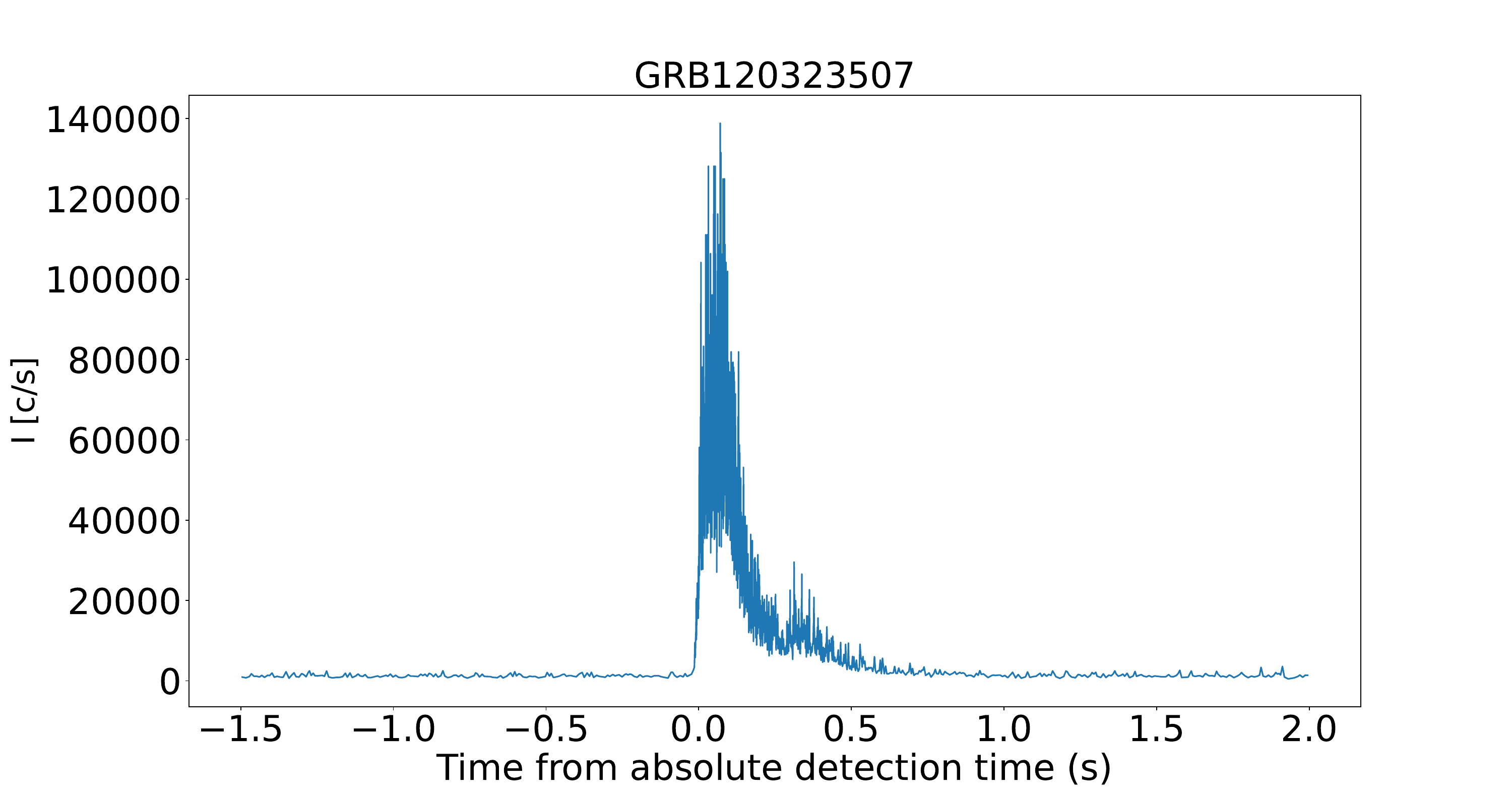}
    \includegraphics[width=0.49\textwidth]{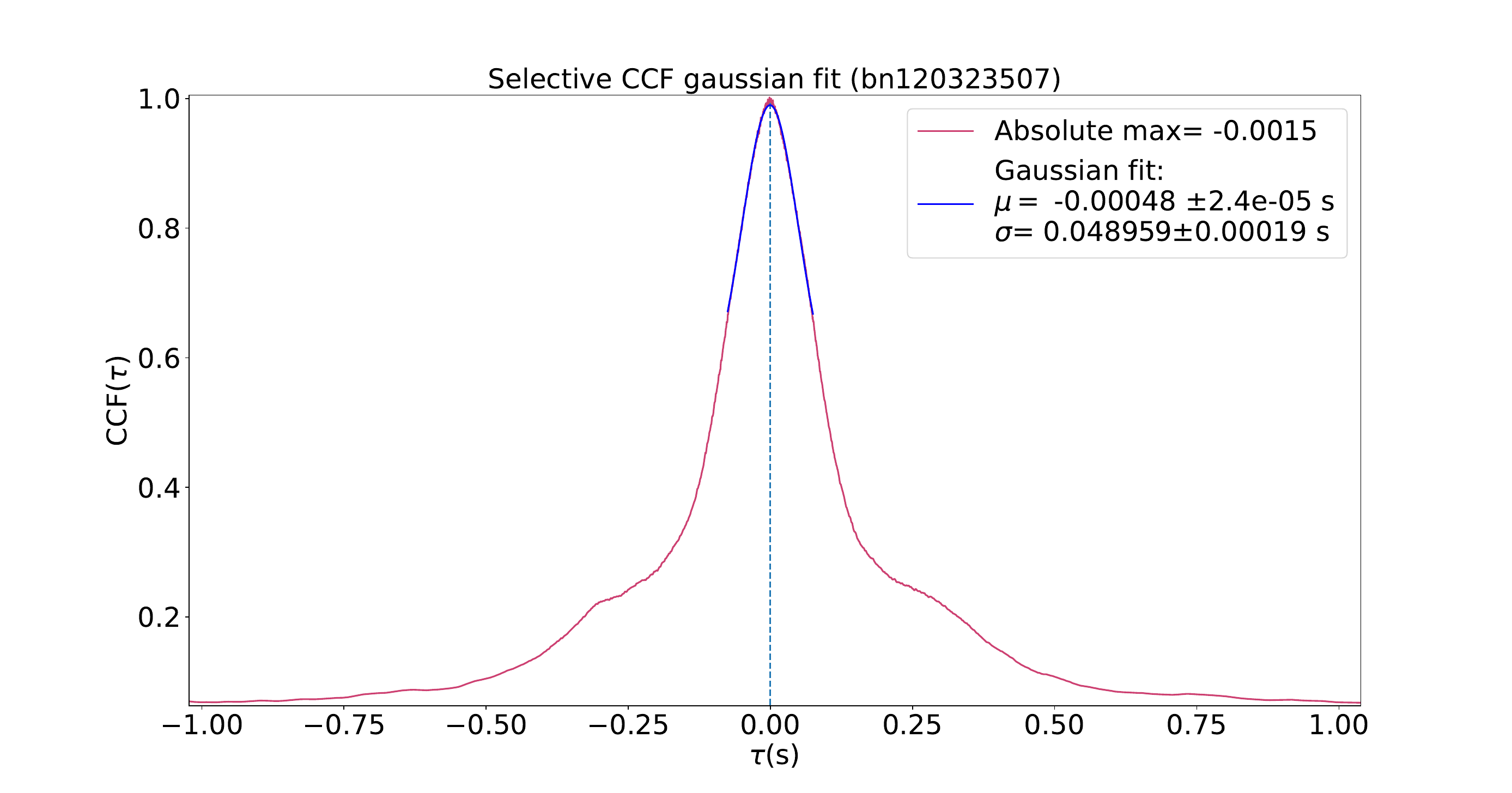}
    \includegraphics[width=0.49\textwidth]{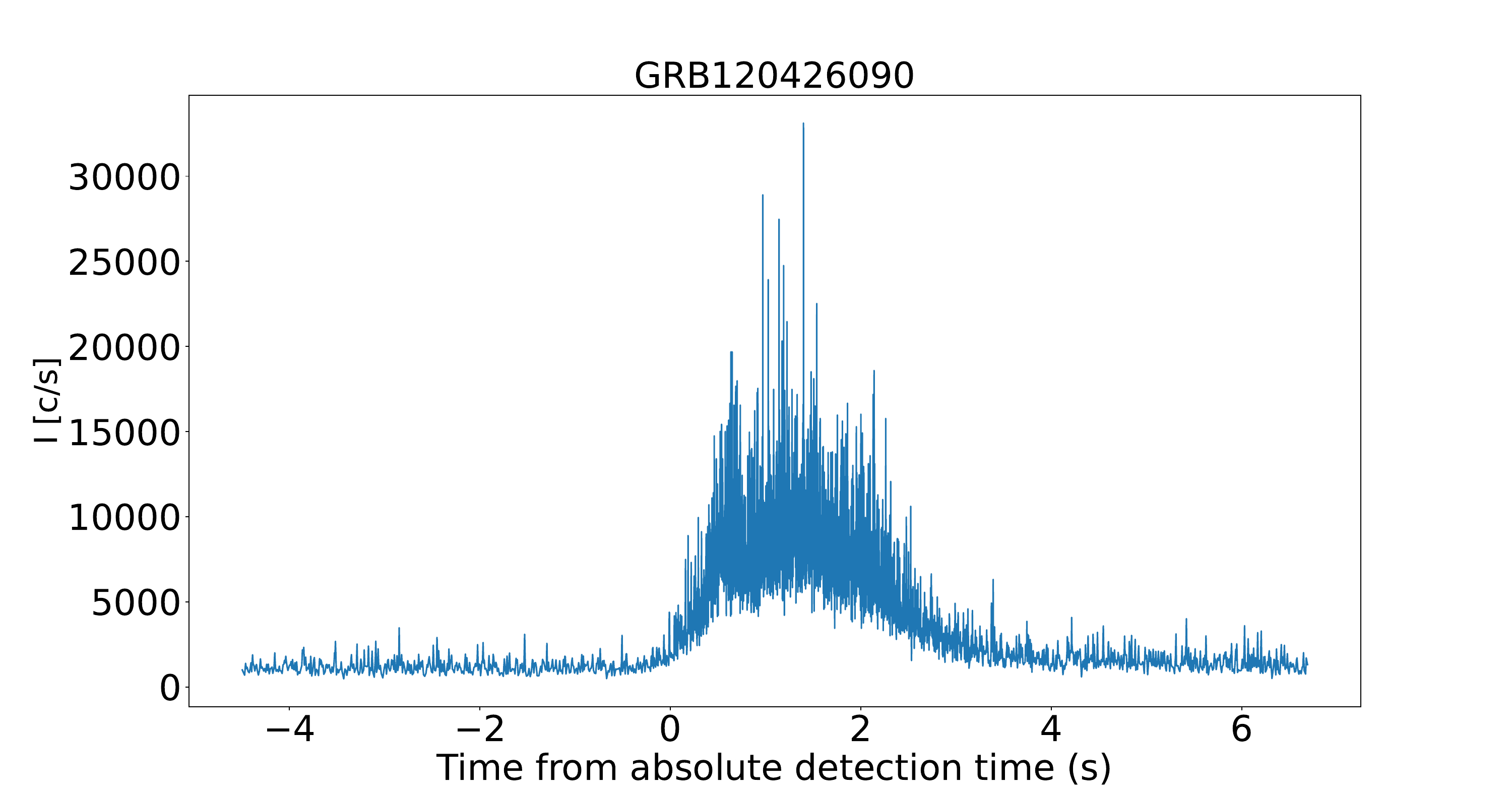}
    \includegraphics[width=0.49\textwidth]{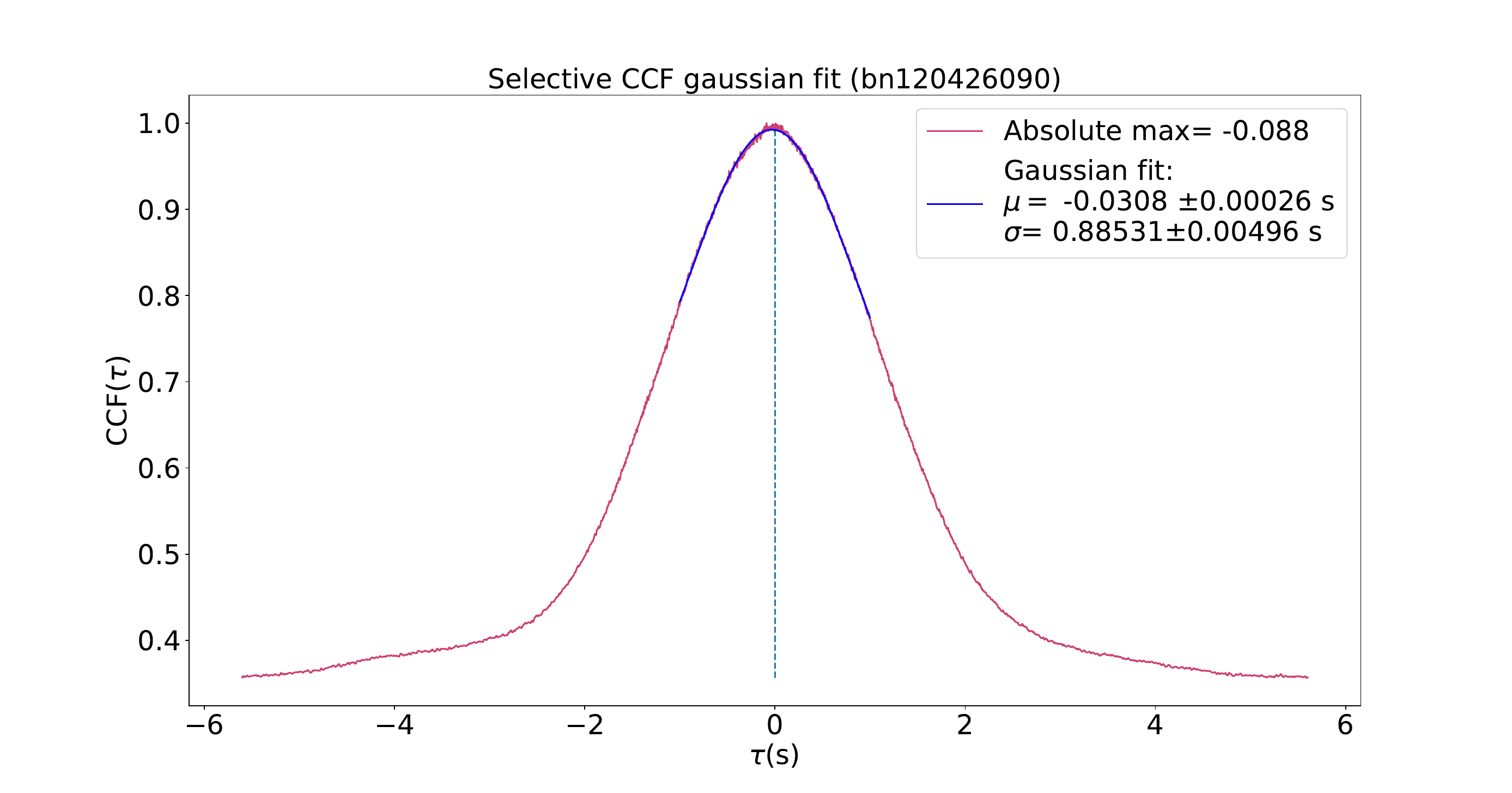}
\end{figure*}

\begin{figure*}[t]
    \centering
    \includegraphics[width=0.49\textwidth]{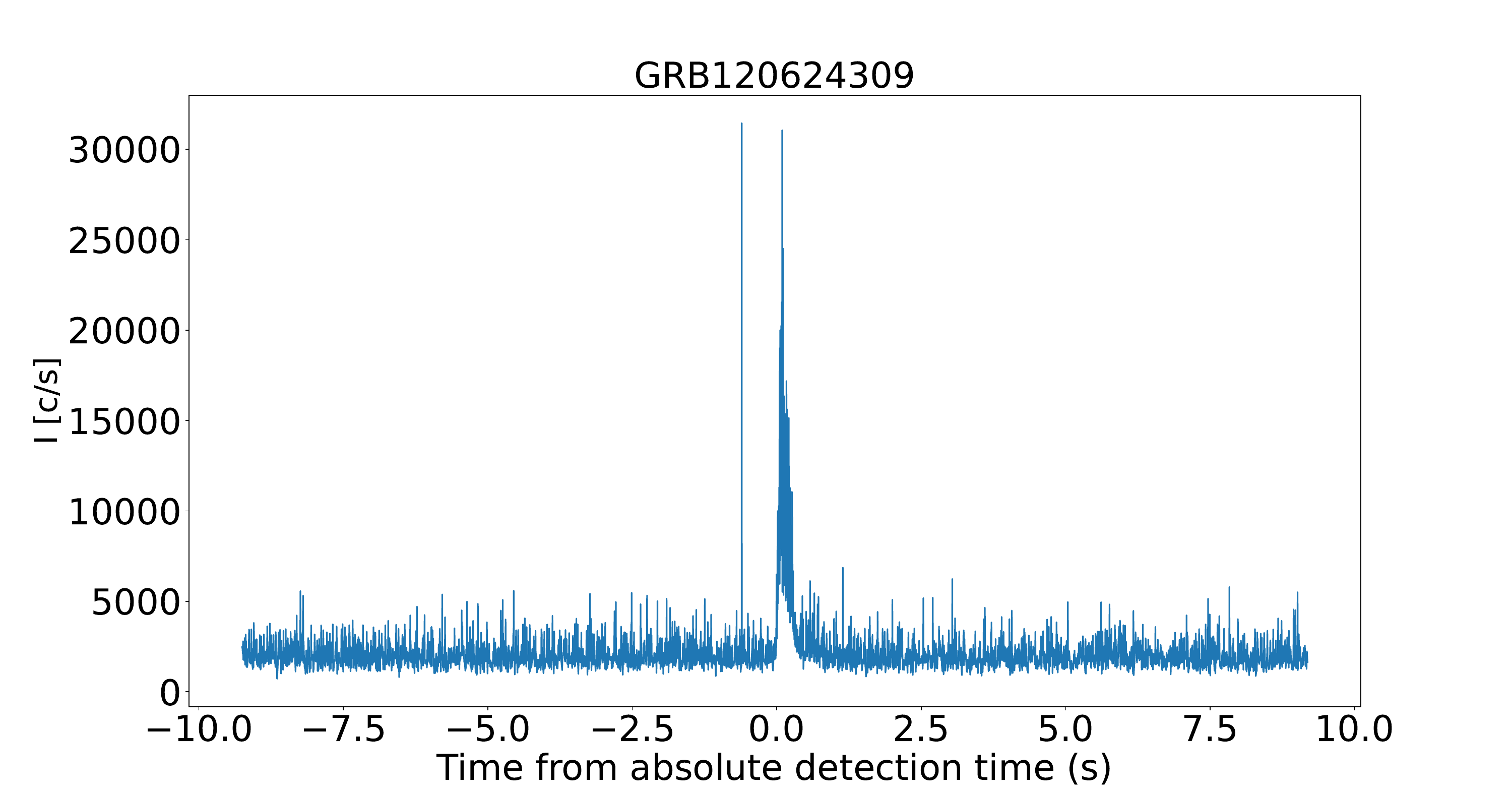}
    \includegraphics[width=0.49\textwidth]{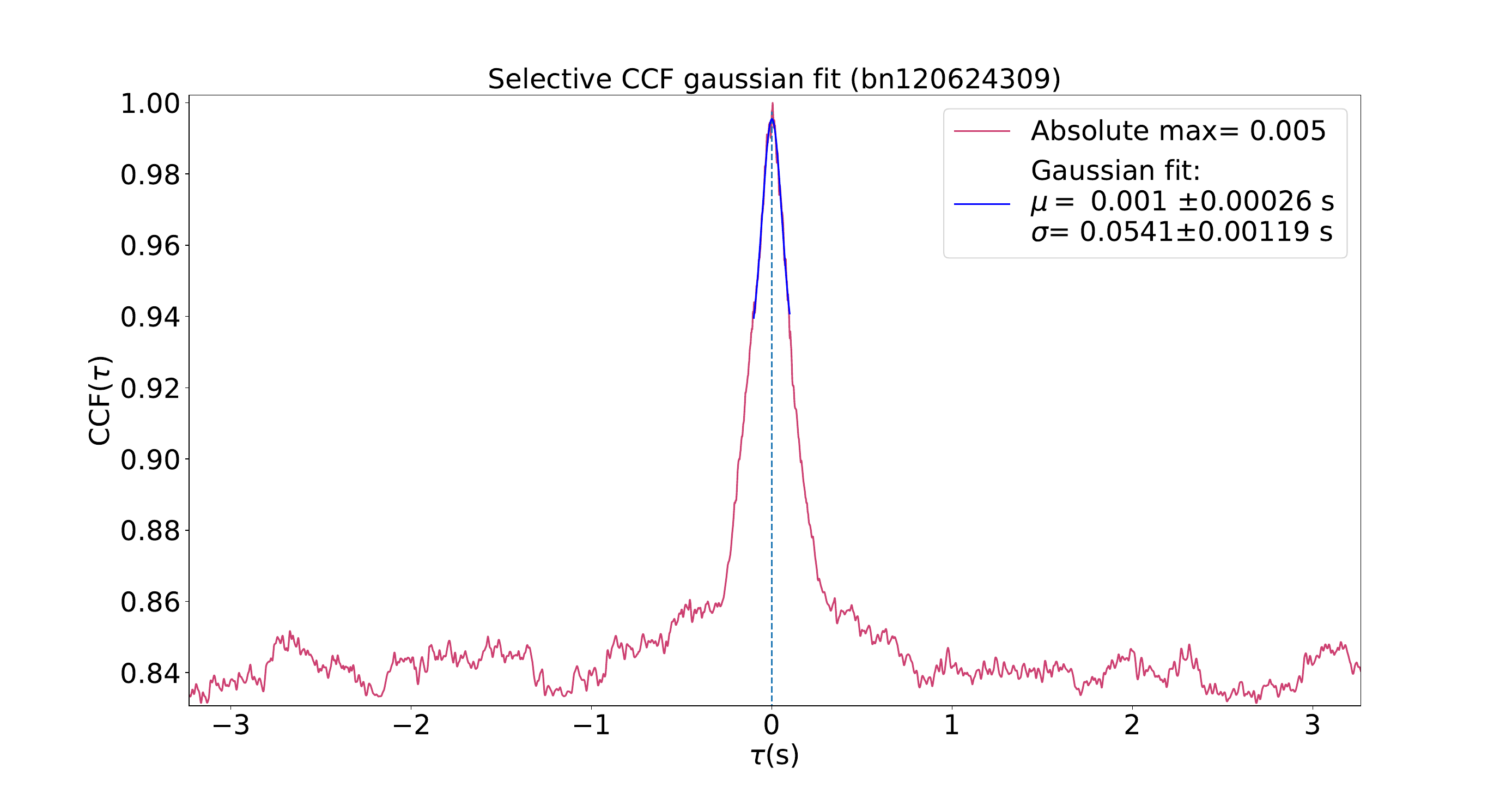}
    \includegraphics[width=0.49\textwidth]{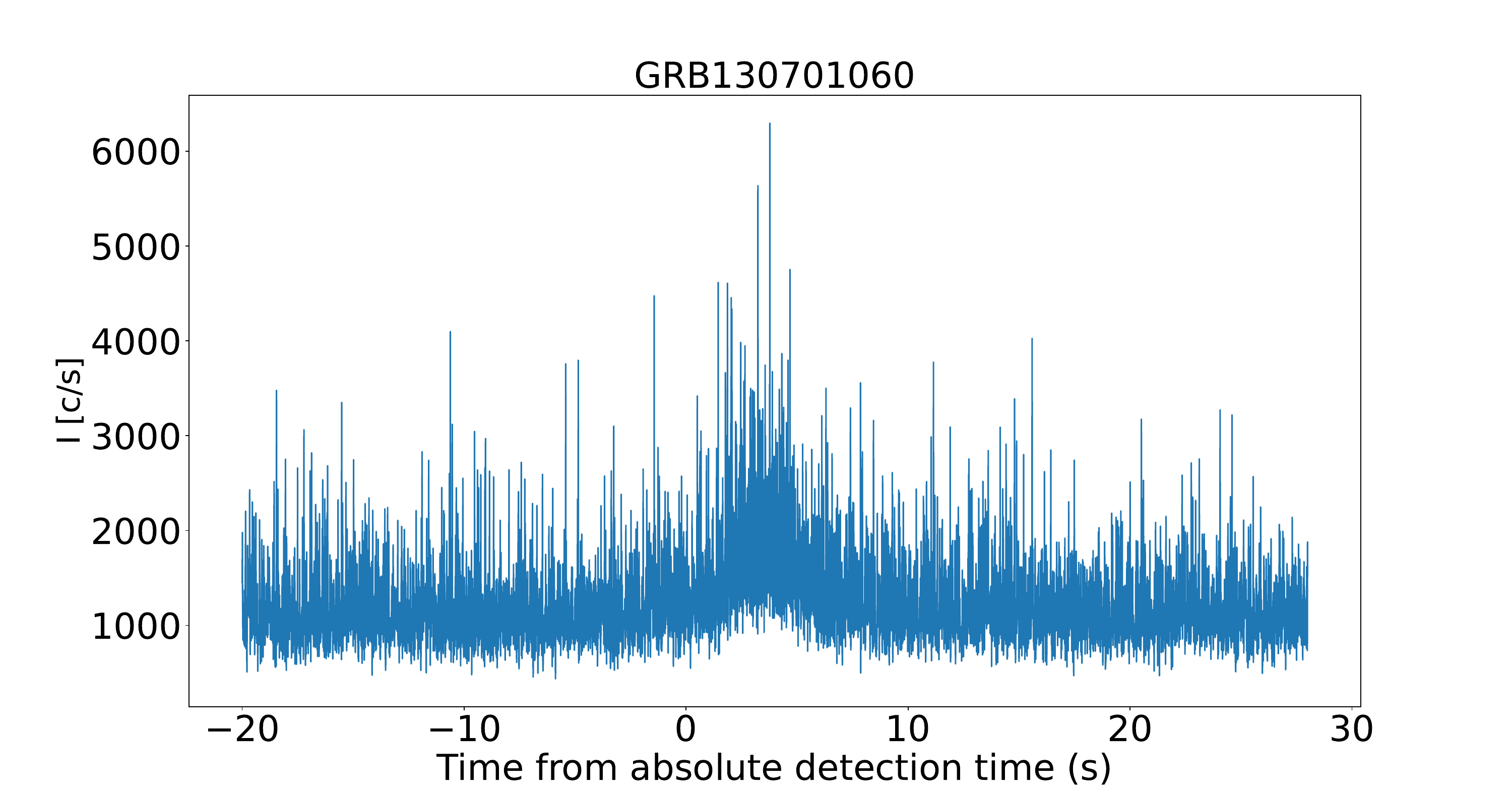}
    \includegraphics[width=0.49\textwidth]{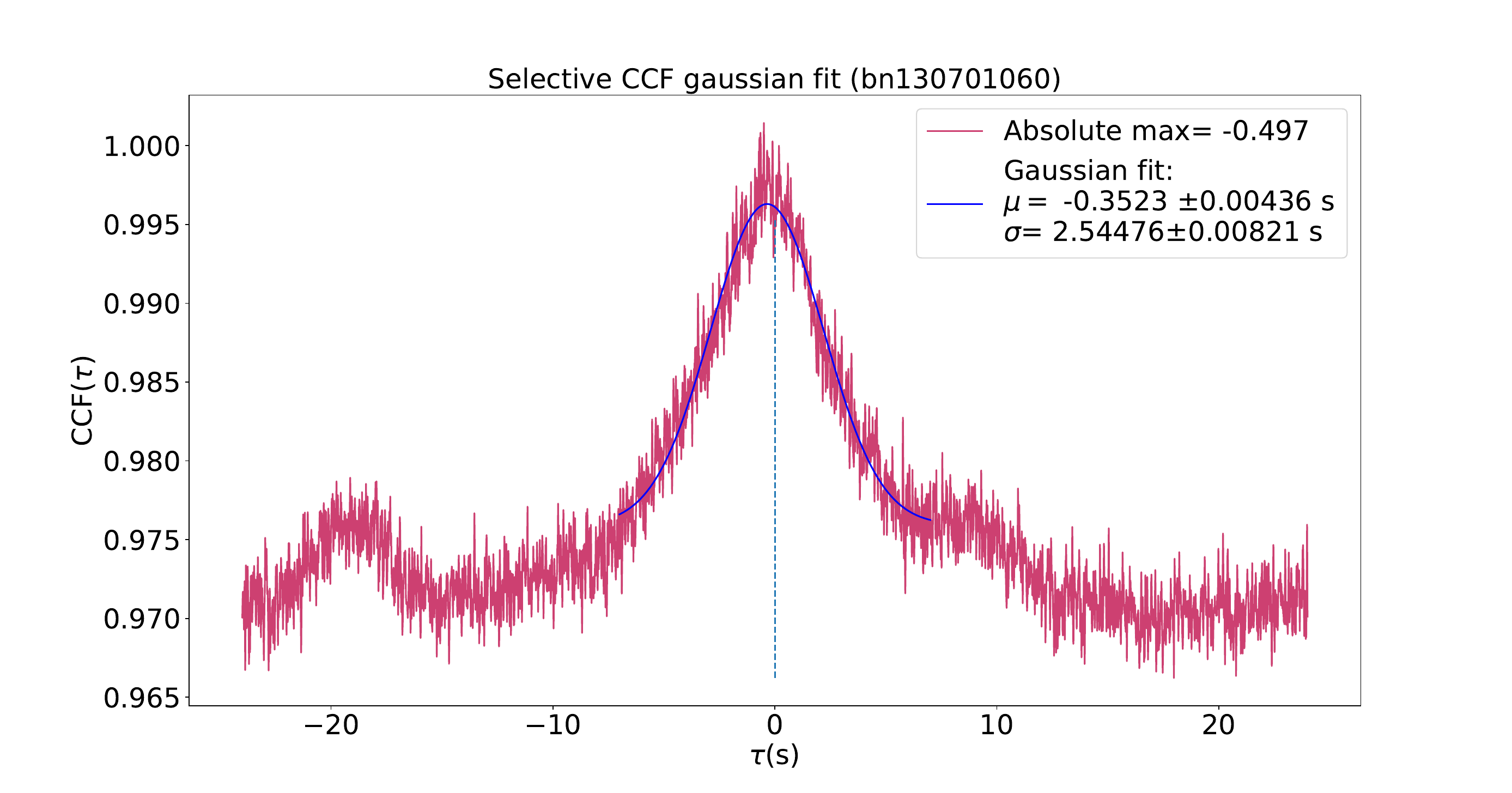}
\end{figure*}

\begin{figure*}[t]
    \centering
    \includegraphics[width=0.49\textwidth]{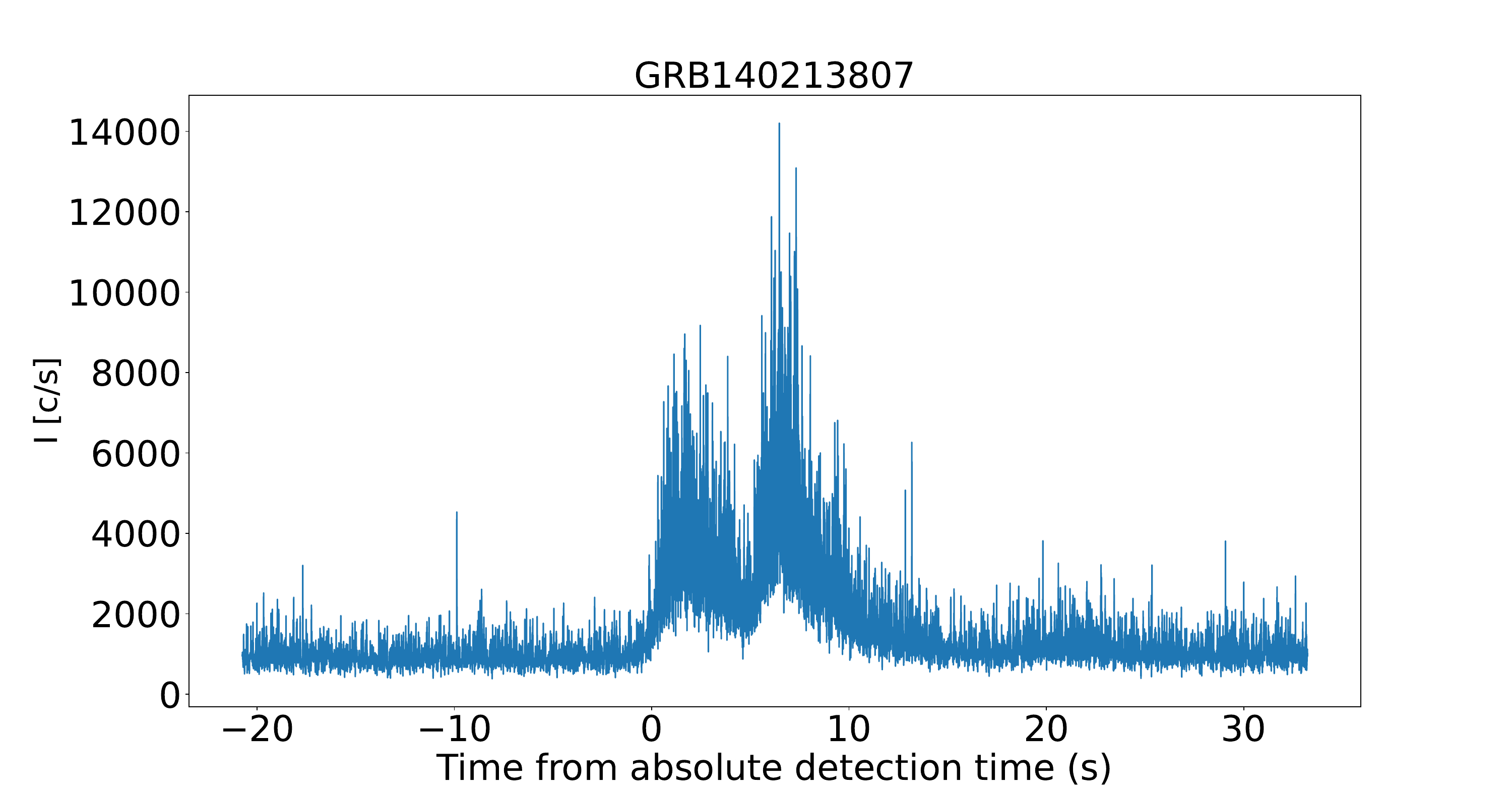}
    \includegraphics[width=0.49\textwidth]{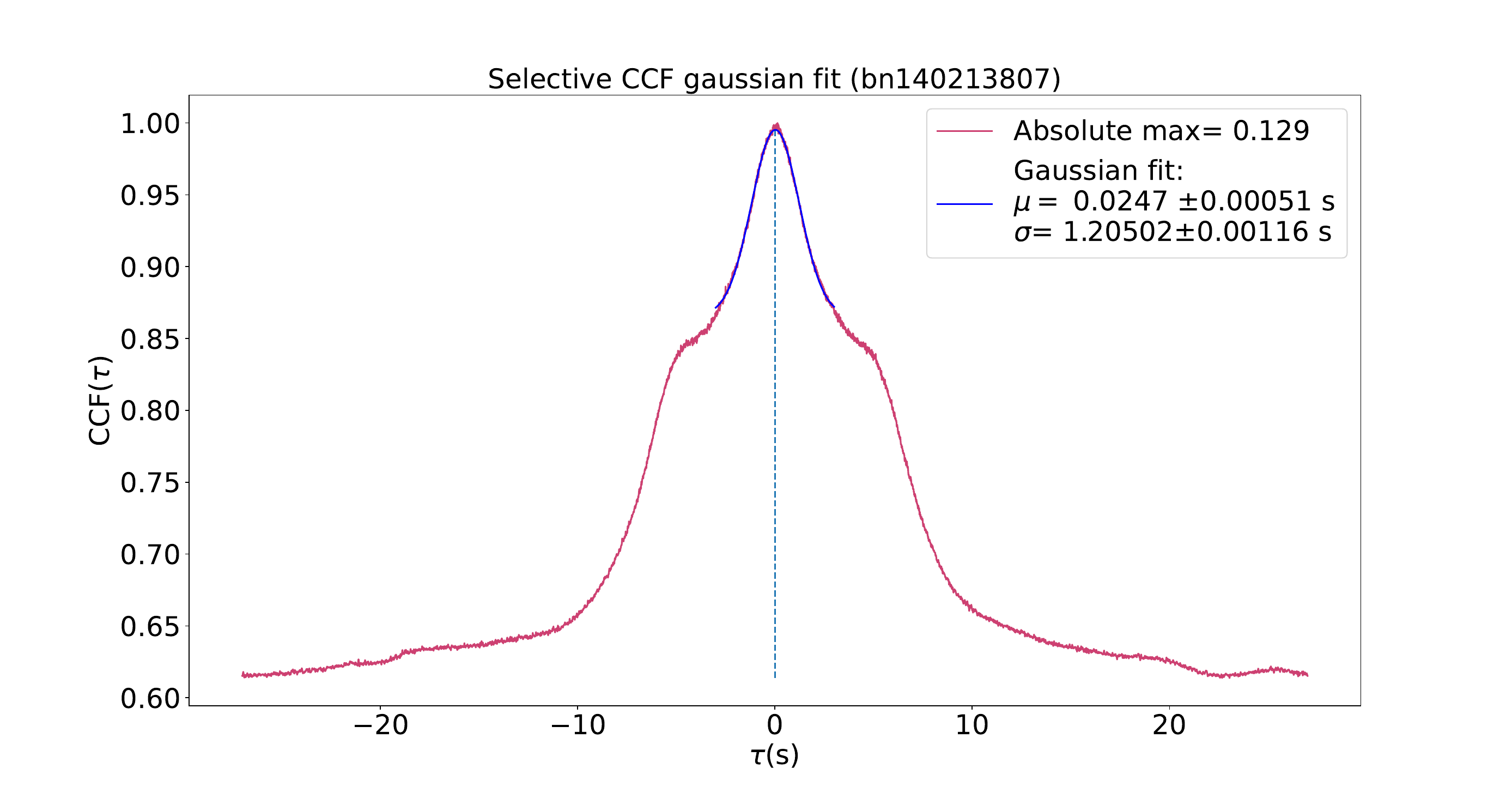}
    \includegraphics[width=0.49\textwidth]{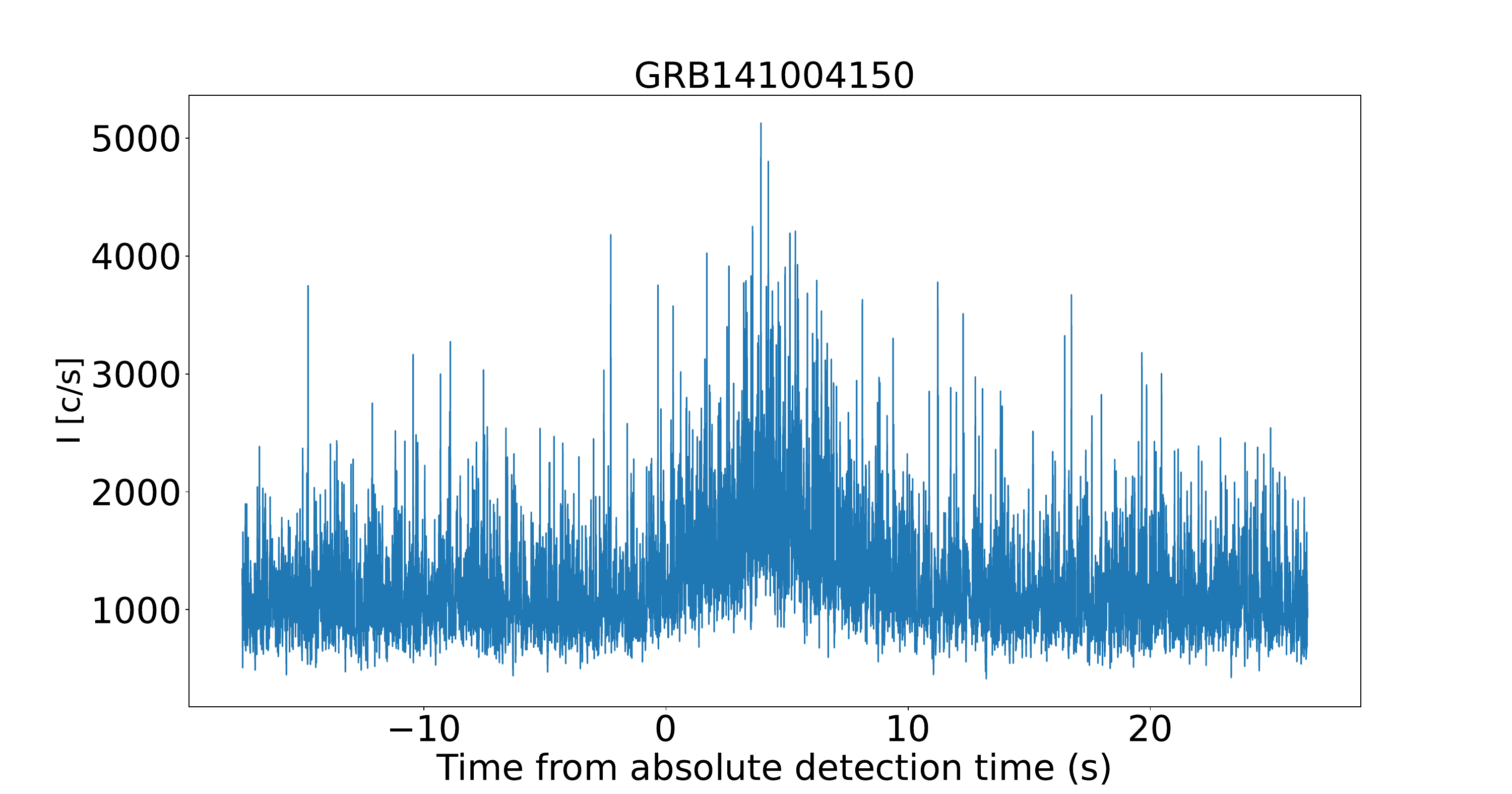}
    \includegraphics[width=0.49\textwidth]{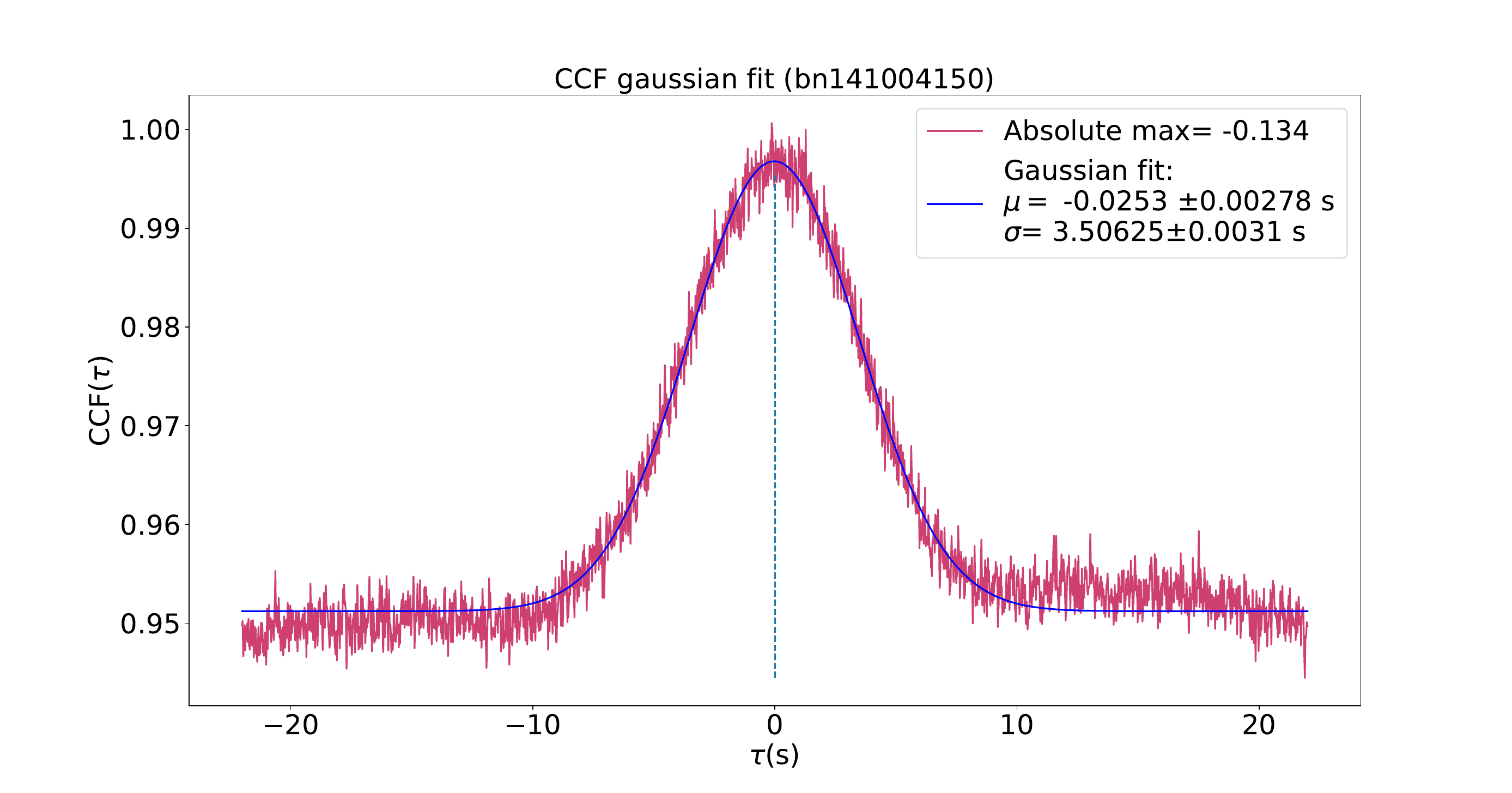}
\end{figure*}

\begin{figure*}[t]
    \centering
    \includegraphics[width=0.49\textwidth]{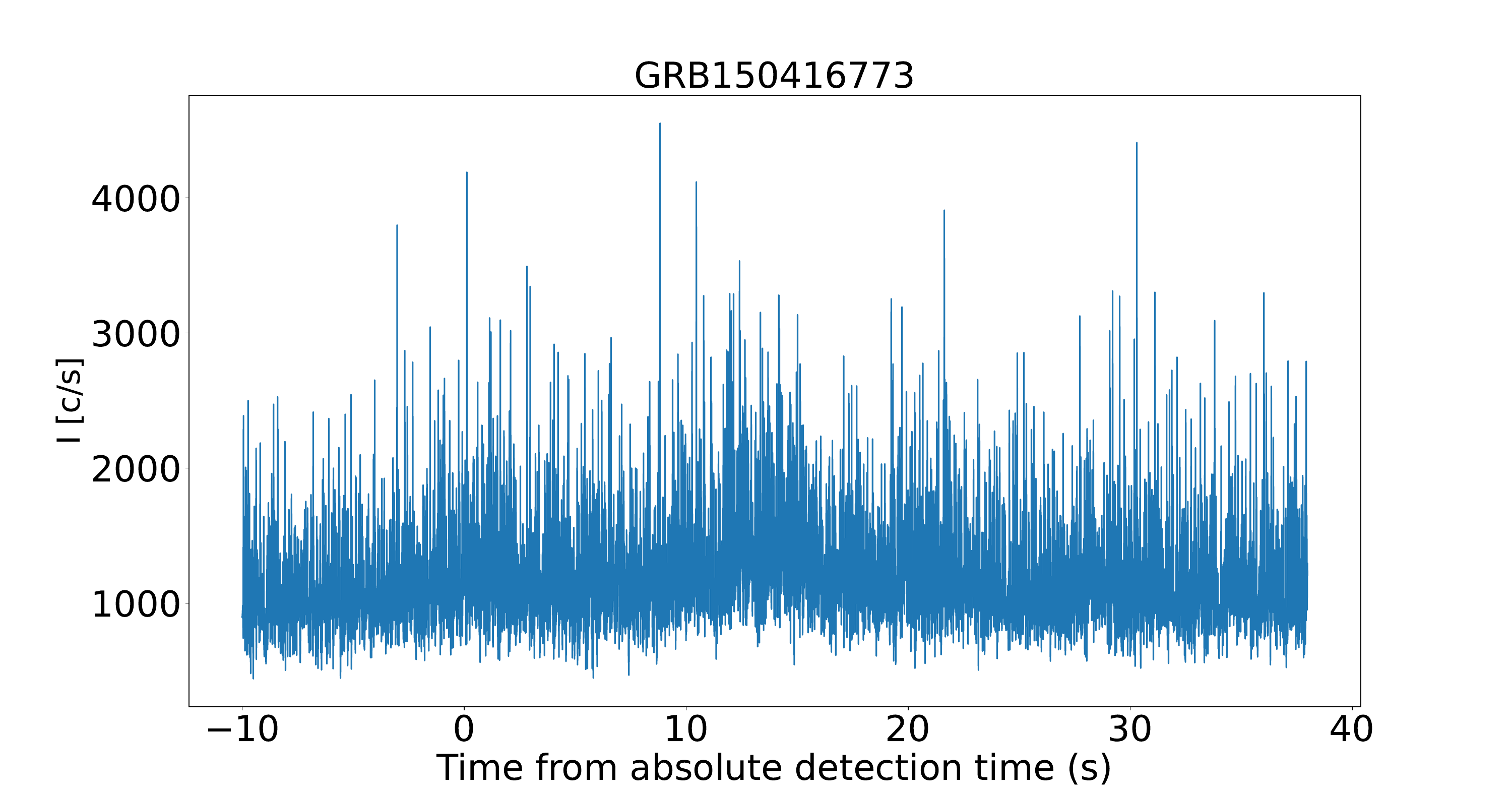}
    \includegraphics[width=0.49\textwidth]{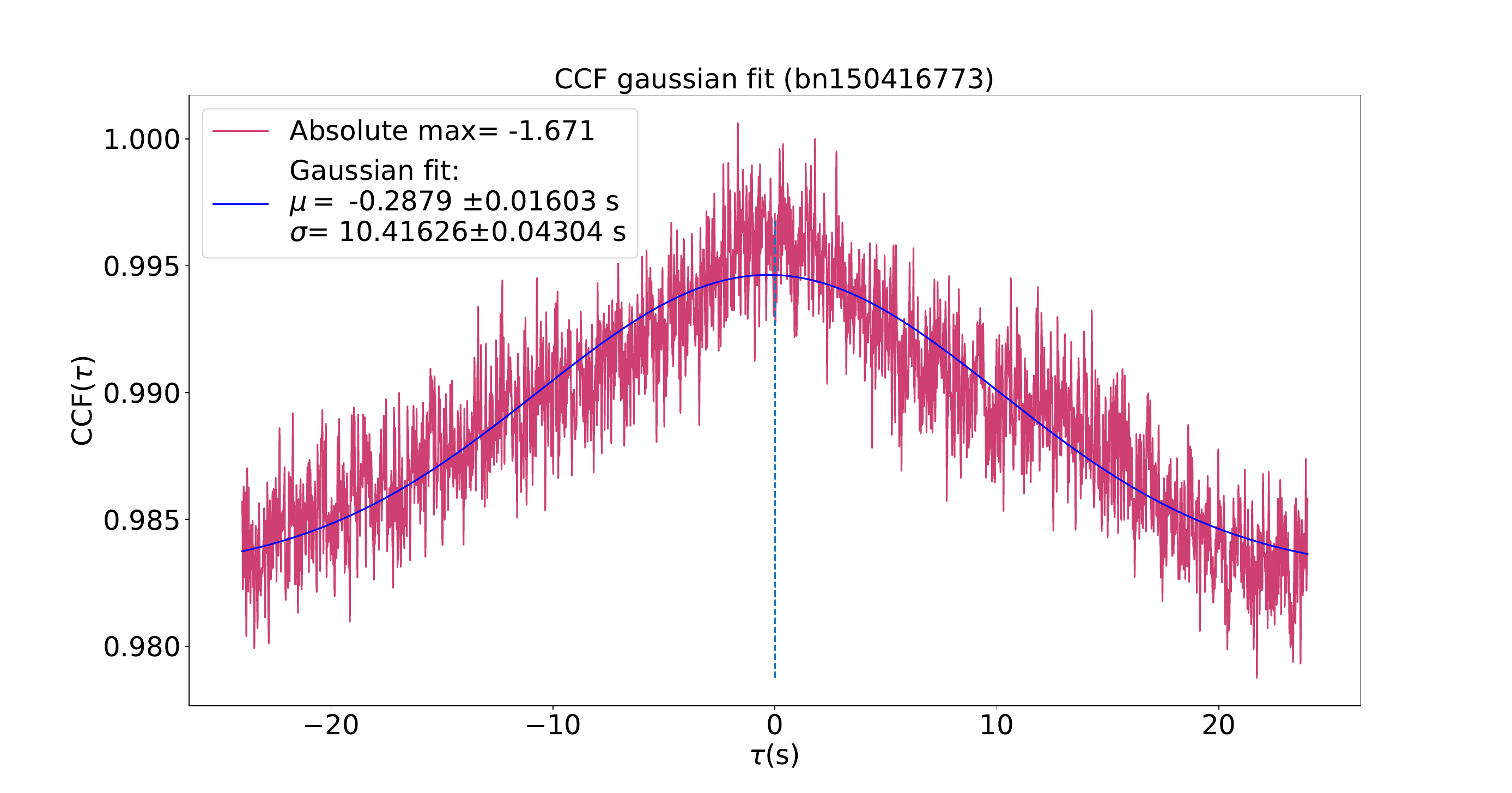}
    \includegraphics[width=0.49\textwidth]{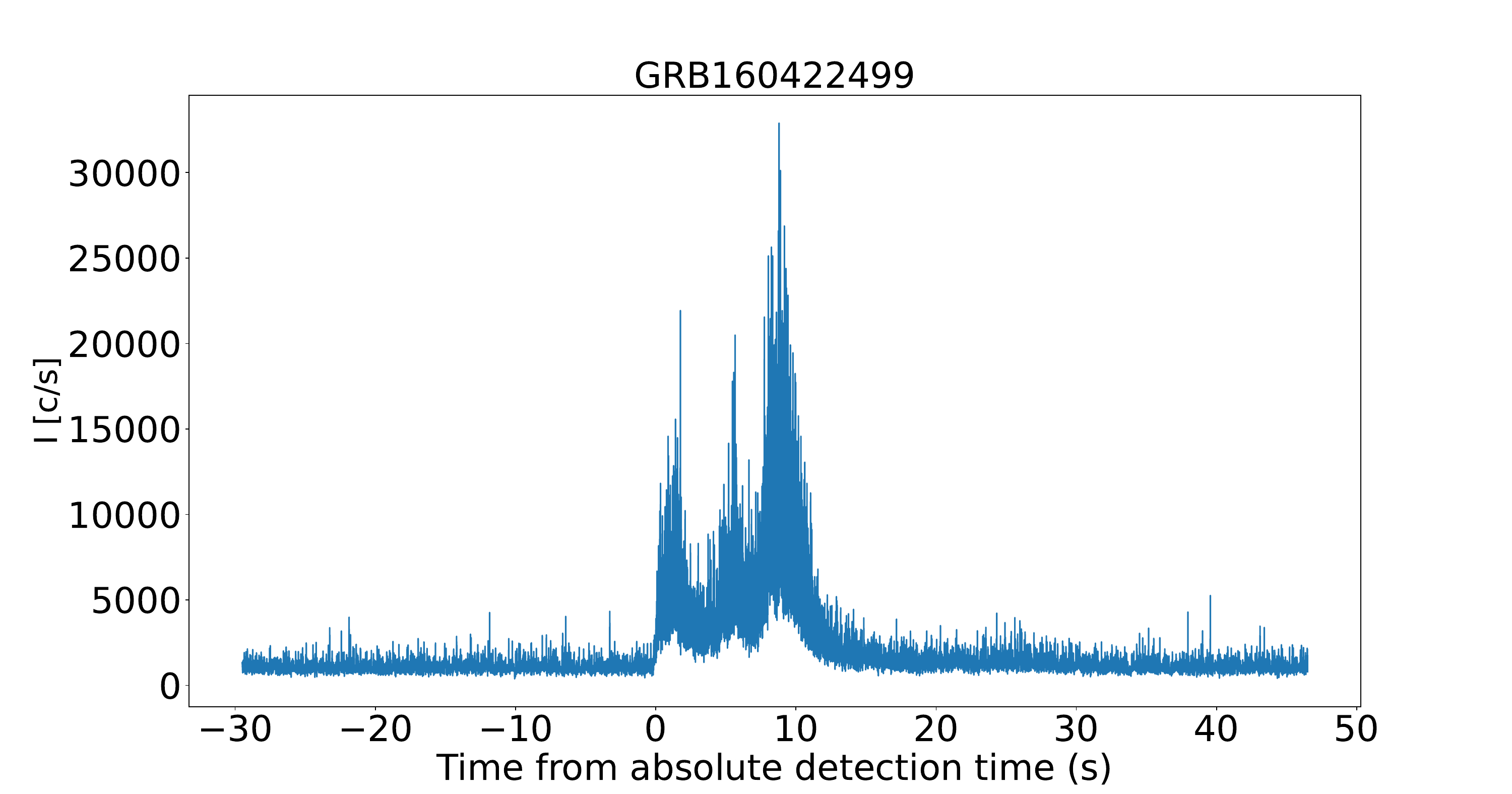}
    \includegraphics[width=0.49\textwidth]{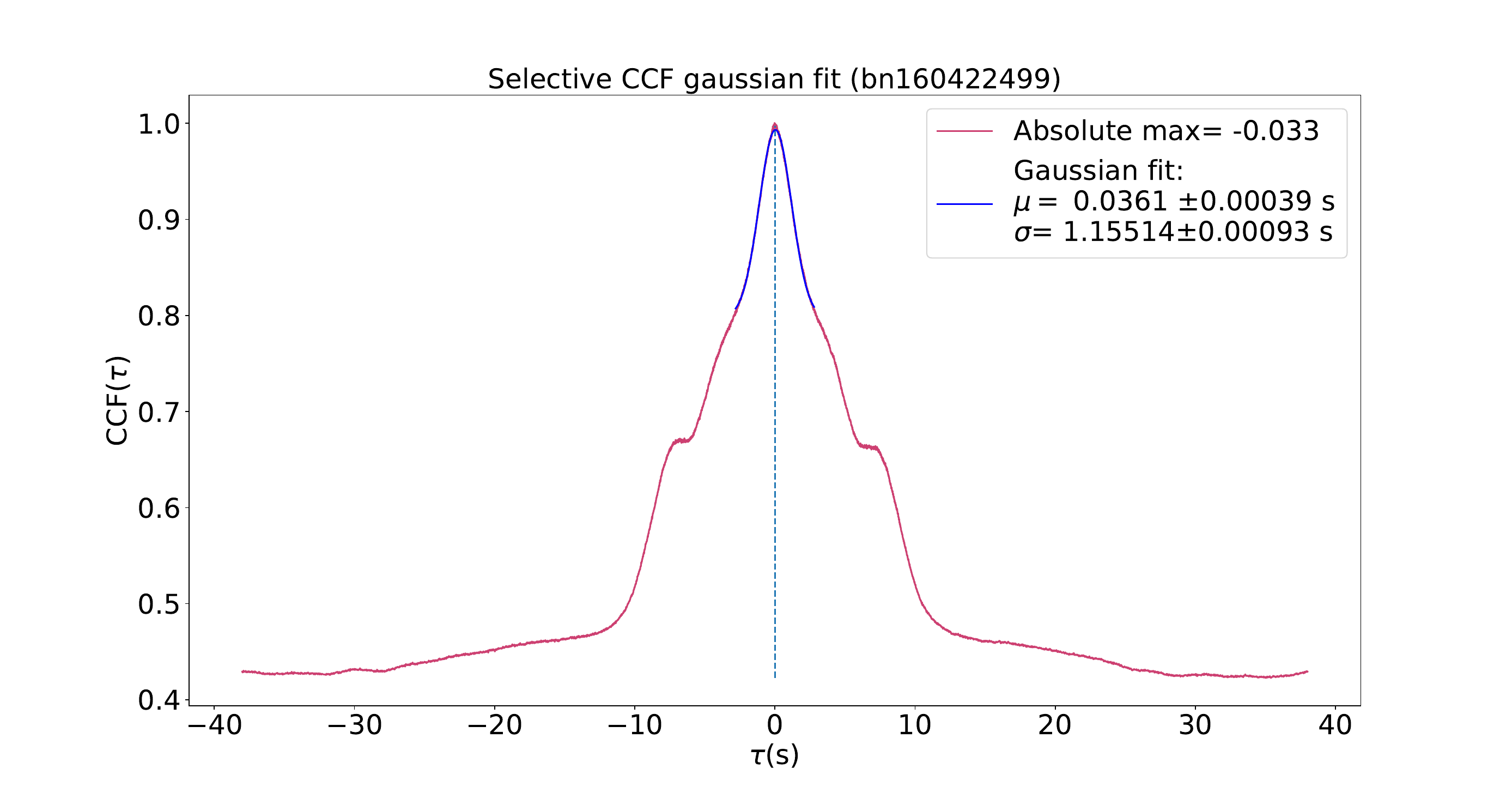}
\end{figure*}

\begin{figure*}[t]
    \centering
    \includegraphics[width=0.49\textwidth]{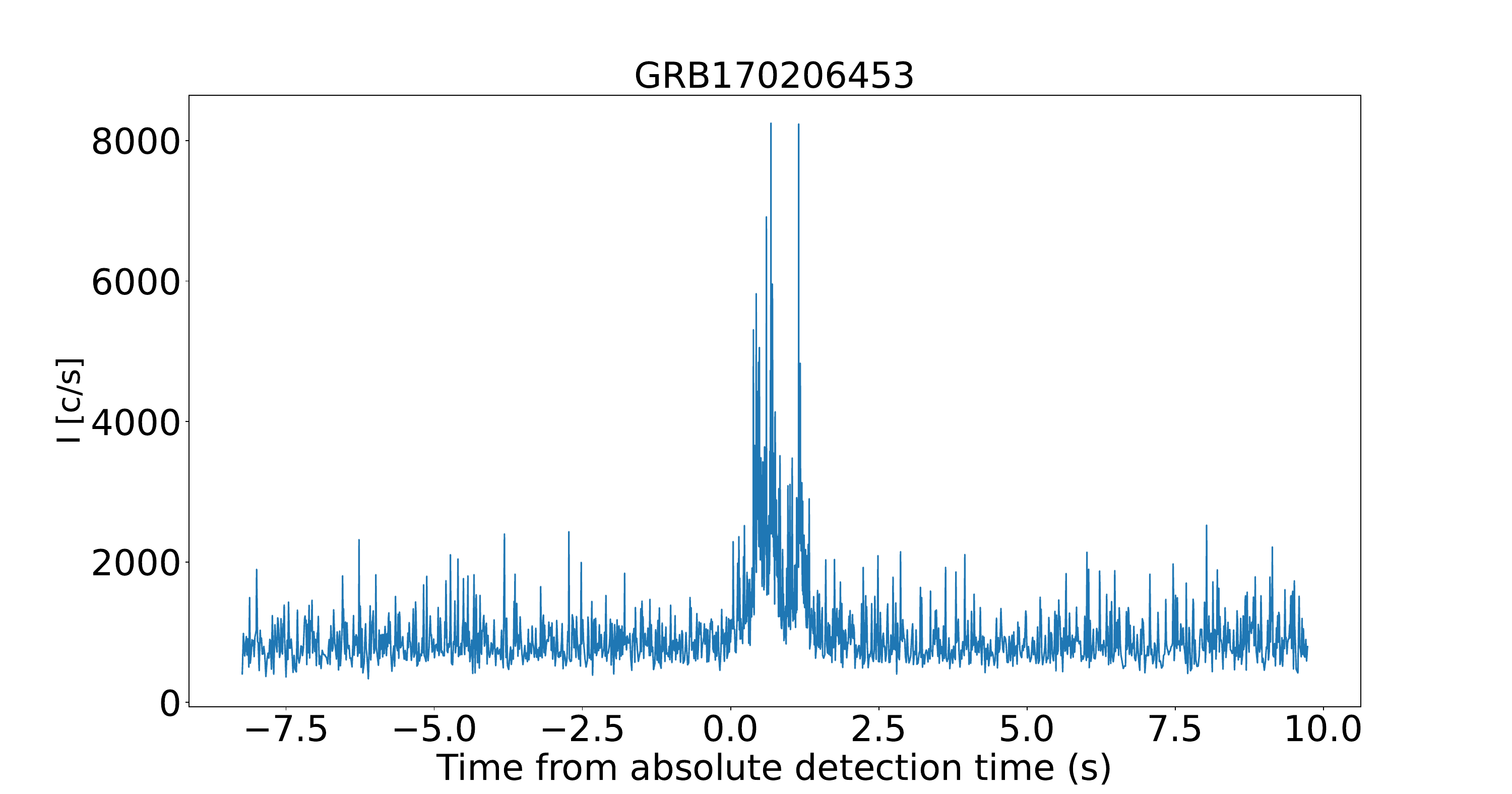}
    \includegraphics[width=0.49\textwidth]{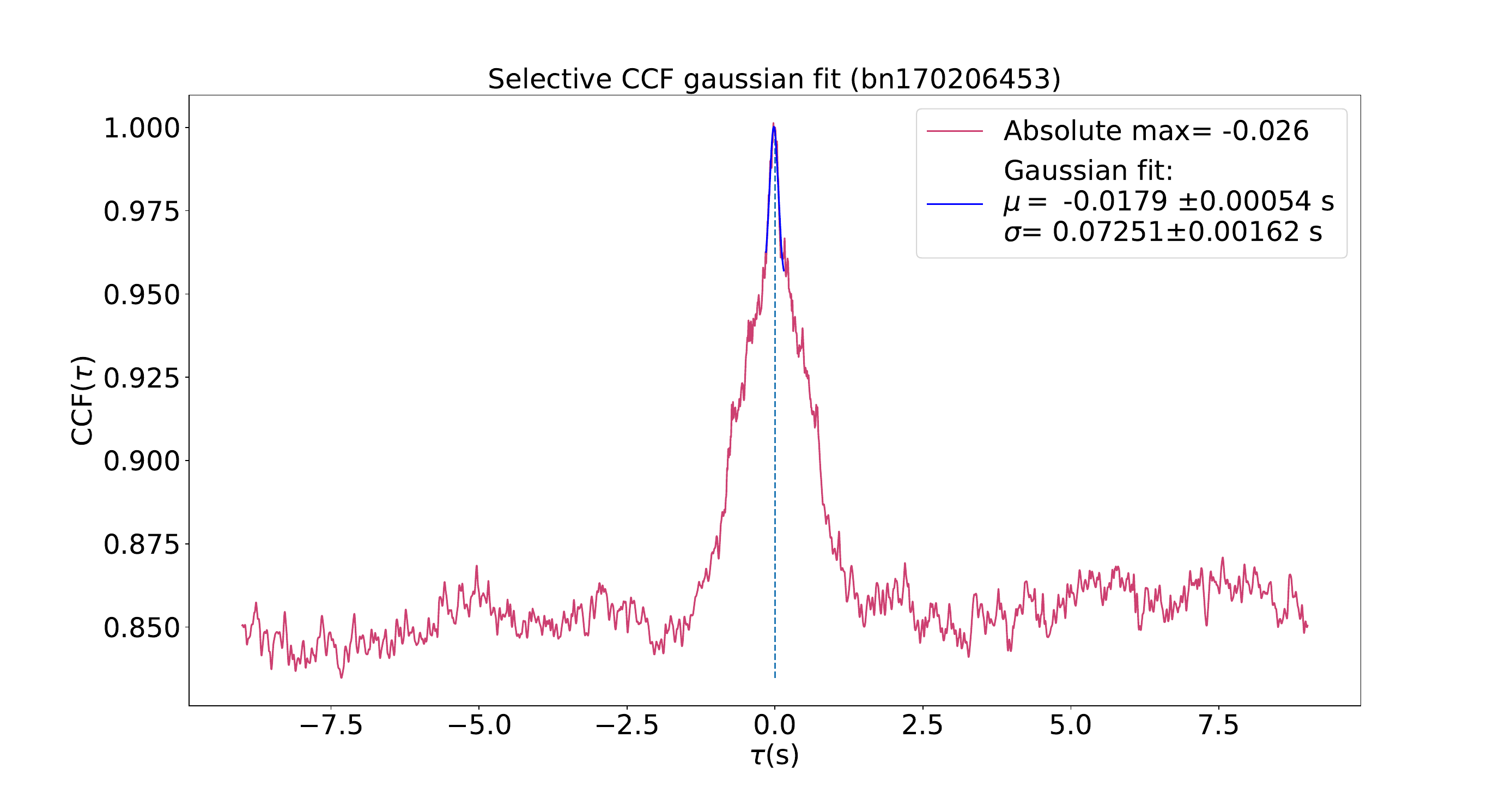}
    \includegraphics[width=0.49\textwidth]{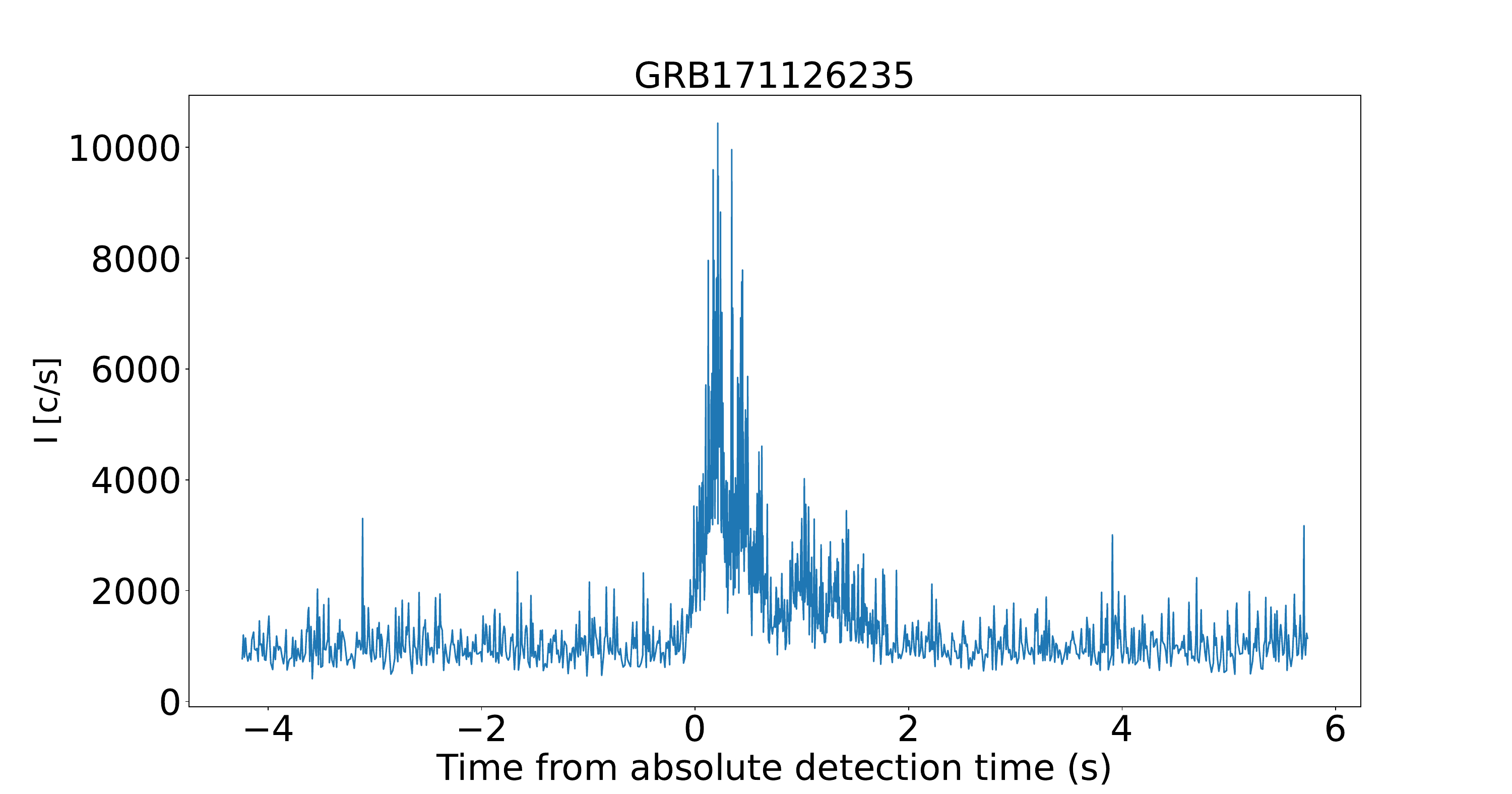}
    \includegraphics[width=0.49\textwidth]{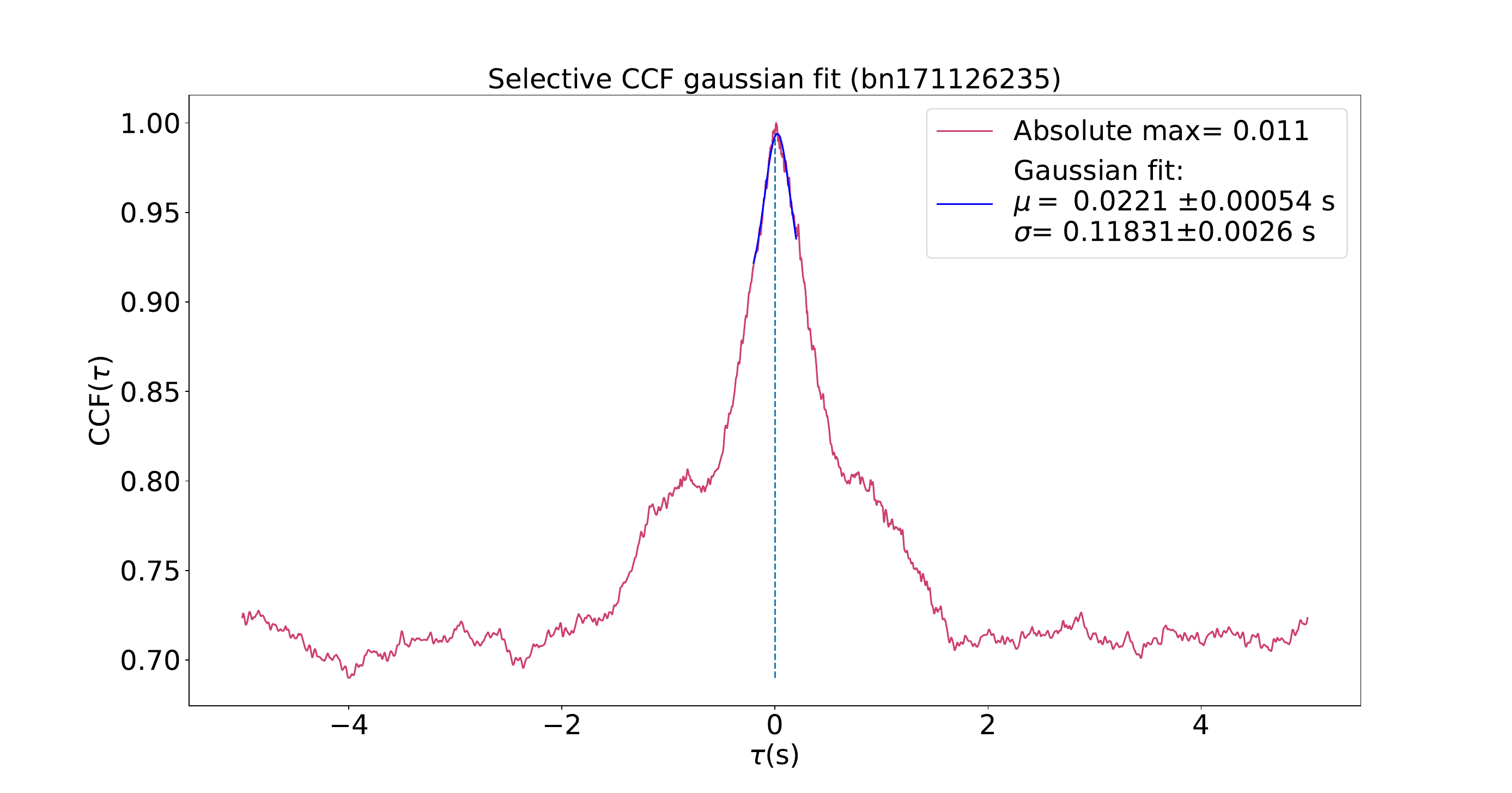}
\end{figure*}

\begin{figure*}[t]
    \centering
    \includegraphics[width=0.49\textwidth]{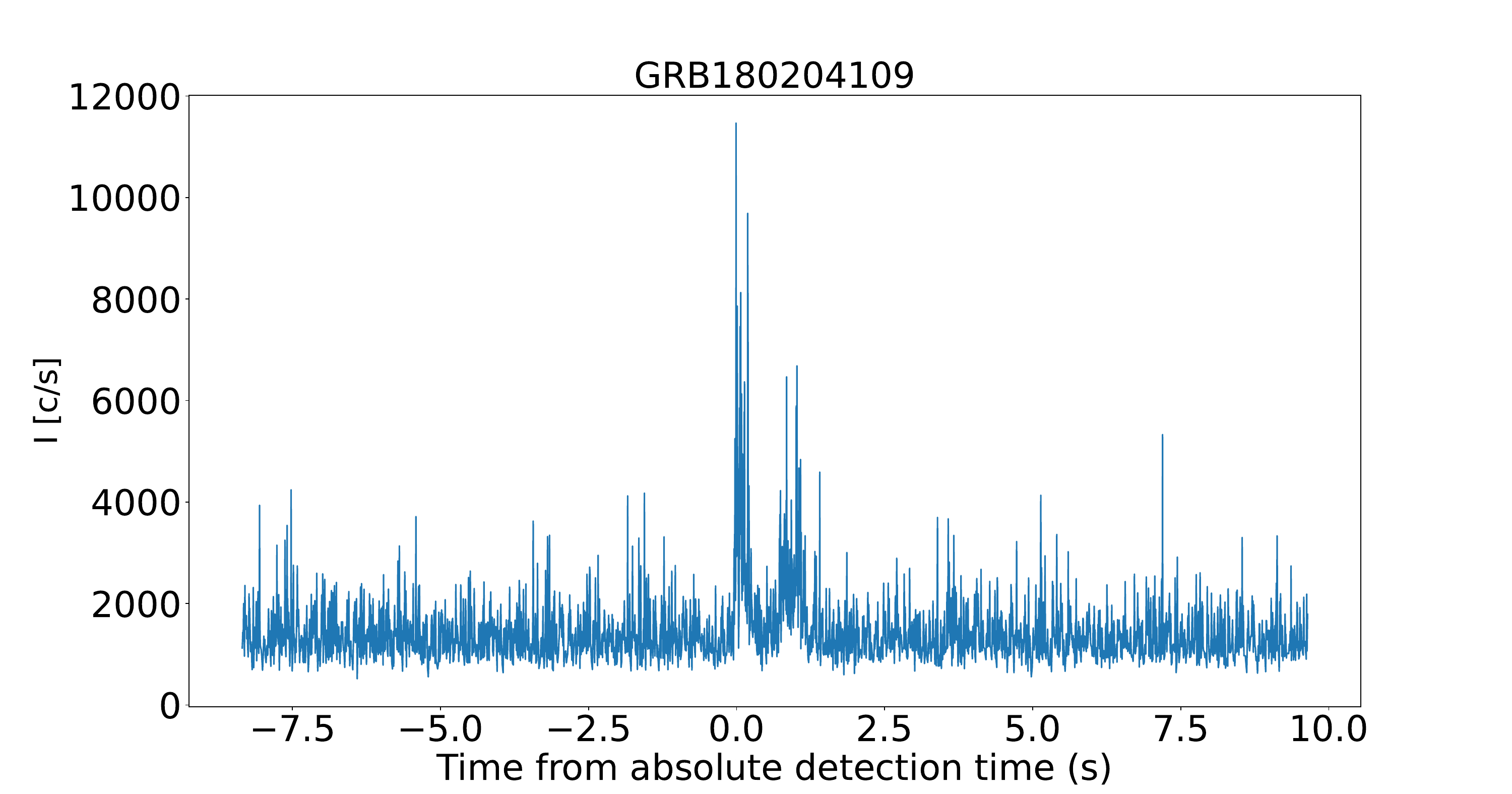}
    \includegraphics[width=0.49\textwidth]{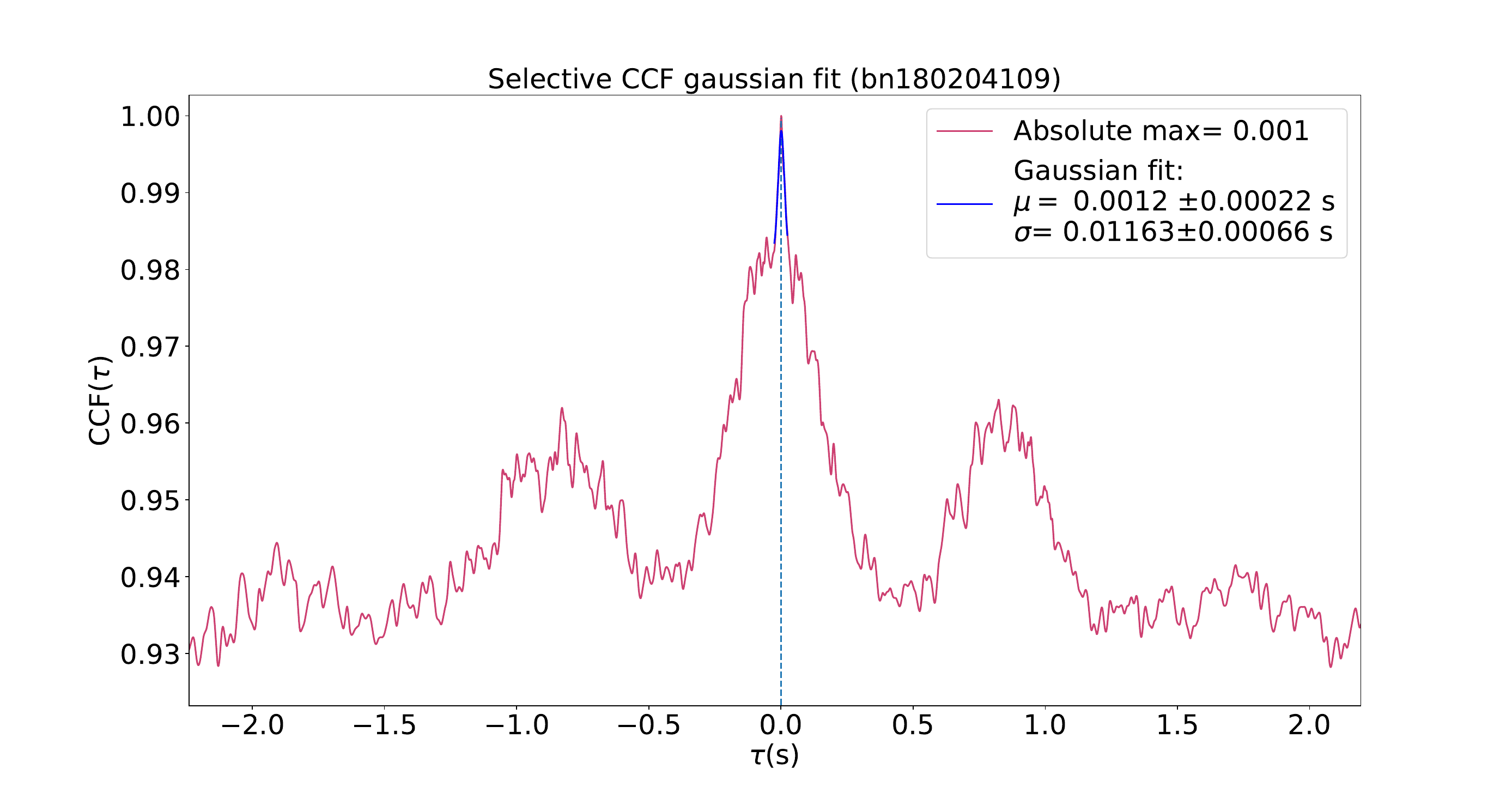}
    \includegraphics[width=0.49\textwidth]{archive/bnexample/bn090820027_lc.pdf}
    \includegraphics[width=0.49\textwidth]{GRB_CCF_examples/CCF_bn090820027.pdf}
\end{figure*}

\begin{figure*}[t]
    \centering
    \includegraphics[width=0.49\textwidth]{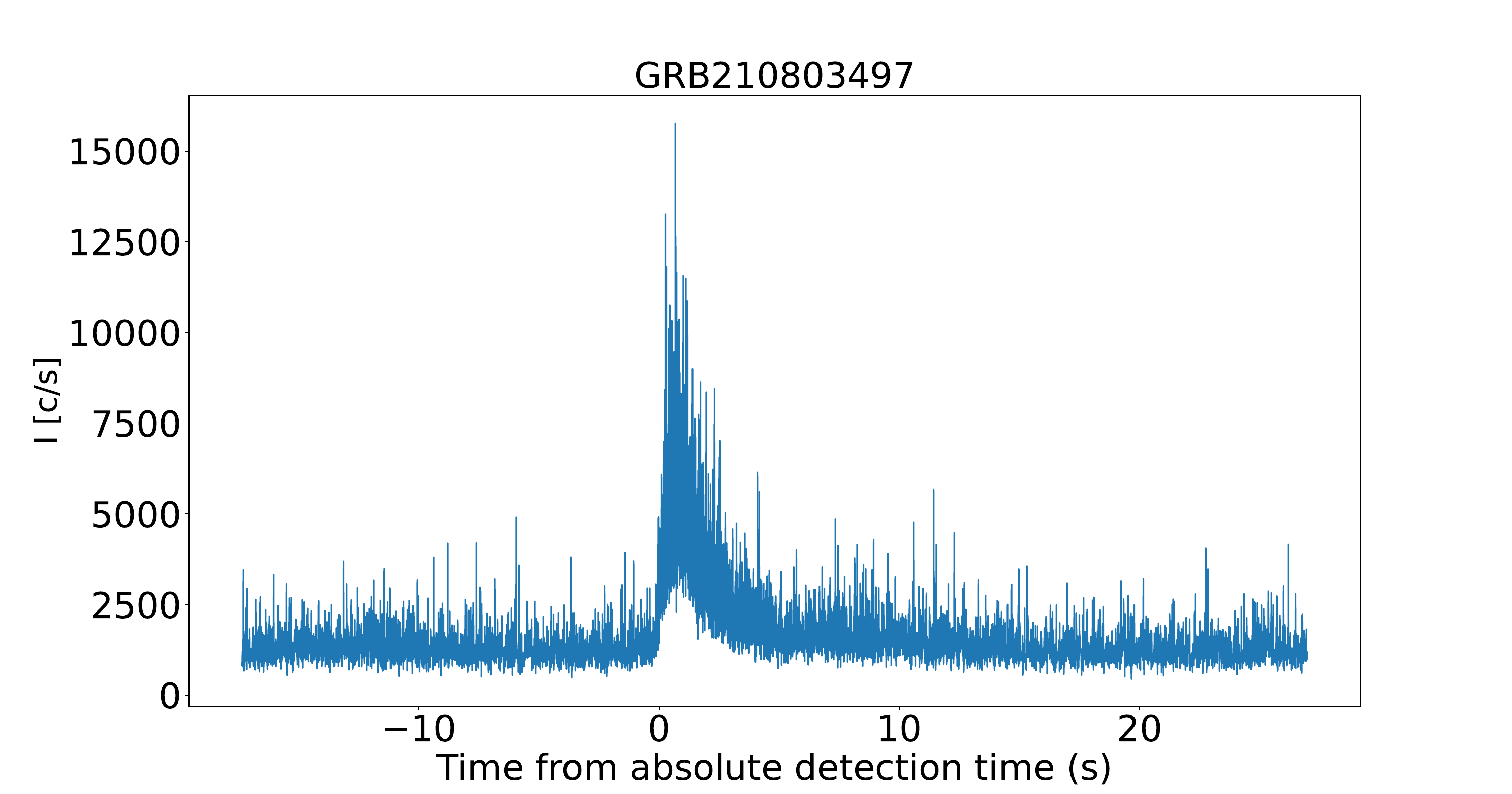}
    \includegraphics[width=0.49\textwidth]{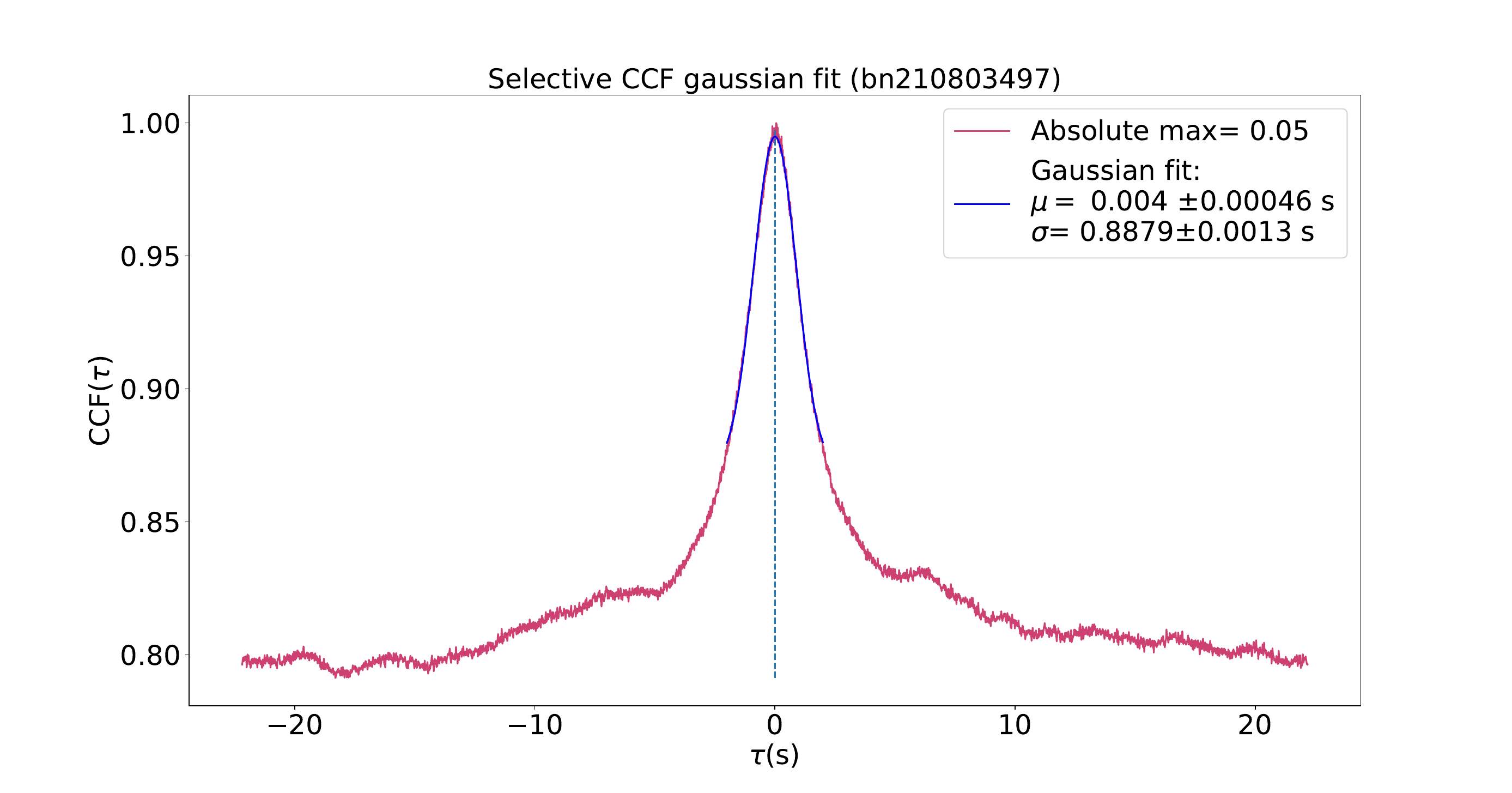}
    \includegraphics[width=0.49\textwidth]{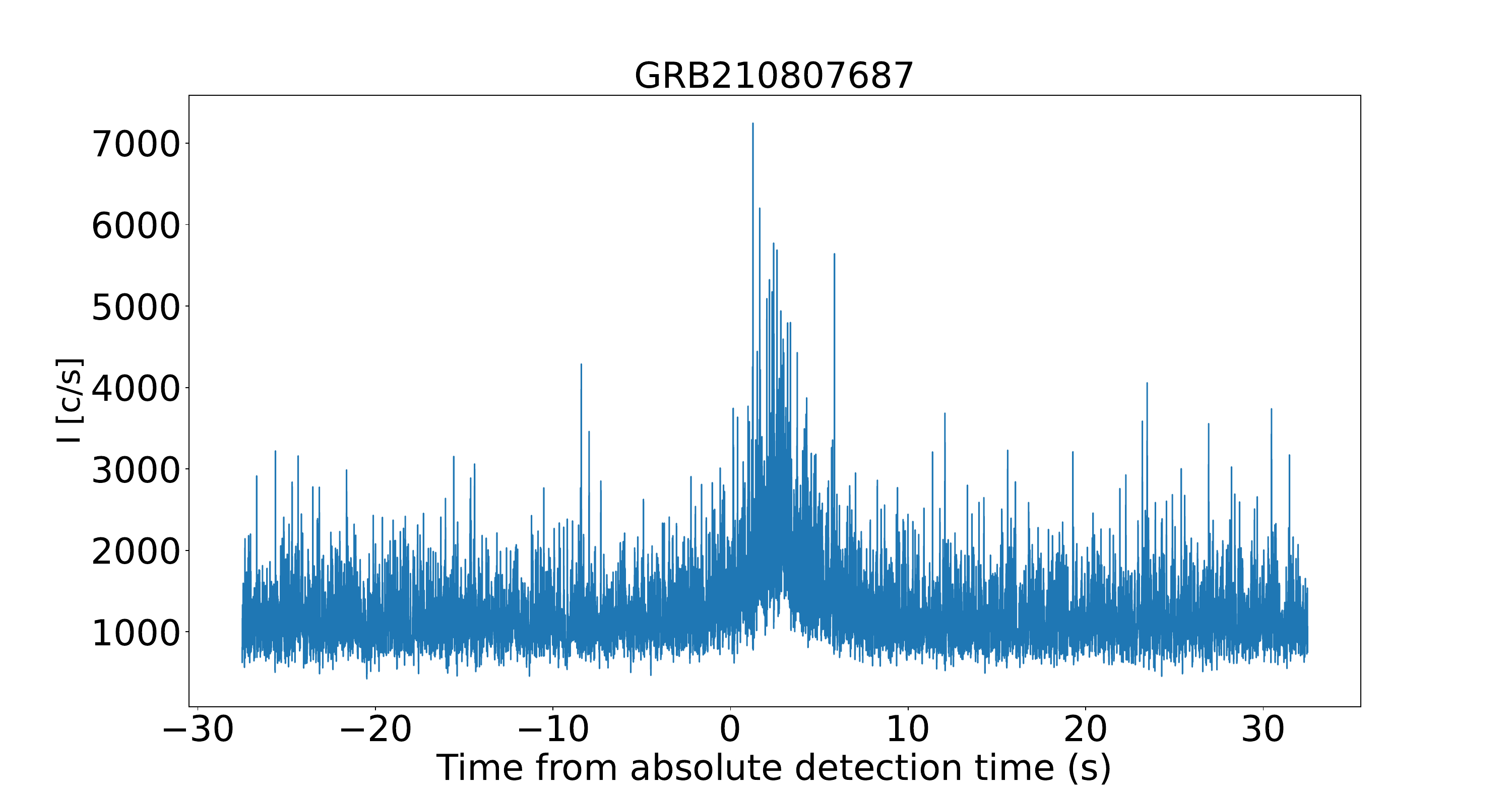}
    \includegraphics[width=0.49\textwidth]{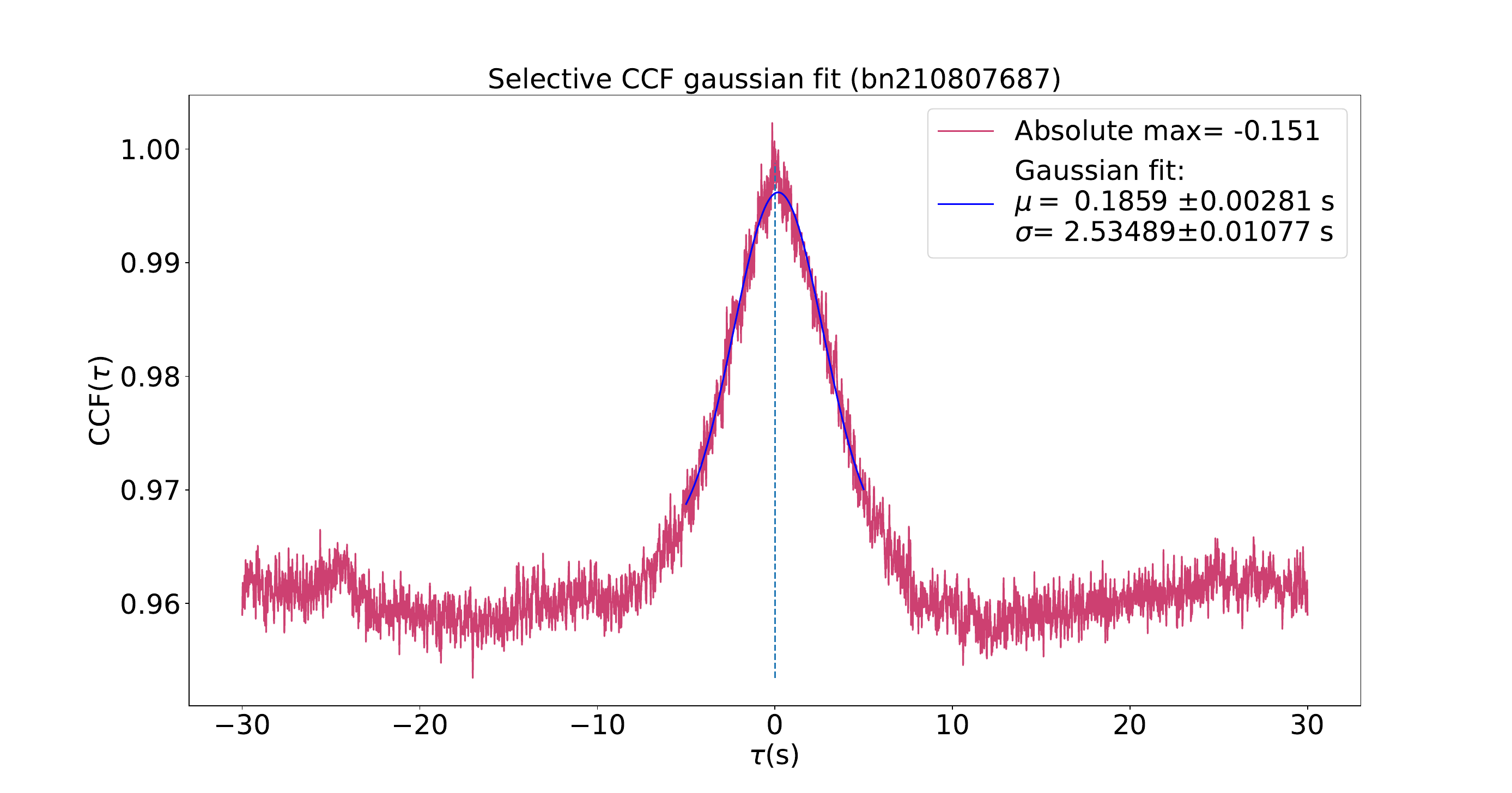}
\end{figure*}

\captionsetup{labelformat=default}

\begin{figure*}[t]
    \centering
    \includegraphics[width=0.49\textwidth]{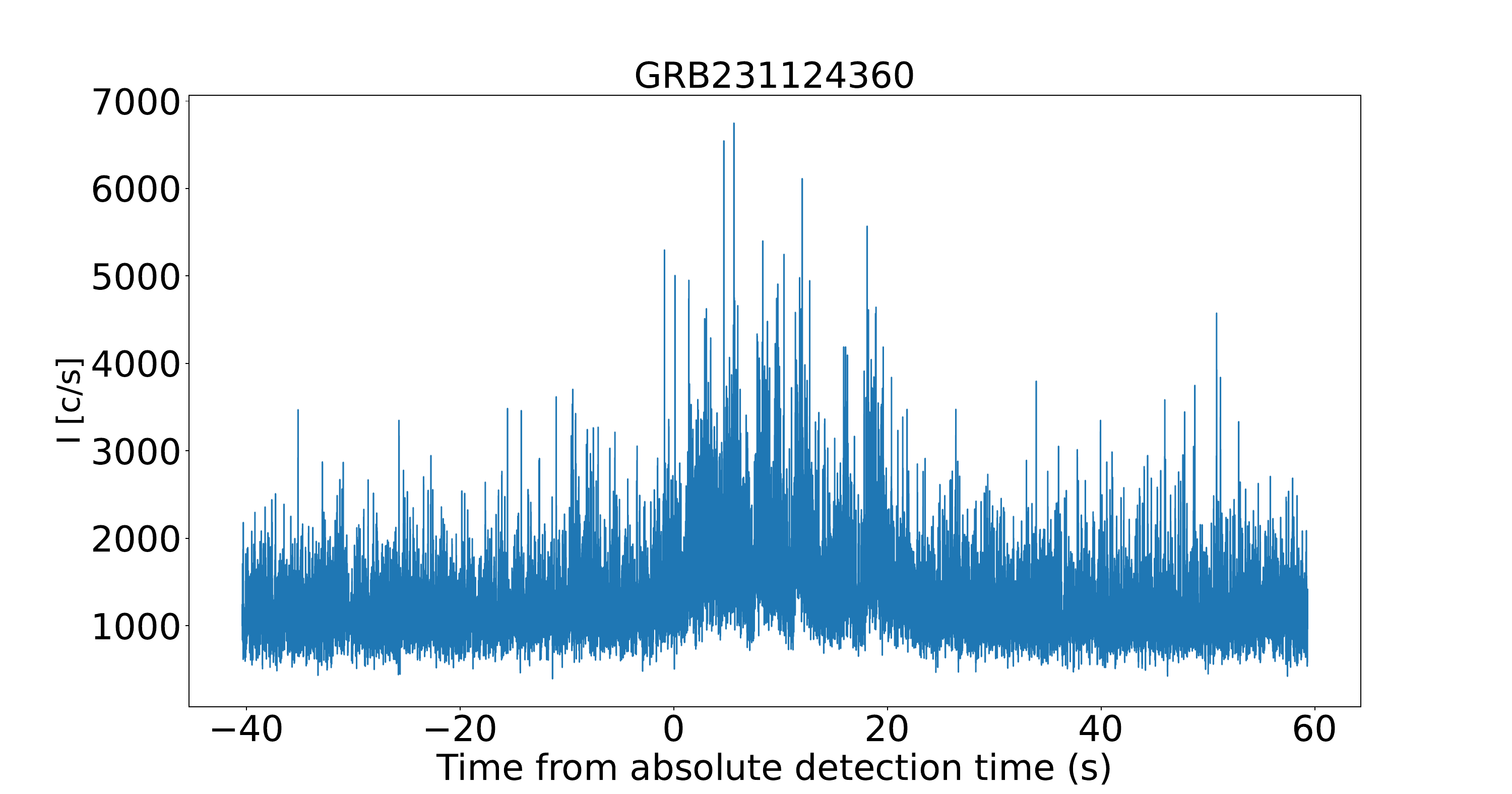}
    \includegraphics[width=0.49\textwidth]{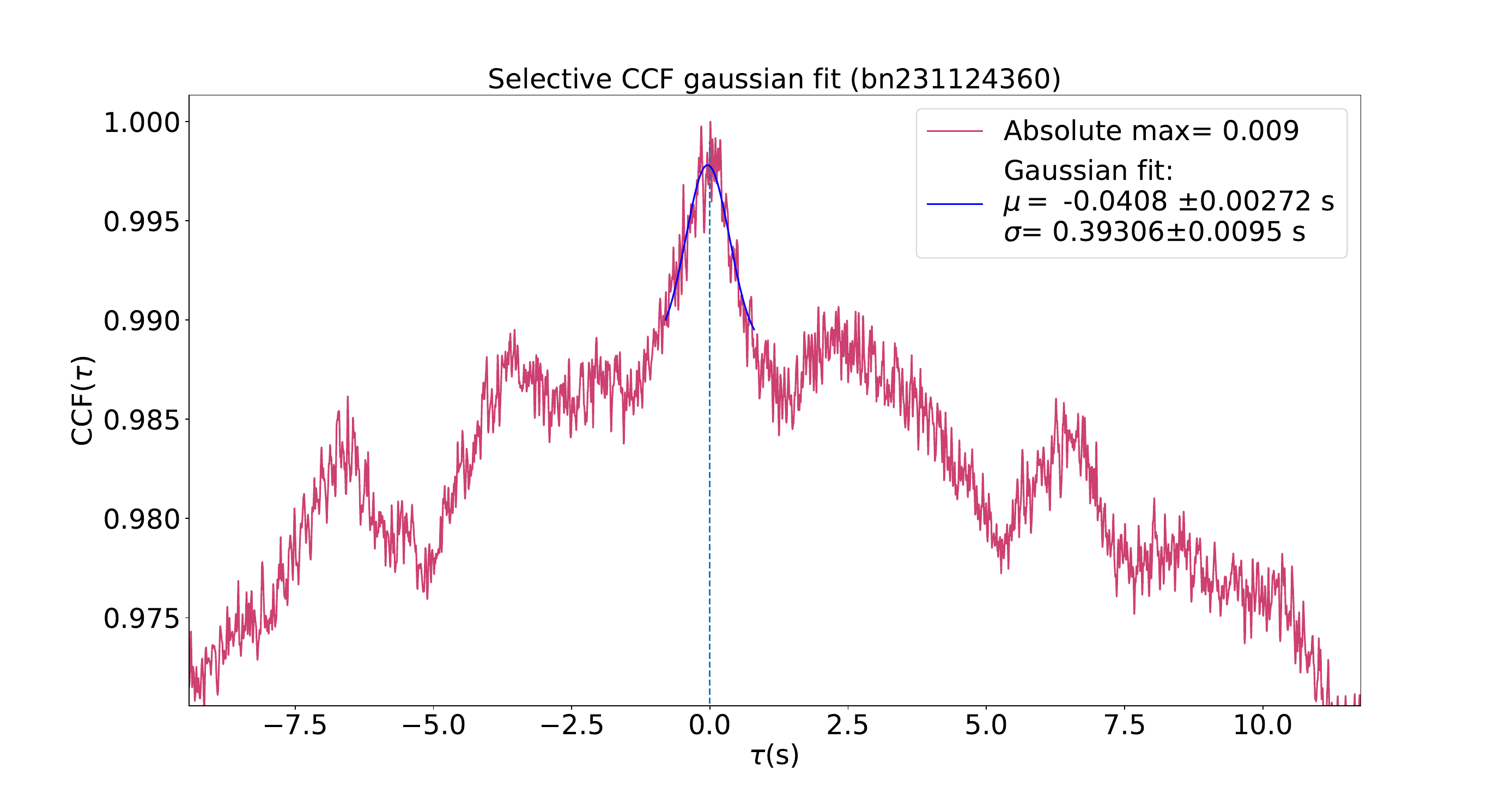}
    \includegraphics[width=0.49\textwidth]{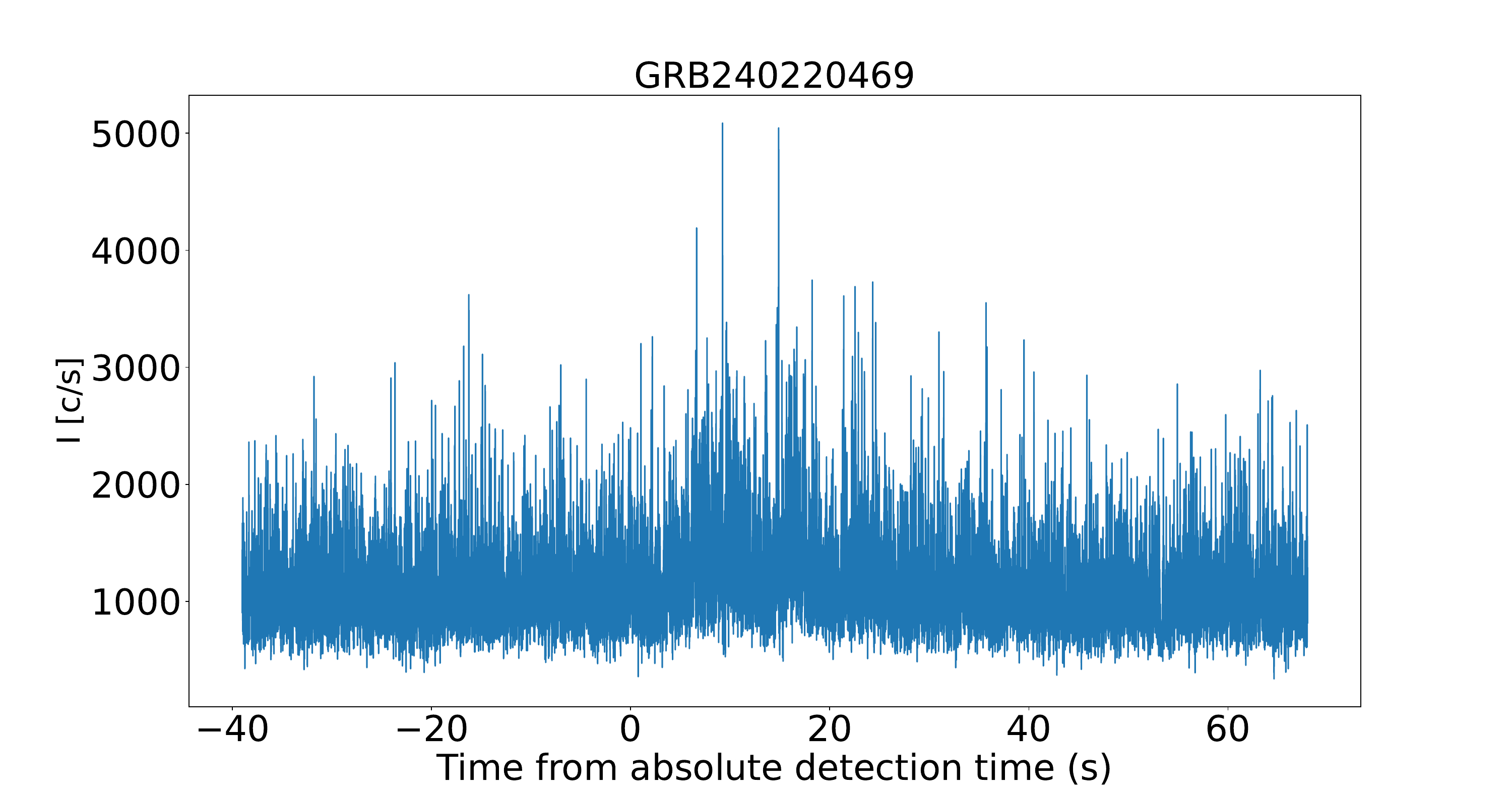}
    \includegraphics[width=0.49\textwidth]{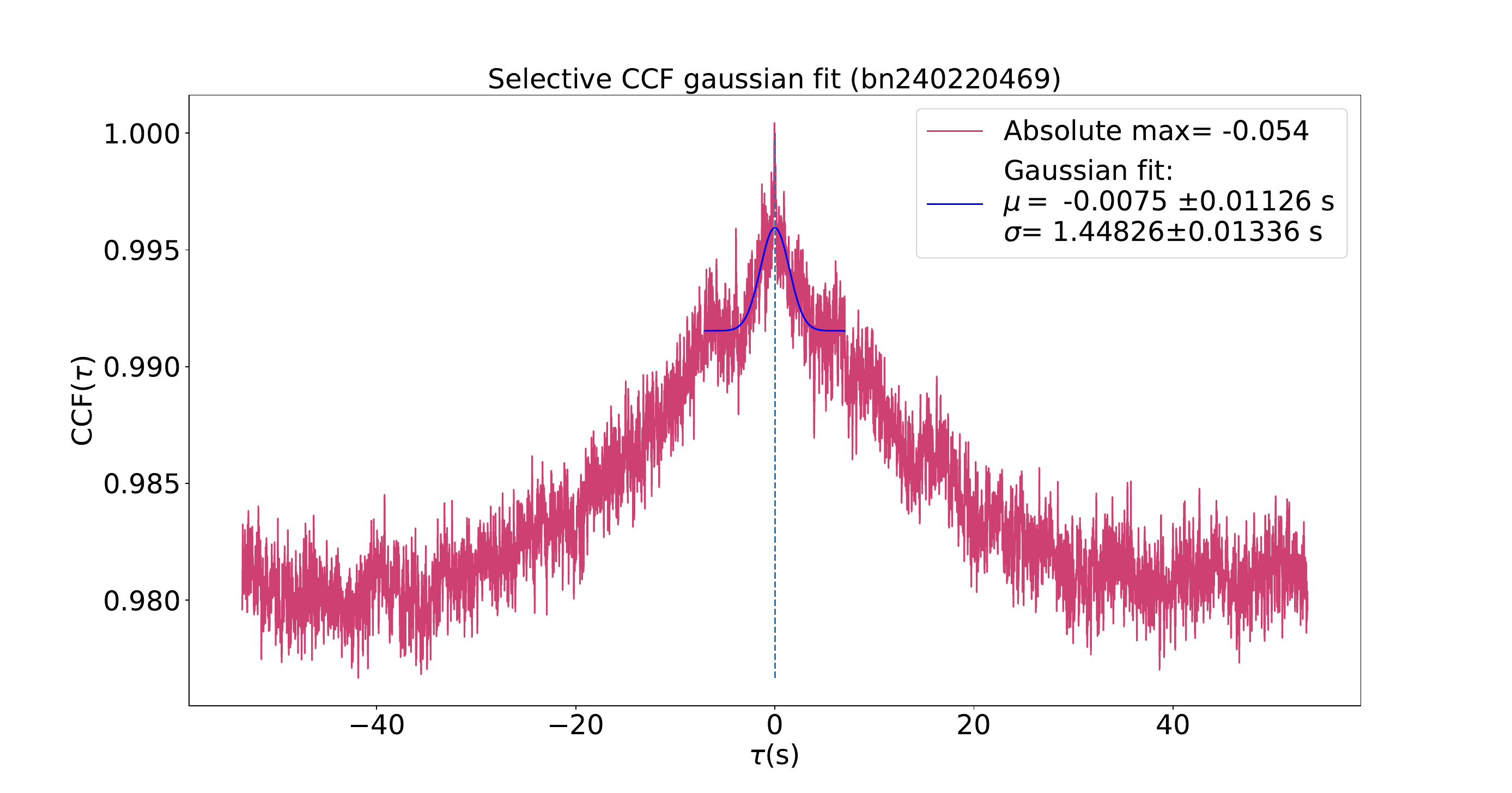}
    \caption{Right panel: Example CCFs performed between ToA lists (see \autoref{subsection:CCF}) obtained via the MDP method (see \autoref{section:MDP_method}). Gaussian fit parameters are highlighted in each plot and fixed for both the MDP and the DP methods testing (see \autoref{section:testing}). ToA lists are retrieved from GRB data as observed by the brightest Fermi/GBM detector monitoring the bursts. Left panel: Light curves from the brightest detector, computed using an adaptive bin size of 10 photons per bin.}
    \label{fig:CCF_grid}
\end{figure*}
\end{appendices}

\end{document}